\documentclass[12pt]{article}

\usepackage{amsfonts}
\usepackage{amssymb}
\usepackage{amsthm}
\usepackage{amsmath}
\usepackage{bbm}
\usepackage{appendix}
\usepackage{array}
\usepackage{booktabs,longtable}
\usepackage{siunitx} 
\usepackage[para,online,flushleft]{threeparttable}
\usepackage{threeparttablex}
\usepackage{color,tikz}
\usepackage{color}
\usepackage{dsfont}
\usepackage{enumerate}
\usepackage{epstopdf}
\usepackage{float}
\usepackage{graphicx}
\usepackage{indentfirst}
\usepackage{longtable}
\usepackage{lscape}
\usepackage{multirow}
\usepackage{multicol}
\usepackage{natbib}
\usepackage{rotating}
\usepackage{setspace}
\usepackage{sectsty}
\usepackage{subcaption}
\usepackage{todonotes}
\usepackage{url}
\usepackage{xcolor}
\usepackage{caption}
\usepackage{subcaption}
\usepackage{titlesec}
\usepackage{times}
\usepackage{mathptmx}
\usepackage{chngcntr}
\usepackage{amsmath,amsfonts, amssymb, dsfont, xcolor, comment}
\usepackage{mathtools}
\usepackage{natbib}
\usepackage{epsfig}
\usepackage{setspace}
\usepackage{graphicx, subcaption}
\usepackage{color}
\usepackage{mdframed}
\usepackage{float}
\usepackage[super]{nth}
\usepackage{tabularx}
\usepackage{ragged2e}
\usepackage{makecell}
\usepackage{placeins}
\usepackage{natbib}
\usepackage{etoolbox}  
\usepackage{setspace}
\usepackage{listings}
\usepackage{changepage}
\newtheoremstyle{boldprop}  
  {\topsep}   
  {\topsep}   
  {\itshape}  
  {}          
  {\bfseries} 
  {.}         
  { }         
  {\thmname{#1} \thmnumber{#2}} 
\theoremstyle{boldprop}

\usepackage{bm}

\newcommand{\Xomit}[1]{}

\usepackage{xcolor}

\usepackage{hyperref}
\usepackage[margin=1.25in]{geometry}
\usepackage[mathscr]{eucal}

\long\def\/*#1*/{}
\hypersetup{colorlinks,linkcolor={blue},citecolor={blue},urlcolor={blue}}

\begin{document}

\newgeometry{top=1.2in, bottom=1.2in, left=0.95in, right=0.95in} 
\title{\LARGE{\textbf{Generative AI for Analysts}}
\thanks{We thank Francesco Grossetti (discussant), Dashan Huang (discussant), Bo Jiang, Da Xu, Tony Xue, attendees at the 2026 European Accounting Association (EAA) Annual Congress and the Machine Learning Application in Financial Economics Conference, and seminar participants at the Xi’an Jiaotong-Liverpool University, Fudan University, and Shanghai Jiaotong University for helpful comments. We thank Xinrui Zhang and Jiaxuan Xiang for their excellent research assistance. Xue acknowledges financial support from the National Natural Science Foundation of China (No.72442014). Zhu acknowledges financial support from Tsinghua University Initiative Scientific Research Program (No.2022Z04W02016) and Tsinghua University School of Economics and Management Research Grant (No.2022051002). All errors are our own.}
}
\vspace{-0.5in}
\date {September 2026}

\author{ \hspace{5mm} 
Jian Xue \footnote{School of Economics and Management, Tsinghua University, xuejian@sem.tsinghua.edu.cn}
\hspace{5mm} Qian Zhang \footnote{School of Management, Xiamen University, qianzh@xmu.edu.cn}
\hspace{5mm} Wu Zhu\footnote{School of Economics and Management, Tsinghua University, zhuwu@sem.tsinghua.edu.cn} 
}
\maketitle
\vspace{-0.3in}

\begin{abstract}
\singlespacing \normalsize
\medskip

{We study how generative artificial intelligence (GenAI) reshapes financial analysts' information production. Using the 2023 integration of GenAI into \textsc{FactSet} as a plausibly exogenous change in AI access, we find that \textsc{FactSet}-associated reports become markedly richer---featuring 26\% more distinct information sources, 24\% broader topical coverage, and 21\% more analytical methods---while also improving timeliness. However, these gains do not uniformly improve decision quality: relative forecast accuracy declines when analysts face greater information-processing demands. Yet, a machine-learning benchmark processing the same observable inputs shows no analogous deterioration, pointing to a human processing constraint rather than poorer underlying information. Placebo tests using other data vendors make a common platform-wide technology trend unlikely. Overall, GenAI relaxes information-acquisition constraints while making human attention a more important bottleneck.}

\bigskip


\end{abstract}

\restoregeometry
\newgeometry{top=1.3in, bottom=1.3in, left=1.1in, right=1.1in} 
\thispagestyle{empty}
\addtocounter{page}{-1}

\newpage
\setstretch{1.4}

\section{Introduction}

Financial analysts play a central role in the functioning of capital markets. As key information intermediaries, they collect, synthesize, and interpret complex and often ambiguous financial data, transforming raw disclosures and market signals into actionable insights for investors and firms. Yet, this critical role is inherently constrained by information processing costs, as analysts must filter vast quantities of data under tight deadlines and cognitive limits. The advent of generative artificial intelligence (GenAI) promises to transform this landscape. By automating data collection and processing, and even report drafting, AI can expand analysts’ access to information and enhance productivity. Indeed, major brokerages have begun deploying AI to streamline workflows, aiming to reduce repetitive tasks and augment analysts’ analytical reach.\footnote{For example, CNBC (January 21, 2025) reports that Goldman Sachs launched an internal GenAI assistant to help analysts and bankers automate data retrieval, financial modeling, and report drafting---reflecting a broader industry shift toward AI-powered research workflows. For details, see \href{https://www.cnbc.com/2025/01/21/goldman-sachs-launches-ai-assistant.html}{the CNBC report}.}
However, whether these tools enhance analysts’ ability to extract value-relevant insights or instead overwhelm them with excessive, unfiltered data remains an open question.

This paper investigates how the integration of GenAI transforms the information production of financial analysts. We ask: \emph{Does GenAI enhance analysts’ productivity and the informational value of their reports, or does it introduce new frictions that erode informational precision?}
This question is both timely and consequential. While practitioners and policymakers celebrate AI as a breakthrough in cognitive automation, regulators have warned that AI could ``introduce new biases in financial decision making'' and obscure accountability in investment advice.\footnote{Former U.S. Treasury Secretary Janet Yellen warned that as artificial intelligence becomes more deeply embedded in the financial services industry, it presents ``significant risks,'' including model opacity, vendor concentration, and the potential for AI to embed or amplify biases in financial decision-making. See the details in \href{see the report}{https://edition.cnn.com/2024/06/05/business/janet-yellen-artificial-intelligence}. The United States Treasury Department announced a call for public comment on the use of AI in financial services, signaling that regulators are intensifying scrutiny of how AI tools may affect market stability and financial-services workflows. See details in \href{see the report} {https://thehill.com/policy/technology/4142101-sec-chair-warns-of-risk-to-financial-systems-from-ai}.} 
Investors, too, are increasingly aware of the risks of relying on AI-generated investment advice \citep{christ2025word, blankespoor2026generative_processing}.
Despite these concerns, prior studies have primarily documented the benefits of AI, showing that it improves information processing across a wide range of market participants \citep{bertomeu2025impact, bradshaw2026generative, cheng2025generative, ecker2025stock, chang2026ai, sheng2026generative}. We aim to bridge the gap between this optimistic tone in the literature and the skepticism expressed by practitioners. To do this, we provide a systematic analysis of how a domain-specific GenAI tool---integrated with proprietary data and tailored to analysts' workflows---affects the structure, quality, and market impact of sell-side research. Our findings reveal a nuanced story: while AI enriches analyst reports, these gains do not uniformly improve decision-making; rather, the net effect depends critically on the analysts' information processing capacity and workload constraints.

Our empirical setting is the December 2023 introduction of \textsc{Mercury}, \textsc{FactSet}'s GenAI platform, which integrates large language models (LLMs) with proprietary financial databases to support auditable natural-language queries, automated visualization and pitchbook creation, and portfolio analysis.\footnote{On November 12, 2024, \textsc{FactSet} announced Intelligent Platform, a further integration of \textsc{Mercury} with other AI-powered solutions to boost efficiency of financial professionals. With \textsc{Mercury} as its core component, the upgrade has become the main offering of \textsc{FactSet} to clients looking to modernize workflows and enhance fact-based decision-making. We provide more institutional background in Section~\ref{sec:institutional_background}.} Because \textsc{Mercury} became available to existing \textsc{FactSet} subscribers at no additional cost, its introduction created a discrete change in the AI capabilities bundled with an established data platform. We exploit this timing in a difference-in-differences design that compares changes in reports associated with \textsc{FactSet} to changes in reports using alternative platforms. Importantly, we observe \textsc{FactSet} citations rather than analysts' direct clicks or prompts in \textsc{Mercury}. Our estimates should therefore be interpreted as the incremental effect of exposure to \textsc{FactSet} after its GenAI integration, rather than as the treatment-on-the-treated effect of verified \textsc{Mercury} use. This distinction also makes our estimates conservative if some control reports use other GenAI tools.

We construct a comprehensive dataset of 46,853 report-firm pairs issued by U.S. analysts between January 2022 and October 2024. To systematically quantify how analysts acquire and process information, we employ \texttt{GPT-4o-mini}, a state-of-the-art language model for the extraction of structured content \citep{hurst2024gpt}. The model parses the full text and embedded objects (tables, figures, and appendices) of each report to identify three key dimensions of information richness: (i) the number and type of distinct information sources (textual, tabular, and visual), (ii) the breadth of topical coverage (firm-, industry-, and macro-level), and (iii) the analytical methods employed (historical analysis, valuation, and forecasting).
Importantly, analysts' explicit source citations allow us to identify whether a report relies on \textsc{FactSet} at the report level. We combine this observable platform citation with the discrete timing of \textsc{Mercury}'s introduction, rather than relying on stylistic AI-detection software to infer GenAI use.\footnote{For simplicity, we sometimes refer to post-introduction reports associated with \textsc{FactSet} as ``AI-assisted reports.'' This label denotes exposure to \textsc{FactSet} after its AI integration, not verified use of \textsc{Mercury} in every report. Some non-\textsc{FactSet} reports may also use generic AI tools such as ChatGPT or Gemini, which would tend to attenuate the estimated contrast. Sections~\ref{sec:institutional_background} and \ref{sec:data}, together with Online Appendix~\ref{app:validate_factset}, discuss and validate this treatment proxy.} The resulting combination of structured report measures and observable platform citations allows us to study comprehensively how access to a domain-specific, AI-integrated data platform changes analysts' information production.

We first describe the adoption patterns and trends of the \textsc{FactSet} platform within our sample period. Prior to its AI integration, \textsc{FactSet} is primarily used by analysts employed at smaller brokerage firms and those with less experience, consistent with \textsc{FactSet} being a budget-friendly choice relative to more expensive vendors such as \textsc{Bloomberg}. Further, analysts who graduated from elite universities and those covering larger portfolios are more likely to use \textsc{FactSet}. After the AI integration, we observe a marked increase in both the prevalence and intensity of \textsc{FactSet} usage, proxied respectively by the percentage of reports citing \textsc{FactSet} and the share of citations to \textsc{FactSet} among all platform citations in the reports. Both these percentages rose from a pre-AI level of approximately 25\% to nearly 50\% following the integration. Notably, analysts from smaller brokerage firms and those working with larger teams disproportionately increased their use of \textsc{FactSet}, while other analyst characteristics remained relatively stable around the integration date.

We next examine how report content changes after \textsc{FactSet}'s AI integration. Our baseline specification absorbs report-date shocks, firm-year and broker-year changes, and time-invariant analyst characteristics. Relative to other reports, post-integration reports associated with \textsc{FactSet} exhibit significant increases in information sources, topical coverage, and analytical methods. The magnitudes range from 21\% to 26\% relative to pre-integration levels, with the largest increases in figure sources, industry topics, and forecasting methods, closely aligned with \textsc{Mercury}'s visualization, information-retrieval, and analytical capabilities. We interpret these estimates as evidence that the AI integration expanded analysts' effective information set and productivity, while recognizing that \textsc{FactSet} association is a proxy for exposure rather than direct observation of \textsc{Mercury} usage.

Does greater informational richness translate into better decisions? Post-integration \textsc{FactSet} reports are more likely to be published within three days of earnings announcements, consistent with faster information production. Average relative forecast accuracy, however, declines. Importantly, this decline is not uniform: the subsequent tests show that it is concentrated among analysts facing greater information-processing constraints. These reports also generate less abnormal trading volume.

We organize the mechanism evidence around a simple idea: GenAI relaxes constraints on information acquisition, but it does not relax the scarcity of human attention needed to verify and process that information. We first show that post-integration \textsc{FactSet} reports contain less aligned positive and negative signals and that lower signal alignment is associated with lower forecast accuracy. We then ask when this information-processing problem should bind. The deterioration in accuracy and signal alignment is concentrated among analysts who cover more complex portfolios, work with smaller teams, or revise multiple firms' forecasts on the same day; the effects are weak or absent when these processing demands are low. Thus, signal misalignment is best viewed as a feature of the information set that makes processing difficult, while limited processing capacity is the underlying economic constraint.

A machine-learning benchmark provides an additional separation between information quality and human processing capacity. We train a random forest to predict earnings using the same observable report signals, firm fundamentals, and recent consensus forecasts available to analysts. The machine performs slightly better than human analysts on average, and, crucially, its forecast accuracy does not deteriorate for post-integration \textsc{FactSet} reports or in the high-workload subsamples where human accuracy falls. Because the machine is not subject to human attention constraints, this result is difficult to reconcile with a simple deterioration in the underlying information. Additional tests also find no systematic increase in textual redundancy or decline in readability, and the accuracy deterioration is not concentrated in the most timely reports. Together, the evidence favors an information-processing channel over mechanical noise or a ``speed-over-review'' explanation.

We supplement our main analyses with several robustness tests. First, we conduct placebo tests using \textsc{Bloomberg}, \textsc{Refinitiv}, and \textsc{Capital IQ} as pseudo-treatments. We find no shift in report content or quality for these alternative platforms. These null effects do not by themselves prove direct \textsc{Mercury} usage, but they make it unlikely that our baseline results simply reflect contemporaneous improvements common to major financial data vendors or concurrent use of alternative platforms.
Second, to further mitigate endogeneity concerns, we replicate our main results by (i) including finer fixed effects—such as analyst-month, firm-month, broker-month, and analyst-firm fixed effects—and using alternative clustering methods; (ii) reweighting \textsc{FactSet} and non-\textsc{FactSet} reports using entropy balancing; (iii) redefining exposure to \textsc{FactSet} at the analyst or broker level instead of at the report level; and (iv) controlling for report length. Third, we show that our main inference remains unchanged across different measures of forecast accuracy.
Taken together, these results show that the documented patterns are robust to alternative specifications, samples, and measurements, and materially reduce---though cannot eliminate---concerns about time-varying confounding and treatment misclassification.

Our paper contributes to three distinct strands of literature. First, we contribute to the growing literature on how GenAI affects information processing in financial markets and knowledge work more broadly \citep{bertomeu2025impact, bradshaw2026generative, blankespoor2026generative_processing, cheng2025generative, cai2025ai, chang2026ai, ecker2025stock,sheng2026generative}. The prevailing evidence in this literature documents the benefits of AI adoption: improved forecast frequency, enhanced trading returns, and greater market efficiency. Our paper offers a counterpoint by showing that these benefits are not unconditional. We focus on a setting where analysts face significant attention constraints and find that, under such conditions, AI can reduce forecast accuracy rather than improve it. A second, and equally important, distinction is that prior studies largely examine generic, widely available AI tools such as ChatGPT. By contrast, we study a domain-specific AI platform integrated with proprietary financial and market data, tailored to the specific needs of sell-side analysts. This is precisely the type of proprietary AI capability that practitioners identify as lacking in their organizations but critically needed to generate differentiated insights \citep{christ2025word}. To our knowledge, we are the first to document the effects of such a domain-specific, data-integrated AI tool on the structural dimensions of analyst research production.

Our findings also help reconcile the seemingly conflicting results in the literature regarding AI's effect on analyst performance: while some studies find that the availability of AI tools improves forecast accuracy and investment returns \citep{bertomeu2025impact, sheng2026generative}, survey evidence indicates that a majority of analysts and investors worry about accuracy and bias in AI-generated outputs \citep{christ2025word, blankespoor2026generative_processing}.\footnote{\citet{christ2025word} find that 58\% of professional analysts remain reluctant to adopt GenAI in their research; crucially, while only 9\% report that using AI improves forecast accuracy, 55\% express ``concerns about accuracy or bias.'' Similar patterns emerge among investors: \citet{blankespoor2026generative_processing} document that 54\% of investors who use GenAI to process information are concerned about reliability and accuracy, and 55\% of non-users are uncertain or skeptical of GenAI's ability to improve processing.} We do not claim, nor do we find, that GenAI uniformly decreases forecast accuracy. Rather, we show that this decline is conditional on analysts facing severe attention constraints. Moreover, while generic AI tools are primarily used to streamline routine tasks like text summarization, we focus on a different type of tool: a domain-specific AI incorporated into proprietary databases that can supply analysts with additional data, aid in information discovery, and thus affect their information processing costs differently. Our findings thus qualify the prevailing optimism in the AI literature and point to the importance of human cognitive capacity as a boundary condition for AI-driven productivity gains.

Second, we contribute to the literature on limited attention and information processing costs in capital markets \citep{sims2003implications, hirshleifer2003limited, blankespoor2020disclosure, hirshleifer2019decision, driskill2020concurrent}. With the popularization of AI, an open question in this literature is whether AI alleviates or exacerbates the cognitive constraints imposed by limited attention. Our findings provide a clear answer: AI does not magically resolve attention constraints. On the contrary, when analysts face high workload and information demands, AI-generated information compounds their cognitive burden, leading to lower forecast accuracy. The same AI tool that enhances report richness and timeliness under routine conditions can become a source of distraction when attention is strained. By documenting this heterogeneity, we show that the net effect of AI on decision quality depends critically on the cognitive state of the user. Moreover, we show that this overload spills over to investors, who exhibit weaker market reactions to AI-assisted analyst reports, consistent with the idea that the cost of processing complex, mixed signals extends beyond analysts to the broader market.

Third, we contribute to the analyst forecasting literature by highlighting a dimension that has received relatively little attention: the role of third-party data platforms in shaping analysts' research production and performance. Classic studies have focused on analyst characteristics such as experience, ability, incentives, and career concerns \citep{clement1999analyst, hong2003analyzing, harford2019analyst}. However, analysts do not work in isolation; they rely on a wide array of technological tools and data platforms to perform their tasks. With the rise of AI, professionals are increasingly looking to third-party vendors to bridge the gap between AI capability and proprietary databases \citep{christ2025word}. Our setting, centered on the introduction of an AI-powered data platform, offers a unique opportunity to examine how these platforms transform analysts' information acquisition, report structure, and ultimately their forecast accuracy and timeliness. We show that data platforms are important determinants of how analysts discover, organize, and communicate insights. By focusing on this overlooked institutional actor, we open a new avenue for understanding the technological infrastructure that underpins financial information intermediation.

Our results also sharpen the managerial implication of AI adoption. As AI lowers the cost of information acquisition, the bottleneck can shift from finding information to verifying, prioritizing, and synthesizing it. The implication is not that managers should use less AI, but that AI deployment should be paired with workflow design that protects scarce human attention: limiting simultaneous coverage demands when possible, strengthening review protocols when AI surfaces conflicting evidence, and designing interfaces that rank and reconcile signals rather than simply returning more of them. In this sense, effective human-AI complementarity requires managing the attention bottleneck created by information overload.

The remainder of the paper proceeds as follows. 
Section~\ref{sec:institutional_background} provides institutional background and discusses related papers.
Section~\ref{sec:data} describes the data, sample construction, and empirical design.
Section~\ref{sec:results} presents the results.
Section~\ref{sec:conclusion} concludes.

\section{Institutional Background}
\label{sec:institutional_background}
\subsection{AI-powered financial data platforms}
\label{sec:institutional_platforms}
Sell-side research is intermediated by a concentrated set of data and analytics vendors---\textsc{Bloomberg}, \textsc{FactSet}, \textsc{Refinitiv}, and \textsc{Capital IQ}---that collectively span the core stages of the analyst workflow (data discovery, modeling, visualization, and report production).
While their offerings overlap, each platform occupies a distinct functional niche.
\textsc{Bloomberg} is the benchmark for real-time market data, fixed-income analytics, and surveillance, and is nearly ubiquitous among large brokerages. 
\textsc{FactSet} offers more affordable data solutions to smaller research firms and has strategically positioned itself as an AI-powered workflow optimizer, exemplified by its \textsc{Mercury} platform. 
\textsc{Refinitiv} and \textsc{Capital IQ}, by contrast, hold comparatively smaller market shares among sell-side analysts, with the former emphasizing unified data synthesis and the latter leveraging deep fundamental datasets for customized solutions.

Since the popularization of LLMs, data vendors have been competing to offer AI-powered solutions to streamline user experience and expand analytical reach, as professionals increasingly view third-party platforms as ``better positioned'' to deliver such proprietary AI capabilities \citep{christ2025word}. 
In practice, the degree of AI integration---whether, and how tightly, LLMs are coupled with audited, finance-specific data and visualization pipelines---varies materially across providers and over time. 
Drawing on product release publicity materials and company announcements, Online Appendix~\ref{app:ai_powered_platforms} summarizes the roll-out of analyst-specific AI features among the major vendors. 
\textsc{FactSet}’s \textsc{Mercury} platform (announced December 14, 2023) marked an early, tightly integrated deployment that combined natural-language querying with automated visualization and pitchbook-creation tools, anchored to \textsc{FactSet}'s audited data. 
Subsequent offerings with comparable workflow depth appeared months later: \textsc{Capital IQ} released \textsc{Chat IQ} in November 2024, emphasizing agentic querying and automation within its fundamentals stack; in April 2025, \textsc{Bloomberg} launched Document Insights, enabling natural-language queries across millions of company documents. 
\textsc{Refinitiv}'s initial steps, by contrast, centered on AI-curated news within external interfaces, with less emphasis on deep workflow integration.

This staggered rollout---with \textsc{Mercury} being the earliest to offer deep integration of LLMs with proprietary financial data and report-production workflows---makes \textsc{FactSet} a particularly informative setting for studying how domain-specific generative AI alters analysts' information acquisition and production. We provide a more detailed discussion of \textsc{Mercury}'s functionality and integration in the next section.

\subsection{\textsc{FactSet Mercury}}
\label{sec:factset_mercury}
On December 14, 2023, \textsc{FactSet} unveiled \textsc{Mercury}, an AI-powered research platform developed under its strategic ``AI Blueprint'' initiative. 
The platform was designed to address a long-standing bottleneck in sell-side research: the time and cognitive effort analysts devote to locating, synthesizing, and visualizing vast quantities of financial data scattered across disparate systems. 
\textsc{Mercury} aims to convert this unstructured search process into an interactive, content-driven workflow. By embedding LLMs directly within \textsc{FactSet}'s audited databases, spanning real-time market feeds, company fundamentals, transactions, and industry benchmarks, the system enables analysts to discover patterns and contextual insights. In doing so, it expands both the breadth and depth of analysts' informational reach.

The initial version of \textsc{Mercury} integrates three core AI capabilities.\footnote{See screenshots of primary product functionalities in Online Appendix~\ref{app:Mercury_demo}. See the product demonstration at \href{https://videos.factset.com/watch/cwZF4FTpkw3KSPECz1j4so?_gl=1}{\textsc{Mercury Demo Video 1}} and \href{https://www.factset.com/marketplace/catalog/product/pitch-creator}{\textsc{Mercury Demo Video 2}}.} 
First, its conversational interface supports natural-language queries, dynamically generating visualizations and contextual explanations, allowing analysts, for example, to identify firms with rising leverage or shifting competitive exposures in seconds. 
Second, its workflow automation suite, including the \textit{Pitch Creator}, links dynamic Excel models with presentation templates to automatically assemble pitchbooks and research exhibits, substantially reducing repetitive formatting and charting tasks. 
Third, portfolio-level tools originally designed for wealth managers can be adapted to help equity analysts benchmark firm performance, evaluate peer groups, and identify potential market movers, thereby strengthening valuation and forecasting exercises. 
Collectively, these features promise to enhance analytical efficiency, streamline data access, and augment the cognitive capacity of research teams. In November 2024, \textsc{FactSet} launched the \textsc{Intelligent Platform}, which integrates \textsc{Mercury} with other AI-powered solutions, including Transcript Assistant and Conversational API, making it the primary interface through which clients interact with \textsc{FactSet}'s AI capabilities.

\textsc{Mercury} is not merely a technological experiment but a strategic priority with demonstrated commercial impact. \textsc{FactSet}'s fiscal 2025 annual report explicitly identifies AI as one of three core strategic pillars, stating that ``AI is embedded across our offerings to enhance data discovery, automate routine workflows and improve the speed and accuracy of client insights'' (FactSet 10-K 2025, p. 5). In the company's Q4 2024 earnings call, CEO Phil Snow emphasized that \textsc{FactSet}'s ``federated approach'' to AI, particularly through its conversational API, has become a key competitive advantage in displacing incumbent providers.\footnote{See \href{https://seekingalpha.com/symbol/FDS/earnings/transcripts}{FactSet Q4 2024 Earnings Call Transcript}.} Industry analysts have echoed this view, with JPMorgan estimating that \textsc{FactSet}'s AI tools could boost EBITDA margins by 3--5\% over the next three years.\footnote{See \href{https://www.ainvest.com/news/factset-q4-revenue-beat-signal-ai-driven-financial-data-demand-2509/}{AInvest Research Report \#1} and \href{https://www.ainvest.com/news/factset-dividend-aristocrat-attractive-valuation-strong-growth-catalysts-2509/}{AInvest Research Report \#2}.} Our interviews with a \textsc{FactSet} manager and an engineer confirm that the \textsc{Mercury} platform is specifically designed for bankers to streamline workflows and expand information access, an area where \textsc{FactSet} has made substantial investments. Unlike generic AI tools that merely summarize publicly available information, \textsc{FactSet Mercury} is integrated with proprietary data to deliver differentiated value that analysts cannot obtain from widely accessible models like ChatGPT or Gemini.

These AI capabilities have generated significant user adoption and revenue. By the end of fiscal 2024, \textsc{FactSet}'s AI product preview program had expanded to over 50 clients across banking, buy-side, and wealth management. Management has reported ``meaningful usage of each [AI product] by clients, which is starting to drive incremental ASV [(annual subscription value)] and improve retention.'' Our interview with a junior analyst who uses \textsc{FactSet} AI daily further corroborates these changes, noting that \textsc{FactSet} AI supports direct data integration into a knowledge base, enabling the analyst to ask specific questions and receive evidence-backed answers. In model-building workflows, where \textsc{FactSet}'s raw disclosure data is considered ``best practice'' within his firm, AI has further accelerated the research process. These qualitative insights align with and reinforce our empirical findings.

\subsection{Analysts' information environment and processing constraints}
\label{sec:processing_constraints}
\textsc{FactSet Mercury} promises to boost analyst efficiency by expanding access to information and automating routine tasks. Yet whether such tools deliver net benefits depends critically on the information environment and processing constraints that analysts actually face. The notion that capital market participants, including analysts, operate under limited attention and processing capacity is well established in the literature. Seminal studies by \citet{hirshleifer2003limited} and \citet{sims2003implications} demonstrate that economic agents are imperfect processors of readily available information. However, if time and attention are costly, such behavior can be rational, a concept that \citet{sims2003implications} terms ``rational inattention.'' Due to limited information-processing capacity, agents optimally process only a subset of available information to maximize their total utility, accounting for the inevitable processing costs.

Building on this framework, empirical research has documented implications that diverge from standard models assuming investors continuously optimize regardless of information presentation \citep{blankespoor2020disclosure}. Analysts are particularly susceptible to limited attention, as they constantly monitor multiple firms and must attend to various tasks, ranging from conducting site visits and maintaining client relations to, notably, producing research reports. \citet{hirshleifer2019decision} document declines in analyst performance as they perform consecutive forecasting tasks within the same day. \citet{driskill2020concurrent} find that analysts' timeliness significantly declines when more than one earnings disclosure by their portfolio firms occurs on the same day. In fact, analysts' attention is stretched so thin that even air pollution has been shown to increase their information processing delay \citep{dong2021air}. Bound by limited capacity, analysts also actively manage cognitive fatigue by prioritizing firms that are important to their career outcomes \citep{harford2019analyst, jiao2024decision}.

Given that the majority of analysts and investors alike remain concerned about the reliability of AI-generated outputs, especially in high-stakes financial decisions \citep{christ2025word, blankespoor2026generative_processing}, information produced by GenAI models will ultimately undergo human review before use or publication. More AI-generated information thus entails greater verification effort. Given the institutional reality that analysts are already constrained in time and attention, it is ultimately an empirical question whether greater informational richness translates into higher forecast accuracy. We posit that when analysts already face an excessive amount of information to process, using AI exacerbates their limited attention problem. They may fail to effectively synthesize all the information returned by AI and consequently become less accurate. Whether the integration of GenAI ultimately improves or impairs analyst decision quality thus remains an open question that we take to the data in the sections that follow.

\section{Data}
\label{sec:data}
We construct a comprehensive panel dataset by combining multiple sources, including (i) manually collected analyst reports, (ii) Institutional Brokerage Estimate System (I/B/E/S), (iii) LinkedIn resumes of analysts, and (iv) Center for Research in Security Prices (CRSP). Below, we describe each dataset in turn.

\paragraph{Analyst reports.}
We manually collect 162,862 equity research reports issued by sell-side analysts on all U.S. public companies between January 1, 2022, and October 31, 2024. Original reports, obtained in PDF format, contain rich qualitative and quantitative content, including textual narratives, figures, tables, and meta-data on brokerage and analysts. Using a state-of-the-art LLM (GPT-4o-mini), we systematically extract structured measures of information sources, topics, and analytical methods (described in Section~\ref{sec:analyst_report_content}). This dataset serves as the core of our analysis, enabling us to quantify various characteristics of report content and track their changes over time.

\paragraph{Forecast revisions.} We obtain I/B/E/S forecast revision data, which contain analyst-level and consensus earnings forecasts. We match our analyst reports to I/B/E/S forecast records following \citet{de2015analyst}, allowing us to link the issuance of the report to contemporaneous changes in earnings expectations (detailed below). I/B/E/S also provides detailed information that allows us to measure the analyst team's overall career length, firm-specific coverage experience, and other traits \citep{clement1999analyst, hirshleifer2019decision}.

\paragraph{Analyst backgrounds.}
We manually compile analyst resumes from LinkedIn, which contain detailed information on education and employment history. From these records, we construct measures of prior employment in IT-related industries and education for each analyst. We then merge the LinkedIn resume dataset with our analyst report data.

\paragraph{Market trading.}
We retrieve daily stock returns from CRSP to compute abnormal trading generated by analyst reports.

\paragraph{Sample construction.}
Table~\ref{tab:sample_construction} details our sample construction process.\footnote{Online Appendix~\ref{app:sample_construction} provides a step-by-step guide of sample construction, including how we identify analyst names in the masked databases of I/B/E/S and merge resume profiles with individual analysts. With an external dataset, we manage to confirm that our match is accurate 98\% of the time.} The initial sample consists of 162,862 unique reports corresponding to 315,499 report-firm pairs, as some reports contain more than one firm. We first merge the manually collected reports to forecast revision records in I/B/E/S.\footnote{A significant portion of reports are reiterating reports, where analysts issue a report without updating outstanding EPS forecasts. I/B/E/S notably does not register a new entry to their database when analysts reissue outstanding forecasts. Prior literature addresses this by matching the report to the most recent revision but imputing zero as the forecast revision, treating it as a control variable \citep{huang2014evidence, de2015analyst}. This approach is appropriate in settings where forecast revision is the outcome of interest. However, our study focuses on forecast \emph{accuracy}, which requires a contemporaneous, updated forecast to be meaningfully assessed. As such, we exclude these reports from our sample. To mitigate the concern that analysts may use the \textsc{FactSet} AI platform differently when producing reiterating versus revising reports, we replicate our main analysis of content richness on a random 10\% subsample of the reiterating reports, and find similar results. See Online Appendix~\ref{app:sample_construction} for detail.}
Then, we restrict the sample to analyst teams that issued at least one report in both the pre- and post-periods of the \textsc{Mercury} launch to ensure comparability, and drop analyst-firm pairs where coverage is terminated in the pre-period to further mitigate confounding factors in the analysts' endogenous coverage decisions. After dropping observations with missing variables, our final sample consists of 46,853 report-firm observations covering 2,443 unique firms and 413 analyst teams employed at 21 brokerages. Online Appendix \ref{app:brokerage} shows the number of analyst teams employed and reports issued by those brokers. Report issuance is highly concentrated among the largest institutions: the top five collectively generate nearly 80\% of all reports. This concentration underscores the institutional dominance of leading brokers in the production of sell‐side research.\footnote{We investigated why certain boutique brokers, notably Truist Securities and Stephens Inc., are underrepresented in our sample. We find that their report coverage in the database dropped to zero during 2021--2023 and resumed thereafter, while top-tier brokers show no such interruption. This pattern cannot be explained by broker mergers or name changes. Instead, we attribute it to a data coverage gap on the part of Mergent Investext from which the reports were collected, rather than to errors in our data collection: we downloaded reports without any broker-level filters, and the coverage gap persisted across two different data portals and collection periods, suggesting a supplier-side issue. Five other boutique brokers experienced similar gaps. Because our empirical design requires analysts to appear in both pre- and post-periods, these brokers are largely excluded from our final sample. We acknowledge this representation issue. Nevertheless, the unaffected brokers still account for the vast majority of observations, mitigating concerns that this gap materially affects our inferences.}

\begin{center}
[Insert Table~\ref{tab:sample_construction} About Here]
\end{center}

The following sections describe the key variables of interest along with the controls (see definitions in Appendix~\ref{sec:variable_definition}). 

\subsection{Analyst reports and structured information}\label{sec:analyst_report_content}

This section details our procedure for extracting structured information from analyst reports using GPT-4o-mini, a state-of-the-art LLM that offers an effective balance between precision, cost, and computational efficiency \citep{hurst2024gpt}.

Our approach proceeds in three steps. First, we segment each analyst report into sub-documents of fewer than 2,000 characters to mitigate the risk that the model omits or misremembers the content when processing lengthy inputs. To preserve context, each sub‐document is defined to contain complete report subsections rather than arbitrary text blocks. 
Second, we instruct GPT-4o-mini to operate as a domain expert in finance, applying a structured chain-of-thought reasoning process to analyze each sub-document. The model is instructed to extract specific structured information while explicitly articulating the rationale for each output, only based on the document provided. This procedure reduces the likelihood of hallucinations and ensures internal consistency.\footnote{Online Appendix~\ref{app:prompt_example} provides a representative prompt. Boilerplate such as disclaimers is omitted using keyword search. We validate our LLM-based measures using three approaches. First, two research assistants independently evaluated 198 randomly selected report segments; the conservative inter-rater agreement yields an error rate of 2.81\%, 10.75\%, and 8.73\% for identifying sources, topics, and methods, respectively, with no significant difference across time or \textsc{FactSet} usage. Second, we benchmark our topic measures against an LDA-based alternative \citep{huang2018analyst} and find a correlation above 0.5. Third, we vary the text segmentation length (1,500, 2,000, and 2,500 characters) and find that richness measures are highly correlated across specifications, with the exception of historical methods. Detailed results are reported in Online Appendix~\ref{app:validate_llm}.} 
Finally, we aggregate the extracted output across all sub-documents to form report-level content richness measures, spanning three dimensions: information sources, information topics, and analytical methods.

\paragraph{Information sources.}
We extract detailed information on the sources that analysts rely on in forming their assessments. The model classifies all sources cited into three categories: (i) \emph{textual sources}, which appear in the narrative sections of the report; (ii) \emph{figure sources}, which are referenced in graphical analyses; and (iii) \emph{table sources}, as cited in the tabular analyses. For each source, the model records the underlying data source (e.g., Bloomberg, FactSet), and a concise description. This classification allows us to quantify the richness and composition of evidence in analyst research and to examine how analysts integrate structured (tables and figures) with unstructured (text) information.

\paragraph{Information topics.}  
We also direct the model to classify the substantive content of each report into topics at three hierarchical levels: (i) \emph{firm‐specific} (e.g., earnings guidance, product launches), (ii) \emph{industry‐level} (e.g., supply chain conditions, regulatory developments, technological changes), and (iii) \emph{macroeconomic} (e.g., monetary policy, inflation, GDP growth). This structured classification enables us to quantify how analysts allocate attention across firm, industry, and macroeconomic domains.

\paragraph{Analytic methods.}  
We further instruct the model to identify and classify the analytical techniques used in each report into three categories: (i) \emph{historical trend analysis} aimed at evaluating the firm's past performance and patterns; (ii) \emph{valuation models} (e.g., discounted cash flow, comparable company multiples); and (iii) \emph{forecasting methods} (e.g., regressions, scenario analysis). This structured classification allows systematic measurement of the methodological toolkit used by analysts and facilitates comparisons in analytical rigor and complexity.

\bigskip
For each item, the LLM also generates an indicator of implied earnings changes (``up,'' ``down,'' or ``unknown''), assigns a confidence score, and provides the underlying reasoning.
To avoid double-counting of the same item across multiple segments of a single report, we consolidate all identical or highly similar items at the report level and remove duplicates. Combining these dimensions yields our information richness measures, which capture the breadth of \emph{unique} sources, topics, and methods incorporated in each report. 
While we do not claim that LLM-based measures are superior to traditional approaches, we believe they are particularly well-suited to our setting for three reasons. First, LLMs offer a unified framework for extracting sources, topics, and methods using standardized prompts, eliminating the need to combine disparate approaches---such as using topic modeling for topical diversity and keyword search for analytical methods. Second, their reasoning capabilities enable scalable inference of earnings implications from tables and numerical descriptions, which tone-based measures alone cannot achieve. Third, LLM outputs are human-readable, facilitating transparent benchmarking against human assessments and enhancing the credibility of our measures.

\begin{center}
[Insert Table~\ref{tab:summary_stats} About Here]
\end{center}

Hereafter, we define \emph{content richness} as the degree of informational variety embedded in an analyst report, reflected in the number of distinct sources, topics, and analytical methods it incorporates.

Panel A of Table~\ref{tab:summary_stats} reports summary statistics on content richness. On average, an analyst report in our sample cites 8.8 sources in textual discussions, 3.5 figure sources, and 7.7 table sources. Textual sources typically consist of analyst commentary, earnings call transcripts, and management discussions. Figures and tables often draw on market research data and company financial statements.

In terms of information topics, a typical report discusses 12.5 firm‐specific topics, 5.6 industry‐level topics, and 6.5 macroeconomic topics. The themes covered most frequently at the firm level are financial performance, revenue growth, and cost management. Industry‐level discussions often focus on competition, regulatory developments, and supply chain conditions, while macroeconomic coverage is most commonly centered on interest rates and inflation.

With respect to analytical methods, the average report employs 1.6 historical analyses, 2.9 valuation methods, and 3.8 forecasting methods. Common historical analyses include trend analysis and comparative analysis. The valuation methods most frequently applied are discounted cash flow, price-earnings multiples, and EV/EBITDA multiples. Forecasting techniques include regression analysis, extrapolation, and exponential smoothing. These statistics highlight both the breadth and depth of structured information contained in analyst reports, providing a systematic benchmark for our empirical analysis.

\subsection{Platform usage and report quality}
\label{sec:analyst_report_quality}

\paragraph{Platform usage.}
We measure the extent to which analysts leverage financial data platforms in preparing their research reports. Specifically, we exploit disclosures of data sources in the main texts, figures, and tables to identify whether a report draws on any of the four leading platforms: \textsc{FactSet}, \textsc{Bloomberg}, \textsc{Refinitiv}, and \textsc{Capital IQ}. For each platform, we construct an indicator variable equal to one if it is cited as a source in the report and zero otherwise. We also compute a usage intensity measure, defined as the proportion of references to each platform in a report. Panel B of Table \ref{tab:summary_stats} presents summary statistics on platform usage. \textsc{Bloomberg} is the most frequently cited platform (43.9\%), followed by \textsc{FactSet} (33.4\%), \textsc{Refinitiv} (14.2\%), and \textsc{Capital IQ} (2.7\%). We also define a binary variable \textit{Post} that equals one if a report is released after the integration of AI by \textsc{FactSet} and otherwise zero. Our key identifying assumption is that reports citing \textsc{FactSet} after the launch of \textsc{Mercury} plausibly reflect the use of its AI capabilities.\footnote{We identify \textsc{FactSet} citations by extracting lines containing ``source'' from PDFs and flagging those that also mention ``factset,'' ensuring we capture actual data citations rather than mere mentions. To validate that post-period \textsc{FactSet} citations reflect AI usage, we draw on institutional evidence showing that AI is a core strategic priority for \textsc{FactSet} with demonstrated commercial adoption (see Section~\ref{sec:institutional_background}), and on textual analysis showing that \textsc{FactSet}-assisted reports exhibit AI-like writing patterns (lower burstiness and higher GPTZero detection rates) after the platform's AI integration. Detailed results are reported in Online Appendix~\ref{app:validate_factset}.}

\paragraph{Report quality.}
With respect to the information quality of analyst reports, we focus on their timeliness, accuracy, and market impact.

First, we measure forecast timeliness ($Prompt$) using an indicator variable of whether the report is issued within three days after the firm's earnings announcement. This measure captures analysts' responsiveness to new public information. Analysts play an important role in interpreting earnings news and disseminating their views to the market, and timely reports following earnings announcements are a key dimension of analyst performance \citep{zhang2008analyst, du2020locked, driskill2020concurrent}. By focusing on the immediate post-announcement window, $Prompt$ reflects the analyst's ability to quickly process and incorporate earnings information into their research output.

Second, we measure forecast accuracy ($Accuracy$) following \citet{hirshleifer2019decision}. We define $Accuracy$ as the difference between the median forecast error of all analysts covering the same firm in the past 90 days and the analyst's own forecast error, scaled by the standard deviation of all analysts' forecast errors over the same window. By benchmarking each analyst's forecast against contemporaneous peer forecasts issued over similar horizons, this measure controls for firm- and time-specific factors that affect forecast difficulty, and facilitates cross-sectional comparisons across firms. Higher values indicate greater relative accuracy.\footnote{To ensure robustness, we also employ three alternative measures of forecast accuracy commonly used in prior literature: (i) forecast error deflated by the firm's beginning-of-year stock price \citep{dechow2012analysts}; (ii) accuracy measured as the maximum forecast error of all analysts following the same firm in the same year less the analyst's own error, scaled by the range of all analysts' forecast errors \citep{clement2005financial}; and (iii) proportional forecast error scaled by the mean forecast error of all analysts following the same firm in the past 90 days, multiplied by $-1$ so that higher values indicate greater accuracy \citep{clement1999analyst}. Our inferences remain unchanged across all measures, as reported in Online Appendix~\ref{app:robustness}.}

Third, we examine the market impact of analyst reports by measuring abnormal trading volume ($AbnVol$) around the report's release. Following \citet{de2015analyst} and \citet{blankespoor2018capital}, we calculate $AbnVol$ as the daily average market-adjusted number of shares traded over the $[-1,1]$ trading-day window surrounding the report release, scaled by total shares outstanding, less the equivalent amount over the $[-40,-11]$ trading-day window. We multiply this variable by 100 for ease of presentation. Higher abnormal trading volume indicates that the report generates greater investor attention and trading activity, reflecting the informational value of the report to market participants.

Panel B of Table ~\ref{tab:summary_stats} shows summary statistics on these variables. The mean of $Prompt$ is 0.294, suggesting that 30\% of our sample reports are issued promptly after earnings announcements, comparable with the 43\% reported in \citet{livnat2012information}.
The mean of $Accuracy$ is 0.435, suggesting that revised forecasts are on average more accurate than those issued over the past quarter.
The daily abnormal trading volume around analyst reports averages at 0.60\% of total outstanding shares, comparable to the 0.47\% reported in \citet{de2015analyst}.

\subsection{Analyst Characteristics}
\label{sec:analyst_characteristics}
\paragraph{Experience and expertise.}
We proxy for experience and expertise of the analyst teams using four measures. Following \cite{clement1999analyst}, we measure general experience and expertise, \textit{Team-Career}, as the number of years since the team first appeared in the I/B/E/S database, and measure firm-specific experience, \textit{Firm-Experience}, as the number of years since the team started covering a focal firm, capturing firm-specific learning. We also construct \textit{IT-experience}, an indicator being one if at least one member of the team has experience in the IT industry, capturing technical familiarity with AI technologies \citep{bertomeu2025impact}. Following \cite{li2023effect}, we construct \textit{Elite-Education}, which takes the value of one if one member of the analyst team studied at an institution ranked among the top 100 in the 2022 QS World University Rankings, or else zero.

\paragraph{Resources.} 
We measure the resources available to the analyst team using two team‐level variables. \textit{Brokerage-Size} captures the scale of the brokerage house, defined as the number of analyst teams employed by the house in the reporting year. Larger brokerages generally provide greater research infrastructure, proprietary data access, and distribution networks \citep{jacob1999expertise}. \textit{Team-Size} captures the number of signatory analysts on the team, reflecting the division of labor and complementary expertise within research groups \citep{brown2009team}. Both measures proxy for organizational resources supporting the production of reports by an analyst. 

\paragraph{Busyness.} 
We further capture analyst busyness using coverage variables. Specifically, we consider \textit{Forecast-Frequency}, which is defined as the number of forecasts issued by the analyst team in the prior fiscal year, and \textit{Firms-Followed}, which is defined as the number of firms simultaneously followed by the analyst team in the prior fiscal year \citep{jacob1999expertise, hirshleifer2019decision}. The former captures the amount of effort devoted to covering individual firms, and the latter represents the potential attention constraints on the analyst team \citep{hirshleifer2019decision}. Both variables are measured in the prior year as current-year measures could be affected by, among others, AI adoption.

Panel C of Table~\ref{tab:summary_stats} reports descriptive statistics on the characteristics of the analyst. On average, analysts in our sample have 16.5 years of general experience and 5.8 years of firm‐specific experience. Approximately 46\% of the analyst teams include a member with IT experience, while 36\% include an analyst who graduated from a top 100 university. The average brokerage employs 101 analyst teams, with a maximum of 199 (J.P.~Morgan), and the average research team consists of 3 signatory analysts, with a maximum of 6. On average, an analyst issues 5 forecast revisions per firm per year and follows 22 portfolio firms in total.

\section{Results}\label{sec:results}
\subsection{Adoption trends and patterns}\label{sec:adoption_patterns}
\begin{figure}[H]
\centering
\caption{Data Platform Usage}
\label{fig:analyst_data_platform}
\captionsetup{justification=justified,singlelinecheck=false}
\captionsetup[sub]{justification=centering}
\caption*{\footnotesize
This figure shows the trends in analysts’ use of major data platforms. Panel A reports the quarterly percentage of reports citing \textsc{FactSet}, \textsc{Bloomberg}, \textsc{Refinitiv}, or \textsc{Capital IQ}. Panel B reports usage intensity weighted by the number of citations to each platform within a report.
}
\vspace{0.5ex}
\begin{minipage}{0.70\textwidth}
  \centering
  \includegraphics[width=\linewidth]{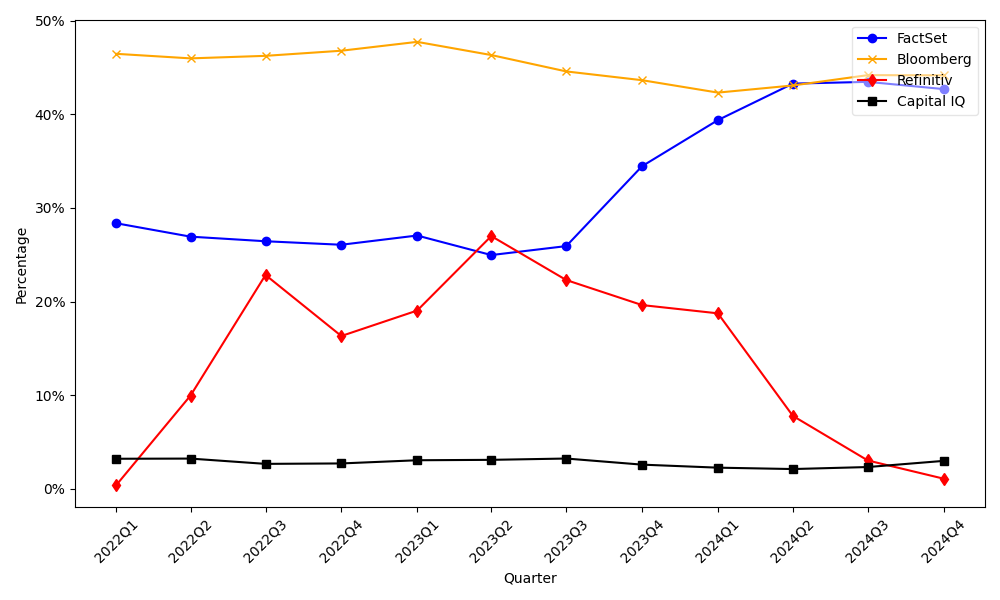}
  \subcaption{Panel A: Usage percentage by platform}
  \label{fig:platform_A}
\end{minipage}
\hfill
\begin{minipage}{0.70\textwidth}
  \centering
  \includegraphics[width=\linewidth]{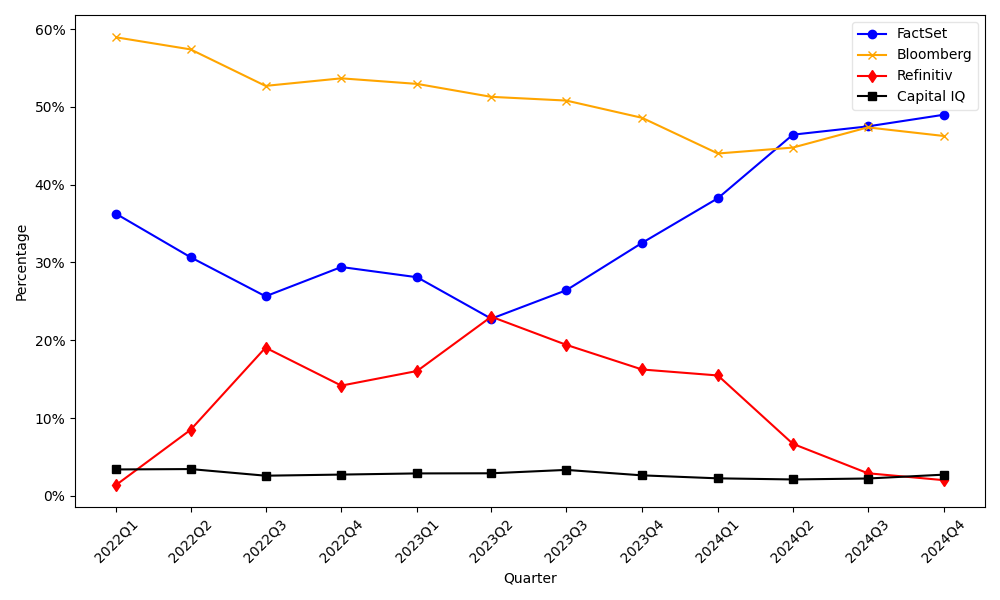}
  \subcaption{Panel B: Usage intensity by platform}
  \label{fig:platform_B}
\end{minipage}
\end{figure}

We first examine the trends and patterns of \textsc{FactSet} platform adoption by our sample analysts. Figure~\ref{fig:analyst_data_platform} illustrates analysts' use of major financial data platforms. Panel A plots the percentage of reports citing each platform between 2022 and 2024, while Panel B plots usage intensity, measured as the average share of citations to each platform. Both panels show a marked increase in \textsc{FactSet} usage beginning in 2023Q4, coinciding with the launch of its AI platform. For example, the proportion of \textsc{FactSet} citations rose to approximately 50\% by the end of 2024, up from around 25\% in 2023Q3. In contrast, \textsc{Refinitiv} exhibits a rapid decline in usage over the same period,\footnote{\textsc{Refinitiv} was acquired by the London Stock Exchange Group (LSEG) on January 29, 2020, before our sample period begins. Despite the acquisition, the data platform is still referred to as \textsc{Refinitiv}, and the parent firm, LSEG, is rarely cited (in less than 1\% of reports) by analysts. We therefore follow industry convention and refer to it as \textsc{Refinitiv}.} \textsc{Bloomberg} shows a steady but gradual decrease, and \textsc{Capital IQ} remains stable, albeit with only a negligible market share throughout the sample period.\footnote{We also examine the usage of other financial data platforms, such as MarketWatch and Yahoo Finance, and find that they are cited in less than 1\% of our sample reports.}

\begin{center}
[Insert Figure~\ref{fig:analyst_data_platform} About Here]
\end{center}

A natural next question is: who uses \textsc{FactSet}? We use our panel data to shed light on the characteristics of analysts adopting \textsc{FactSet} and how its user base has shifted after the AI integration. We estimate the following specification:

\vspace{-0.25cm}
\begin{equation}\label{eq:factset-usership}
FactSet_p = \alpha + (Post_p \times \bm Z_{pf})' \delta + \bm Z_{pf}' \beta 
+ \text{Day FE} + \text{Firm-Year FE} + \varepsilon_{pf},
\end{equation}

\noindent where $FactSet_p$ denotes the use of \textsc{FactSet} in report $p$. The vector $\bm Z_{pf}$ includes the standard controls at the analyst and analyst-firm levels. Following \citet{clement1999analyst}, analyst characteristics are standardized by subtracting the firm-year minimum and dividing by the within-firm range. The report-date fixed effect absorbs economy-wide shocks and technological advances that affect all analysts simultaneously, including the nationwide rollout of generic AI tools, and the firm-year fixed effect holds time-varying firm characteristics (e.g., management changes, information environment shifts, and investor base) constant, allowing us to compare analysts covering the same firm in the same fiscal year. We also include a binary variable each for the use of other platforms. We cluster standard errors by analyst and by firm to account for heteroskedasticity in our panel data.

\begin{center}
[Insert Table~\ref{tab:ai_adopting_analysts} About Here]
\end{center}

Table~\ref{tab:ai_adopting_analysts} reports the results. Column (1) relates to the binary variable (citation = 1, no citation = 0) and column (2) relates to the continuous variable (proportion of citations). There are fewer observations in column (2) because reports with no data platform citations are dropped.

Turning first to the non-interactive coefficients, which reflect baseline adoption patterns prior to the \textsc{Mercury} launch, we find that \textsc{FactSet} is disproportionately used by younger analysts and those with degrees from elite universities. Adoption is also more prevalent among analysts in smaller brokerage firms, consistent with \textsc{FactSet}'s greater affordability relative to competitors such as \textsc{Bloomberg}. Finally, analysts facing greater workload, measured by forecast frequency and coverage portfolio size, are more likely to use \textsc{FactSet}. By contrast, IT experience and team size do not consistently predict adoption.

We next turn to the interactive terms. Following the launch of \textsc{Mercury}, analysts from smaller brokerage firms and those working with larger teams disproportionately increase their use of \textsc{FactSet}. Beyond these two dimensions, however, there is otherwise no significant shift in the user base around the launch event. This suggests that the increase in \textsc{FactSet} adoption observed in Figure~\ref{fig:analyst_data_platform} is primarily driven by analysts who were already using \textsc{FactSet} or those with characteristics similar to incumbent users, which helps alleviate, though does not eliminate, concerns about self-selection bias.

We acknowledge that analysts endogenously choose whether to use \textsc{FactSet}, and that this preference may also shift in response to the platform's AI integration. In the sections that follow, we employ econometric strategies---including more granular fixed effects, placebo tests, entropy balancing, and alternative treatment definitions---to mitigate endogeneity concerns to the extent possible with observable data. Nevertheless, as with any study of technology adoption, we cannot completely rule out all forms of endogenous selection. We therefore interpret our estimates as associations that plausibly reflect the causal effects of AI adoption under the identifying assumptions we make explicit.

\subsection{AI adoption and report richness}\label{sec:consequence_richness}

\begin{figure}[htbp]
\centering
\caption{Sources, Topics, and Methods in Analyst Reports}
\label{fig:content_richness_trend}
\captionsetup{justification=justified,singlelinecheck=false}
\captionsetup[sub]{justification=centering}
\caption*{\footnotesize
The figure plots the quarterly trends in the number of sources, topics, and methods extracted from \textsc{FactSet}-assisted reports and non-FactSet reports, respectively. 
Each observation is the quarterly average after removing monthly effects, with confidence intervals at the 95\% level.
Panels A--C relate to sources, topics, and methods, respectively.
}

\vspace{0.5ex}
\begin{minipage}{0.65\textwidth}
  \centering
  \includegraphics[width=\linewidth]{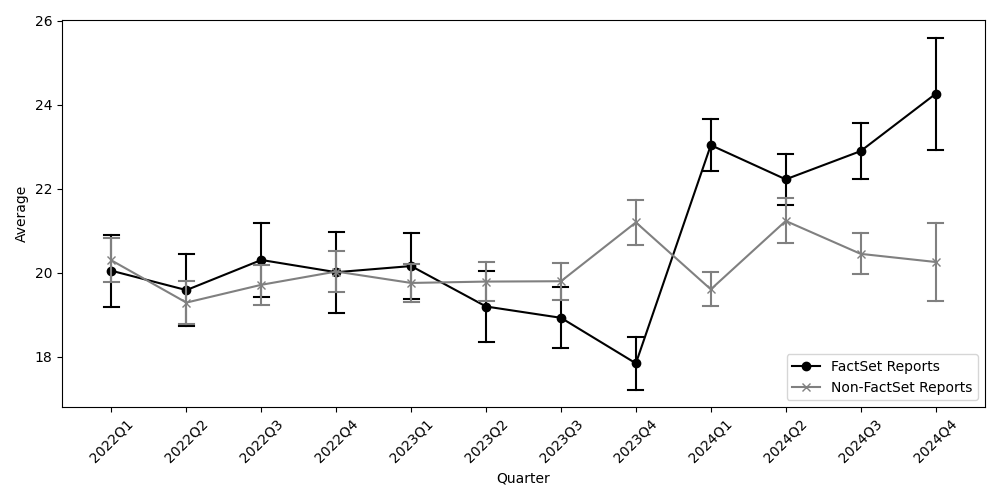}
  \subcaption{Panel A: Sources}
  \label{fig:richness_A}
\end{minipage}
\begin{minipage}{0.65\textwidth}
  \centering
  \includegraphics[width=\linewidth]{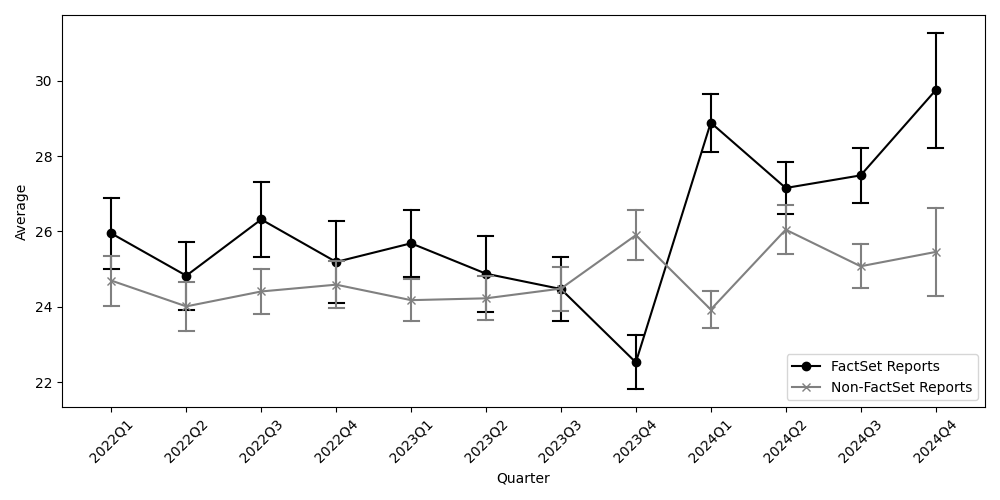}
  \subcaption{Panel B: Topics}
  \label{fig:richness_B}
\end{minipage}
\begin{minipage}{0.65\textwidth}
  \centering
  \includegraphics[width=\linewidth]{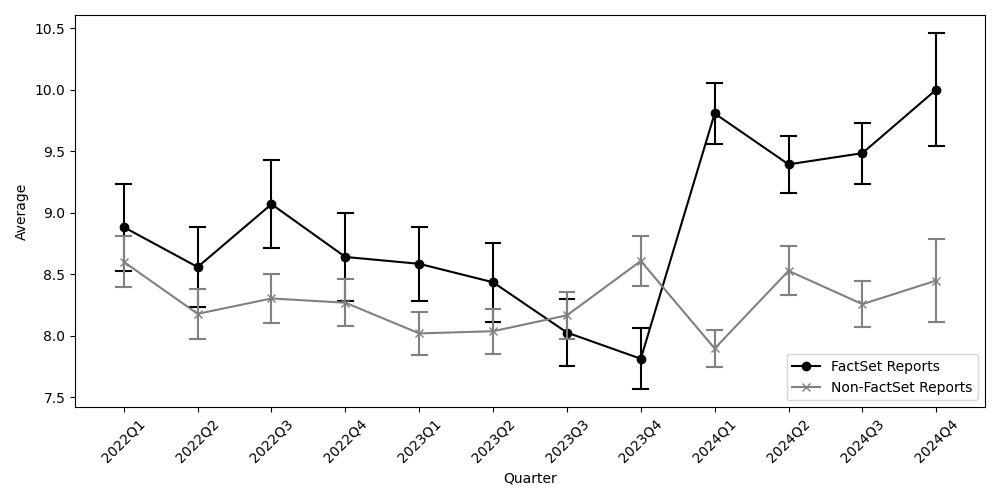}
  \subcaption{Panel C: Methods}
  \label{fig:richness_C}
\end{minipage}
\end{figure}

We now examine how AI adoption affects the informational richness of analyst reports. Concurrent with the rising usage of the AI-powered platform, Figure~\ref{fig:content_richness_trend} plots the quarterly average content richness of reports developed with and without \textsc{FactSet}, separately, from 2022 to 2024. To control for seasonality, we remove month fixed effects prior to averaging.\footnote{Reports issued in certain months of the year, for example around earnings announcements, could be systematically different in content from those issued in other months. To remove the monthly effect, we subtract the difference between the average of all reports issued in a given month and the sample average from each report issued in that month to derive the ``de-trended'' value of the measure.} Panels A through C present the number of sources, topics, and methods, respectively, with confidence intervals at the 95\% level. Across all panels, the richness measures of \textsc{FactSet} reports remain stable before 2023Q4, followed by a pronounced increase beginning in 2024Q1, while non-\textsc{FactSet} reports show largely flat trends within our sample period. In terms of magnitude, the number of sources in \textsc{FactSet} reports increased from 19 to 24 after the \textsc{Mercury} launch, topics increased from 24 to 29, and methods increased from 8 to 10. Taken together, these descriptive patterns suggest that the launch of \textsc{FactSet Mercury} could be a potential driver of the observed rise in the richness of analyst reports, motivating a closer examination using multivariate regressions.

\begin{center}
[Insert Figure~\ref{fig:content_richness_trend} About Here]
\end{center}

We estimate the following difference-in-differences specification:

\vspace{-0.75cm}
\begin{equation}\label{eq:baseline}
\begin{split}
y_{p} &= \alpha + \delta \,(FactSet_p \times Post_p) + \lambda \, FactSet_p + \bm Z_{pf}'\beta \\
&+ \text{Day FE} + \text{Firm-Year FE} + \text{Analyst FE} + \text{Broker-Year FE} + \varepsilon_{pf}, 
\end{split}
\end{equation}

\noindent where $y_{p}$ denotes a content richness measure, as described in Section~\ref{sec:analyst_report_content}. We include the same standard controls as before, and further include the analyst fixed effect and the broker-year fixed effect. This combination of fixed effects allows us to identify the effect of AI platform use from variation in \textsc{FactSet} citations across reports written by the same analyst for the same firm within the same year, after stripping out common time trends, firm-specific shocks, and brokerage-level changes. Standard errors are clustered by analyst and by firm. The variable of interest is the interaction term $FactSet_p \times Post_p$, which we expect to be significantly positive if AI adoption enhances report richness.

\begin{center}
[Insert Tables~\ref{tab:baseline_sources} -- \ref{tab:baseline_methods} About Here]
\end{center}

Tables~\ref{tab:baseline_sources}, \ref{tab:baseline_topics}, and \ref{tab:baseline_methods} report the results for sources, topics, and methods, respectively. First, the coefficient on $FactSet$ is significantly positive across all nine columns, suggesting that the use of \textsc{FactSet} is associated with greater information richness. More importantly, the coefficient on $FactSet \times Post$ is also significantly positive across all columns, suggesting that after the introduction of \textsc{Mercury}, the use of \textsc{FactSet} is associated with an even greater increase in report richness. In terms of controls, larger analyst teams and teams with elite-educated analysts produce informationally richer reports.

Turning to the components of report richness, analysts exhibit the largest relative increase in their use of visual information following AI adoption. AI platform adoption is associated with a 29\% ($= 0.977 \div 3.362$) increase in figure sources relative to the pre-period, compared to a 27\% ($= 2.273 \div 8.431$) increase in text sources and a 23\% ($= 1.690 \div 7.438$) increase in table sources. These findings highlight that analysts benefit most from AI through improved visualization and communication of data-driven insights, rather than through simply adding more text-based commentary.

The expansion in topical coverage is most pronounced for industry-level topics at 26\% ($= 1.394 \div 5.385$), whereas firm-specific and macroeconomic topics rise more modestly by 24\% ($= 2.874 \div 12.012$) and 24\% ($= 1.506 \div 6.322$), respectively. While analysts tend to focus primarily on firm-level discussions, the AI platform enhances their ability to incorporate broader, cross-firm, and macroeconomic perspectives into their reports.

In terms of analytical methods, the most pronounced increase occurs in forecasting techniques, the domain where AI algorithms, particularly machine learning models, excel at detecting complex non-linear patterns \citep{gu2020empirical}. The use of analytical methods rises by 11\% ($= 0.169 \div 1.607$) for historical trend analysis, 19\% ($= 0.533 \div 2.783$) for valuation, and 27\% ($= 1.010 \div 3.682$) for forecasting. This pattern is consistent with the view that AI-powered platforms primarily enhance data-driven forecasting rather than valuation tasks that rely heavily on analyst judgment and domain expertise.

While this setting provides a plausibly exogenous shock to analysts' access to generative AI technology, the treatment inevitably captures only a subset of all AI-assisted research activity. In particular, \textsc{FactSet} remains a multi-purpose data vendor, and analysts may adjust their platform choices for non-AI reasons such as pricing, licensing policies, or internal brokerage IT integration. Conversely, some analysts in the control group may independently use alternative AI tools (for example, internal copilots, ChatGPT, or early-stage AI features offered by \textsc{Bloomberg} or \textsc{Capital IQ}). Both forms of misclassification---\textsc{FactSet} adoption unrelated to AI, and AI usage outside \textsc{FactSet}---tend to attenuate our estimated treatment effects toward zero. Hence, our results likely represent a conservative lower bound on the true impact of generative AI adoption on the information production and performance of financial analysts.

Table~\ref{tab:dynamic_effects} examines the dynamic effects of AI adoption on report richness by decomposing $Post$ into quarterly dummies spanning 2023Q1 through 2024Q4, with 2022 serving as the benchmark. All other specifications remain unchanged. The estimates reveal largely no significant pre-trends across reports developed with and without \textsc{FactSet}, reassuring us that the effect is likely attributable to the launch of the AI platform.\footnote{Nevertheless, we still observe significantly negative coefficients on $\#MacroTopics$ and $\#HistMethods$. We attribute this to confounding firm-level shocks that are not fully absorbed in the current fixed-effect design: once we further control for the more granular firm-quarter fixed effects (rather than firm-year fixed effects), as reported in Online Appendix~\ref{app:robustness}, the negative coefficients become less significant. Without risking absorbing too much variation and to prevent further sample loss, we retain our main design, which already yields an overall satisfactory parallel trend.}

\subsection{AI adoption and report quality}\label{sec:consequence_quality}

We next examine how report quality changes around \textsc{FactSet}'s AI integration. As described in Section~\ref{sec:analyst_report_quality}, we focus on timeliness, accuracy, and market impact. Table~\ref{tab:ai_report_quality} reports the results. The first column relates to timeliness; the coefficient on $FactSet \times Post$ is significantly positive, suggesting that the AI data platform enables analysts to respond more quickly to earnings news by expanding information access and streamlining report production.

The second column relates to forecast accuracy. The coefficient on $FactSet \times Post$ is $-0.157$ and statistically significant at the 1\% level. This implies that post-integration \textsc{FactSet}-associated reports exhibit 0.157 standard deviations lower relative forecast accuracy than the comparison group. To put this magnitude into perspective, \citet{hirshleifer2019decision}, in their study of how analyst performance declines as fatigue builds up from issuing multiple forecasts on the same day, find that accuracy decreases by 0.225 standard deviations relative to peer forecasts with a one-unit increase in forecast rank (see their Table~3). Thus, the effect we document is economically meaningful and comparable in scale to well-established behavioral effects in the analyst forecasting literature.

The third column examines market reaction. The coefficient on $FactSet \times Post$ is $-0.127$ and statistically significant at the 1\% level, indicating that AI-assisted reports generate 21\% ($= 0.127 \div 0.602$) less abnormal trading volume relative to the sample mean. This finding echoes \citet{bradshaw2026generative}, who document that AI-generated reports published by non-professional analysts elicit weaker market responses. 

Collectively, these results reveal a key trade-off: the AI integration is associated with richer and faster report production, but also with lower average relative forecast accuracy. However, several questions remain: Does the decline in accuracy manifest uniformly across all analysts or only in a subset? What drives the decline: limited information processing capacity on the human side, or pure AI hallucination? We explore these mechanisms in the next section.

\begin{center}
[Insert Table~\ref{tab:ai_report_quality} About Here]
\end{center}

\subsection{Mechanisms}\label{sec:mechanisms}

As discussed in Section~\ref{sec:processing_constraints}, our mechanism is a constraint on human information processing. GenAI can expand the set of information available to analysts, but the additional evidence still must be verified, prioritized, and synthesized into a single forecast. When analyst attention is scarce, a richer but less internally aligned information set may therefore reduce decision precision even if the underlying signals remain informative. We organize the evidence in three steps: we characterize the information set, identify when human processing constraints bind, and benchmark human forecasts against a machine processor that does not face human attention limits. Table~\ref{tab:mechanism_tests_p1} reports these tests.

\begin{center}
[Insert Table~\ref{tab:mechanism_tests_p1} About Here]
\end{center}

If information signals are not fully verified or organized due to limited processing capacity, we would expect that signals are less aligned when \textsc{FactSet} AI is involved in report production, and that lower signal alignment correlates with lower forecast accuracy. We report these tests in Panel A. When parsing the reports, we instructed GPT to classify each signal as positive, negative, or neutral based on whether it predicted an earnings increase, decrease, or no change or unknown (see Section~\ref{sec:analyst_report_content}). We assign a score of 1 to positive signals, $-1$ to negative signals, and 0 to neutral signals, and calculate the standard deviation of the signal sentiment scores ($SignalSD$) as a proxy for how misaligned the signals are. As reported in the first column of Panel A, the coefficient on $FactSet \times Post$ is significantly positive ($t = 2.33$), indicating that post-integration \textsc{FactSet} reports contain less aligned signals. In the second column, the coefficient on $SignalSD$ is significantly negative ($t = -3.48$): reports with more internally conflicting signals are associated with lower forecast accuracy. We view this result as characterizing the information environment created by broader information access; by itself, it does not identify limited attention. The cross-sectional tests that follow ask whether the accuracy loss emerges specifically when processing capacity is more likely to bind.

We next examine whether the decline in accuracy and signal alignment is more pronounced when analyst attention is more strained. We consider three proxies for attention constraints, following prior literature \citep{clement1999analyst, fang2021analyst, hirshleifer2019decision}. Our first proxy is the analyst's portfolio complexity, measured as the number of firms followed (\textit{Firms-Followed}). Analysts who cover more firms have less time and attention for each individual firm and may suffer from information overload, resulting in lower accuracy \citep{clement1999analyst}. The second proxy is the analyst's team size (\textit{Team-Size}), as analysts who work with a smaller team (or on their own) have less internal support, especially for information verification and report drafting, subjecting them to greater processing capacity constraints. For these two proxies, we partition the sample based on the median and examine the coefficient difference between high- and low-groups. The final proxy is a binary variable that equals one if the analyst revises forecasts for two or more companies on a given day and zero otherwise (\textit{Busy-Day}), motivated by \citet{hirshleifer2019decision}, who find that when analysts issue multiple forecasts on the same day, their accuracy suffers due to decision fatigue.

Panel B of Table~\ref{tab:mechanism_tests_p1} reports the subsample analyses for both forecast accuracy and signal misalignment along these three dimensions. Coefficient differences are tested using a full interaction model that reports one-tail $p$-values. As Panel B shows, we consistently find that GenAI use decreases accuracy and increases signal misalignment significantly when analysts face attention constraints, while the effects are largely insignificant when attention constraints are low. This supports the idea that AI-generated information exacerbates analyst information overload and leads to lower accuracy.

If the decline in accuracy reflects a human processing constraint rather than deterioration in the underlying information, the same observable inputs should not generate an analogous post-integration accuracy loss when processed by an algorithm that does not face human attention constraints. We therefore construct a machine-learning benchmark.
We obtain 45 individual information signal measures from each report (the number of each of the 9 categories of information signals across 5 sentiment types: high-confidence positive, low-confidence positive, neutral, low-confidence negative, and high-confidence negative, where sentiment and confidence levels are obtained from the GPT model during parsing), 3 composite signal measures (percentage of positive and negative signals as well as textual tone), 16 historical firm metrics (prior-year EPS, beginning-of-year stock price, total assets, net income, sales revenue, book equity, market equity, age, return volatility, return on assets, book-to-market, institutional holdings, Herfindahl index, sales growth, leverage, and most recent quarterly EPS), and 2 consensus forecast measures (median and mean of forecasts issued in the prior 90 days). These are data accessible to analysts at the point they derive their estimates.

We feed these variables to a random forest model with standard parameters and use actual EPS as the target variable for the prediction. We hold out 30\% of the sample as the testing set and use the remaining 70\% as training. Critically, we partition the sample by firm-years, not by individual reports, to ensure that the same firm-year does not appear in both the training and testing sets, which would reveal true patterns of earnings and historical signals to the model and result in inflated out-of-sample accuracy. We repeat the random sample splits 100 times and for each iteration obtain (i) the root mean squared error (RMSE) of the out-of-sample tests; (ii) the coefficients on $FactSet \times Post$ using human accuracy and machine accuracy,\footnote{This is measured in the same fashion as $Accuracy$: the median forecast error of all analysts in the past 90 days less the machine's forecast error, scaled by the standard deviation of all analyst forecasts in the past 90 days.} respectively, as the dependent variable; and (iii) the coefficients on $FactSet \times Post$ using machine forecast accuracy in six different subsamples as defined in Panel B (high versus low attention constraints). We use the 100 Monte Carlo iterations to conduct our statistical inference.

Panel C shows that across the 100 iterations, the random forest model on average performs slightly better than human analysts (lower RMSE). Importantly, while the average coefficient on $FactSet \times Post$ is still significantly negative for human analyst accuracy, it is insignificant for machine-generated forecasts. This contrast is informative because the machine processes essentially the same observable information set without human attention constraints: unlike human forecasts, its accuracy does not deteriorate for post-integration \textsc{FactSet} reports. In Panel D, the cross-sectional pattern from Panel B likewise largely disappears for machine forecasts. We interpret these findings as evidence favoring a human processing-capacity channel over a simple deterioration in the informativeness of the available inputs. The exercise does not establish limited attention as the sole mechanism, but it provides a useful benchmark that separates the quality of the information set from the capacity of the processor.\footnote{We acknowledge several caveats. First, we do not claim that limited attention is the sole mechanism at play; other factors may also contribute, though we conduct additional tests to rule out several alternative explanations. Second, the machine-learning model we adopted, while carefully designed, may still suffer from look-ahead bias by using future information to predict the present. This is unavoidable because if we were to use the 2022--2023 sample as the training set, we would not be able to have a pre-period sample in the testing set to replicate the regression. Nevertheless, this analysis remains qualitatively insightful, showing that the decline is not driven by the information signals themselves but by how the information is processed (or not) by analysts.}

These results also help address the apparent conflict between our finding of decreased accuracy and the improvements documented in prior studies \citep{bertomeu2025impact}. First, \citet{bertomeu2025impact} examine how domestic analysts' accuracy changed during a temporary one-month ban of ChatGPT in Italy. Their setting captures the effect of removing a generic AI tool that primarily streamlines routine tasks and frees up analyst attention. In contrast, our setting examines the introduction of a domain-specific AI platform that supplies analysts with additional information, which can exacerbate rather than alleviate attention constraints (see Section~\ref{sec:mechanisms}). Second, because our sample period is longer than the one-month ban, we are able to observe the effects of AI under both unconstrained and attention-constrained conditions. The conditional nature of our findings helps reconcile our results with the broader literature, which typically documents the benefits of AI in more routine contexts.

\paragraph{Alternative explanations.}
In this section we rule out alternative explanations for the decline in accuracy in AI-assisted reports. First, the information itself could deviate from fundamentals if AI hallucinates. Second, AI could be generating low-quality text that adds noise to analysts' judgment. Third, analysts could be using AI to expedite report production without properly reviewing the content. We perform tests to rule out these possibilities in turn.

First, we test whether the quality of information contained in analyst reports deteriorates following the adoption of the AI-powered platform. Specifically, we examine whether the signals extracted from report content remain predictive of firms' subsequent earnings realizations. The regression specification is as follows:

\vspace{-0.5cm}
\begin{equation}\label{eq:information_quality}
\begin{split}
EpsChange_{pf}
&= \alpha + \delta_1 \, SignalPos_{p} + \delta_2 \, SignalPos_{p} \times FactSet_{p} \times Post_{p} \\
&+ \theta_1 \, SignalNeg_{p} + \theta_2 \, SignalNeg_{p} \times FactSet_{p} \times Post_{p} \\
&+ \bm Z_{pf}'\beta + \text{Day FE} + \text{Firm FE} + \text{Analyst FE} + \text{Broker-Year FE} + \varepsilon_{pf}, 
\end{split}
\end{equation}

\noindent where the dependent variable is either a continuous variable equal to the one-year-ahead actual earnings less the prior-year earnings, scaled by the stock price at the beginning of the prior year ($EpsChange$), or a dummy variable that takes the value of 1, 0, or $-1$ for an increase, no change, or decrease in earnings ($EpsChangeDum$). $SignalPos$ ($SignalNeg$) is the proportion of information sources in the report that predict an increase (decrease) in earnings. Other interactive terms, such as $SignalPos \times FactSet$, are all included in the model but omitted for brevity. We include firm fixed effects instead of firm-year fixed effects, as the latter would be collinear with the dependent variable. All other specifications are identical to the baseline.

We report the results in Panel A of Table~\ref{tab:alt_explanations}. The variables of interest are the three-way interactions between signal sentiment scores, $FactSet$, and $Post$. As the results show, $SignalPos$ and $SignalNeg$ are correlated with actual earnings changes in the predicted directions. However, the coefficients on the three-way interactions are both insignificant, suggesting that the signals in AI-assisted reports remain equally informative about firm fundamentals as reports developed without \textsc{FactSet} AI. Therefore, AI adoption has not diminished the intrinsic informational value of report content, further substantiating the processing constraint channel.

Second, we show that AI adoption leads to no systematic increase in information redundancy or complexity in analyst reports, which also alleviates the concern that our richness measures are driven by AI-generated boilerplate. We quantify redundancy following \citet{crossley2016tool} by calculating the type-token ratio of nouns and content words in each report, while ignoring function words such as ``the'' and ``of.'' Using their open-source algorithm, all tokens (nouns or content words, depending on the measure) are first reduced to their respective root forms, and then the number of unique tokens (``type'') is divided by the total number of tokens to calculate the type-token ratio (denoted \textit{NounTTR} and \textit{ContentTTR}). A lower type-token ratio, usually resulting from repetition of certain words, indicates lower lexical diversity and higher redundancy. To quantify textual complexity, we calculate two standard readability measures following prior research \citep{de2015analyst}: the Gunning-Fog index ($Fog$), which equals (words per sentence + percentage of complex words) $\times$ 0.4, and the Flesch Reading Ease score ($FRE$), which equals $1.015 \times$ words per sentence $+ 84.6 \times$ syllables per word $-$ 206.835. We multiply both measures by $-1$ so that higher values indicate greater readability. The results are reported in Panel B of Table~\ref{tab:alt_explanations}. The coefficient on $FactSet \times Post$ is insignificant in all columns, suggesting that there is no systematic shift in redundancy or textual complexity in analyst reports developed using generative AI.

Finally, we find no evidence suggesting that the decline in accuracy is more pronounced in timelier reports developed using the AI platform. We interact AI adoption with $Prompt$ and re-estimate the regression on forecast accuracy. If analysts use AI to expedite report release at the expense of human review, the coefficient on the interaction term $Prompt \times FactSet \times Post$ should be significantly negative. We report the regression results in Panel C of Table~\ref{tab:alt_explanations} and find insignificant results. Together, our results consistently rule out the potential deterioration in information quality and lack of human review as the mechanism for the accuracy decline, highlighting limited processing capacity as the plausible mechanism.

\begin{center}
[Insert Table~\ref{tab:alt_explanations} About Here]
\end{center}

\subsection{Robustness tests}\label{sec:robustness}

\paragraph{Placebo tests.}
A potential concern with our identification is that analysts often rely on multiple data vendors when preparing research reports, and other platforms may offer similar tools that enhance information retrieval and visualization. If so, the observed improvements in report richness and efficiency could reflect contemporaneous upgrades across the data-vendor ecosystem rather than the specific impact of \textsc{FactSet}'s AI-powered \textsc{Mercury} platform. To address this concern, we conduct a set of placebo tests that substitute \textsc{FactSet} with other major data providers. We define the indicator variable \textit{OtherPlatforms} as equal to one if a report cites any non-\textsc{FactSet} data vendor---namely, \textsc{Bloomberg}, \textsc{Refinitiv}, or \textsc{Capital IQ}---and zero otherwise. We repeat all main regressions replacing $FactSet$ with $OtherPlatforms$, and report the results in Table~\ref{tab:placebo_tests}. The coefficients on $OtherPlatforms \times Post$ are insignificant across all columns, confirming that our baseline effects are specifically attributable to the AI-powered \textsc{Mercury} platform rather than to general improvements across data vendors.

\begin{center}
[Insert Table~\ref{tab:placebo_tests} About Here]
\end{center}

\paragraph{Alternative measures of forecast accuracy.}
To ensure that our finding of declining forecast accuracy is not sensitive to how accuracy is measured, we complement our main analysis with three alternative measures commonly used in prior literature. First, we follow \citet{dechow2012analysts} and deflate the forecast error by the firm's beginning-of-year stock price, multiplied by 100 for ease of presentation. Second, we measure accuracy as the maximum forecast error of all analysts following the same firm in the same year less the analyst's own forecast error, scaled by the range of all analysts' forecast errors \citep{clement2005financial}. Third, we follow \citet{clement1999analyst} and compute proportional forecast error as the analyst's forecast error scaled by the mean forecast error of all analysts following the same firm in the past 90 days, then multiply by $-1$ so that higher values indicate greater accuracy. As reported in Online Appendix~\ref{app:robustness}, the coefficient on $FactSet \times Post$ remains negative and statistically significant across all three specifications. Moreover, the magnitudes are economically comparable to benchmark estimates documented in prior studies. These results confirm that our inference on forecast accuracy is robust to measurement choices.

\paragraph{Report length.}
Report length may play an important role in assessing report content richness. To address this concern, we conduct additional tests by first including $Length$, measured as the natural log of the number of words in the report, and then by including length percentile fixed effects to compare reports within the same length percentile. Online Appendix~\ref{app:robustness} reports the results. Our inference remains unchanged under both specifications: the coefficients on $FactSet \times Post$ remain positive and statistically significant across most richness dimensions.\footnote{We note, however, that the coefficient magnitudes are somewhat smaller when length is included as a control, because part of the effect of AI on report richness is absorbed by report length. This attenuation is expected: as GenAI expands analysts' information reach, reports naturally become lengthier to incorporate the increased sources and topics. Thus, to avoid understating the full effect of AI on report content richness, we refrain from controlling for length in our main specifications, allowing us to capture the total effect of AI use.}

\paragraph{Entropy balancing.}
Although our main specifications already include a rich set of fixed effects that absorb a wide range of confounding factors, we further address potential selection bias by replicating our main results using entropy balancing. This approach reweights non-\textsc{FactSet} reports to have covariate distributions that match those of \textsc{FactSet} reports in their first moments, effectively approximating the effect of one-on-one propensity score matching. As reported in Online Appendix~\ref{app:robustness}, the results remain qualitatively similar to our baseline estimates, confirming that our findings are not driven by observable differences between \textsc{FactSet}- and non-\textsc{FactSet}-assisted reports.

\paragraph{Alternative fixed effects and clustering.}
We also examine the sensitivity of our results to alternative fixed-effects structures and standard error clustering. First, we replace year-level fixed effects with month-level fixed effects to absorb more granular time-varying confounders at the firm, analyst, and brokerage levels. Second, we include analyst-firm fixed effects to absorb time-invariant, pair-specific advantages. As for clustering, we adopt two-way clustering at the analyst and firm levels throughout our main analyses. We also estimate our specifications with broker- and firm-level clustering as a robustness check. Online Appendix~\ref{app:robustness} shows that our inference remains qualitatively unchanged.

\paragraph{Alternative treatment definitions.}
Finally, we address potential measurement error in our treatment variable by redefining exposure to \textsc{FactSet} at the analyst and broker levels, rather than at the report level. Specifically, we define \textit{FactSet} as equal to one for analysts (or brokerage firms) who have cited \textsc{FactSet} at least once during the sample period, and zero otherwise, after excluding switching analysts and brokers. Online Appendix~\ref{app:robustness} reports the results, which are qualitatively similar to those obtained from our main tests.

\section{Conclusion}\label{sec:conclusion}
This paper studies how financial analysts' information production changes when a major data platform integrates GenAI into its workflow. Following the December 2023 introduction of \textsc{FactSet Mercury}, reports associated with \textsc{FactSet} become richer across information sources, topical coverage, and analytical methods, and are produced more promptly. Because our treatment is based on observable \textsc{FactSet} citations rather than direct logs of \textsc{Mercury} usage, we interpret these estimates as the effect of exposure to \textsc{FactSet} after its AI integration. Placebo tests using other vendors make a common data-platform trend unlikely, while a broad set of alternative specifications yields similar conclusions.

The central economic insight is that relaxing information-acquisition constraints can shift the bottleneck to human information processing. Average relative forecast accuracy declines after the AI integration, but the deterioration is concentrated among analysts facing greater processing demands. Post-integration \textsc{FactSet} reports also contain less aligned signals, and signal misalignment is associated with lower accuracy. A random-forest benchmark processing the same observable report signals, firm fundamentals, and recent consensus forecasts shows no analogous deterioration, including in the high-workload subsamples where human accuracy falls. Together with evidence against poorer signal quality, textual redundancy, and a ``speed-over-review'' channel, these results favor an interpretation in which AI expands information supply faster than scarce human attention can verify and integrate it.

The managerial implication is therefore not simply to deploy more or less AI. As information acquisition becomes cheaper, organizations may need to redesign workflows around the new scarce input: human attention devoted to validation, prioritization, and synthesis. Interfaces that reconcile conflicting evidence, review protocols that intensify under high workloads, and staffing choices that protect processing capacity may determine whether AI-generated information translates into better decisions. More broadly, our findings suggest that the value of GenAI in professional information intermediation depends not only on what the technology can retrieve or generate, but also on the capacity of human users to absorb it.

\clearpage

\newgeometry{top=3cm, bottom=3cm, left=2cm, right=2cm}
\setlength{\bibsep}{0pt plus 0.3ex}
\makeatletter
\patchcmd{\thebibliography}
  {\settowidth}
  {\setlength{\itemsep}{0pt} \setlength{\parskip}{0pt} \settowidth}
  {}{}
\makeatother

\bibliographystyle{apalike}
\bibliography{reference}
\restoregeometry

\section{Figures \& Tables}

\newgeometry{left=1.2cm, right=1.2cm, top=2cm, bottom=2cm}
\renewcommand{\arraystretch}{1} 
\newlength{\tablewidth}

\begin{table}[htbp]
\centering
\setlength{\tabcolsep}{6pt}
\caption{Sample Construction}
\label{tab:sample_construction}
\caption*{\footnotesize
This table reports the sample construction procedure. The sample consists of U.S. equity research reports from January 2022 to October 2024. 
The left column presents the number of reports, and the right column presents the number of report-firm pairs after expanding reports that cover multiple firms.
}
\vspace{1em}
\begin{tabular}{lrr}
\toprule
Sample Selection & Reports & Report-Firm Pairs \\
\midrule
U.S. analyst equity research reports from January 2022 to October 2024 & 162,862 & 315,499 \\
Drop reports that cannot be matched to I/B/E/S (including reiterating reports) & (106,914) & (247,924) \\
Drop analysts that do not appear in both pre- and post-periods & (11,907) & (14,945) \\
Drop analyst-firm pairs where coverage is terminated in pre-period & (4,511) & (5,335) \\
Drop observations with main variables missing & (354) & (422) \\
\midrule
Main sample & 39,176 & 46,853 \\
\bottomrule
\end{tabular}
\end{table}

\begin{table}[htbp]
\centering
\setlength{\tabcolsep}{4pt}
\caption{Descriptive Statistics}
\label{tab:summary_stats}
\caption*{\footnotesize
This table reports summary statistics for the main sample of 46,853 report-firm observations. Variable definitions are provided in Appendix~\ref{sec:variable_definition}. Panel A relates to content richness, Panel B reports other report attributes, and Panel C reports analyst team attributes.
}
\vspace{1em}
\begin{tabular}{lcccccccc}
\toprule
Variable & {N} & {Mean} & {SD} & {Min} & {$p_{25}$} & {$p_{50}$} & {$p_{75}$} & {Max} \\
\midrule
\multicolumn{9}{c}{\textit{Panel A: Content richness}} \\
\midrule
\#TextSources & 46,853 & 8.805 & 6.789 & 1.000 & 5.000 & 7.000 & 11.000 & 37.000 \\
\#FigureSources & 46,853 & 3.489 & 2.990 & 0.000 & 2.000 & 3.000 & 5.000 & 16.000 \\
\#TableSources & 46,853 & 7.710 & 5.031 & 1.000 & 4.000 & 6.000 & 9.000 & 28.000 \\
\#FirmTopics & 46,853 & 12.504 & 8.423 & 3.000 & 7.000 & 10.000 & 14.000 & 46.000 \\
\#IndustryTopics & 46,853 & 5.609 & 4.297 & 0.000 & 3.000 & 4.000 & 7.000 & 23.000 \\
\#MacroTopics & 46,853 & 6.461 & 4.810 & 0.000 & 3.000 & 5.000 & 8.000 & 25.000 \\
\#HistMethods & 46,853 & 1.622 & 1.070 & 0.000 & 1.000 & 1.000 & 2.000 & 5.000 \\
\#ValMethods & 46,853 & 2.866 & 2.047 & 0.000 & 2.000 & 2.000 & 4.000 & 11.000 \\
\#FcMethods & 46,853 & 3.839 & 2.911 & 0.000 & 2.000 & 3.000 & 5.000 & 15.000 \\
\midrule
\multicolumn{9}{c}{\textit{Panel B: Other report attributes}} \\
\midrule
Post & 46,853 & 0.395 & 0.489 & 0.000 & 0.000 & 0.000 & 1.000 & 1.000 \\
FactSet & 46,853 & 0.334 & 0.472 & 0.000 & 0.000 & 0.000 & 1.000 & 1.000 \\
Bloomberg & 46,853 & 0.439 & 0.496 & 0.000 & 0.000 & 0.000 & 1.000 & 1.000 \\
Refinitiv & 46,853 & 0.142 & 0.349 & 0.000 & 0.000 & 0.000 & 0.000 & 1.000 \\
CapitalIQ & 46,853 & 0.027 & 0.162 & 0.000 & 0.000 & 0.000 & 0.000 & 1.000 \\
Prompt & 46,853 & 0.294 & 0.456 & 0.000 & 0.000 & 0.000 & 1.000 & 1.000 \\
Accuracy & 36,967 & 0.435 & 1.498 & -4.910 & -0.211 & 0.278 & 0.987 & 9.426 \\
AbnVol & 44,518 & 0.602 & 1.104 & -1.185 & -0.008 & 0.261 & 0.783 & 7.153 \\
\midrule
\multicolumn{9}{c}{\textit{Panel C: Analyst team attributes}} \\
\midrule
Team-Career & 46,853 & 16.529 & 11.493 & 0.362 & 6.855 & 13.123 & 23.578 & 40.940 \\
Firm-Experience & 46,853 & 5.755 & 5.768 & 0.000 & 1.477 & 3.740 & 8.156 & 23.660 \\
IT-Experience & 46,853 & 0.457 & 0.498 & 0.000 & 0.000 & 0.000 & 1.000 & 1.000 \\
Elite-Education & 46,853 & 0.359 & 0.480 & 0.000 & 0.000 & 0.000 & 1.000 & 1.000 \\
Brokerage-Size & 46,853 & 100.534 & 70.725 & 7.000 & 37.000 & 70.000 & 178.000 & 199.000 \\
Team-Size & 46,853 & 2.853 & 1.281 & 1.000 & 2.000 & 3.000 & 4.000 & 6.000 \\
Forecast-Frequency & 46,853 & 4.903 & 3.121 & 0.000 & 3.000 & 5.000 & 7.000 & 14.000 \\
Firms-Followed & 46,853 & 21.826 & 9.909 & 0.000 & 15.000 & 22.000 & 28.000 & 46.000 \\
\bottomrule
\end{tabular}
\end{table}

\begin{table}[htbp]
\centering
\setlength{\tabcolsep}{4pt}
\caption{Characteristics of AI Platform Adopters}
\label{tab:ai_adopting_analysts}
\caption*{\footnotesize
This table examines the characteristics of \textsc{FactSet}-adopting analysts around the introduction of the AI-powered platform.
The first column reports the coefficients and $t$-statistics from regressing the indicator of whether \textsc{FactSet} is cited in the report on the regressors. 
The second column reports the results for usage intensity, defined as the number of \textsc{FactSet} citations scaled by the total number of data platform citations within the report.
All regressions include report date and firm-year fixed effects, with standard errors clustered at the analyst and firm levels.
$t$-statistics are reported in parentheses. 
$^{***}$, $^{**}$, and $^{*}$ denote significance at the 1\%, 5\%, and 10\% levels, respectively.
}
\vspace{1em}
\begin{tabular}{l*{4}{S[table-format=3.3]@{\,}l}}
\toprule
& \multicolumn{2}{c}{FactSet} & \multicolumn{2}{c}{\%FactSet} \\
\midrule
Post$\times$Team-Career    & 0.061          & (0.98)         & 0.019          & (0.64)        \\
Post$\times$Firm-Experience & -0.062         & (-1.41)        & 0.029          & (1.14)        \\
Post$\times$IT-Experience   & -0.022         & (-0.55)        & 0.013          & (0.63)        \\
Post$\times$Elite-Education & 0.037          & (0.97)         & -0.004         & (-0.22)       \\
Post$\times$Brokerage-Size  & -0.115$^{**}$  & (-2.24)        & -0.092$^{***}$ & (-4.31)       \\
Post$\times$Team-Size       & 0.080$^{**}$   & (2.06)         & 0.077$^{***}$  & (4.18)        \\
Post$\times$Forecast-Frequency & 0.036       & (0.69)         & 0.022          & (0.77)        \\
Post$\times$Firms-Followed  & -0.080         & (-1.41)        & -0.051         & (-1.47)       \\
Team-Career                & -0.329$^{***}$ & (-5.47)        & -0.111$^{***}$ & (-3.37)       \\
Firm-Experience             & -0.082$^{*}$   & (-1.95)        & -0.047$^{*}$   & (-1.87)       \\
IT-Experience               & -0.008         & (-0.19)        & 0.022          & (1.06)        \\
Elite-Education              & 0.055$^{*}$    & (1.71)         & 0.044$^{***}$  & (2.59)        \\
Brokerage-Size              & -0.586$^{***}$ & (-8.89)        & -0.485$^{***}$ & (-10.57)      \\
Team-Size                   & 0.061$^{**}$   & (2.06)         & 0.017          & (1.30)        \\
Forecast-Frequency          & 0.110$^{**}$   & (2.54)         & 0.037$^{*}$    & (1.67)        \\
Firms-Followed              & 0.159$^{**}$   & (2.03)         & 0.070$^{**}$   & (2.11)        \\
Bloomberg                   & -0.137$^{***}$ & (-4.46)        & -0.571$^{***}$ & (-17.47)      \\
Refinitiv                   & -0.327$^{***}$ & (-2.81)        & -0.604$^{***}$ & (-4.64)       \\
CapitalIQ                   & -0.058$^{*}$   & (-1.66)        & -0.568$^{***}$ & (-20.03)      \\
\midrule
Report Date FE & \multicolumn{2}{c}{Yes} & \multicolumn{2}{c}{Yes} \\
Firm-Year FE   & \multicolumn{2}{c}{Yes} & \multicolumn{2}{c}{Yes} \\
\midrule
N               & \multicolumn{2}{c}{41,631} & \multicolumn{2}{c}{33,642} \\
Adjusted $R^{2}$ & \multicolumn{2}{c}{0.487} & \multicolumn{2}{c}{0.831} \\
\bottomrule
\end{tabular}
\end{table}

\begin{table}[htbp]
\centering
\setlength{\tabcolsep}{4pt}
\caption{AI Platform Adoption and Information Sources in Analyst Reports}
\label{tab:baseline_sources}
\caption*{\footnotesize
This table reports the effect of \textsc{FactSet} AI platform adoption on information sources in analyst reports. 
The dependent variables are the number of textual sources (\textit{\#TextSources}), figure sources (\textit{\#FigureSources}), and table sources (\textit{\#TableSources}).
All regressions include report date, firm-year, analyst, and broker-year fixed effects, with standard errors clustered at the analyst and firm levels.
$t$-statistics are reported in parentheses below coefficient estimates.
$^{***}$, $^{**}$, and $^{*}$ denote significance at the 1\%, 5\%, and 10\% levels, respectively.
}
\vspace{1em}
\begin{tabular}{lccc}
\toprule
& \#TextSources & \#FigureSources & \#TableSources \\
\midrule
FactSet & 1.721$^{***}$ & 0.790$^{***}$ & 1.760$^{***}$ \\
& (5.82) & (6.14) & (7.57) \\
FactSet$\times$Post & 2.273$^{***}$ & 0.977$^{***}$ & 1.690$^{***}$ \\
& (4.73) & (4.87) & (4.76) \\
Team-Career & -0.205 & -0.143 & -0.915 \\
& (-0.21) & (-0.38) & (-1.47) \\
Firm-Experience & -0.533$^{**}$ & -0.027 & -0.121 \\
& (-2.45) & (-0.29) & (-0.70) \\
IT-Experience & 0.625$^{*}$ & 0.231$^{*}$ & 0.124 \\
& (1.70) & (1.82) & (0.62) \\
Elite-Education & 0.603$^{*}$ & 0.255$^{*}$ & 0.462$^{**}$ \\
& (1.67) & (1.96) & (2.22) \\
Brokerage-Size & -1.060$^{*}$ & -0.336 & -0.276 \\
& (-1.69) & (-1.25) & (-0.61) \\
Team-Size & 0.474$^{*}$ & 0.051 & 0.366$^{**}$ \\
& (1.76) & (0.46) & (2.02) \\
Forecast-Frequency & -0.073 & -0.163$^{*}$ & -0.101 \\
& (-0.30) & (-1.69) & (-0.59) \\
Firms-Followed & -0.181 & 0.090 & 0.248 \\
& (-0.42) & (0.50) & (0.85) \\
\midrule
Report Date FE & Yes & Yes & Yes \\
Firm-Year FE & Yes & Yes & Yes \\
Analyst FE & Yes & Yes & Yes \\
Broker-Year FE & Yes & Yes & Yes \\
\midrule
N & 41,391 & 41,391 & 41,391 \\
Adjusted R$^{2}$ & 0.372 & 0.398 & 0.415 \\
\bottomrule
\end{tabular}
\end{table}

\begin{table}[htbp]
\centering
\setlength{\tabcolsep}{4pt}
\caption{AI Platform Adoption and Topic Coverage in Analyst Reports}
\label{tab:baseline_topics}
\caption*{\footnotesize
This table reports the effect of \textsc{FactSet} AI platform adoption on topic coverage in analyst reports.
The dependent variables are the number of firm-level topics (\textit{\#FirmTopics}), industry-level topics (\textit{\#IndustryTopics}), and macroeconomic topics (\textit{\#MacroTopics}).
All regressions include report date, firm-year, analyst, and broker-year fixed effects, with standard errors clustered at the analyst and firm levels.
$t$-statistics are reported in parentheses below coefficient estimates.
$^{***}$, $^{**}$, and $^{*}$ denote significance at the 1\%, 5\%, and 10\% levels, respectively.
}
\vspace{1em}
\begin{tabular}{lccc}
\toprule
& \#FirmTopics & \#IndustryTopics & \#MacroTopics \\
\midrule
FactSet & 2.293$^{***}$ & 0.921$^{***}$ & 1.265$^{***}$ \\
& (6.36) & (5.38) & (6.88) \\
FactSet$\times$Post & 2.874$^{***}$ & 1.394$^{***}$ & 1.506$^{***}$ \\
& (4.67) & (5.09) & (5.16) \\
Team-Career & 0.055 & -0.135 & -0.075 \\
& (0.04) & (-0.23) & (-0.11) \\
Firm-Experience & -0.701$^{***}$ & -0.219$^{*}$ & 0.033 \\
& (-2.67) & (-1.77) & (0.23) \\
IT-Experience & 0.377 & 0.350$^{*}$ & 0.280 \\
& (0.96) & (1.78) & (1.38) \\
Elite-Education & 1.041$^{***}$ & 0.452$^{**}$ & 0.464$^{**}$ \\
& (2.59) & (2.16) & (2.19) \\
Brokerage-Size & -1.068 & -0.385 & -0.439 \\
& (-1.39) & (-0.92) & (-1.00) \\
Team-Size & 1.077$^{***}$ & 0.377$^{**}$ & 0.409$^{**}$ \\
& (3.18) & (2.24) & (2.31) \\
Forecast-Frequency & -0.154 & -0.218 & -0.373$^{**}$ \\
& (-0.51) & (-1.47) & (-2.37) \\
Firms-Followed & -0.177 & -0.198 & -0.166 \\
& (-0.34) & (-0.71) & (-0.52) \\
\midrule
Report Date FE & Yes & Yes & Yes \\
Firm-Year FE & Yes & Yes & Yes \\
Analyst FE & Yes & Yes & Yes \\
Broker-Year FE & Yes & Yes & Yes \\
\midrule
N & 41,391 & 41,391 & 41,391 \\
Adjusted R$^{2}$ & 0.382 & 0.386 & 0.422 \\
\bottomrule
\end{tabular}
\end{table}

\begin{table}[htbp]
\centering
\setlength{\tabcolsep}{4pt}
\caption{AI Platform Adoption and Analytic Methods in Analyst Reports}
\label{tab:baseline_methods}
\caption*{\footnotesize
This table reports the effect of \textsc{FactSet} AI platform adoption on analytic methods used in analyst reports.
The dependent variables are the number of historical trend analysis methods (\textit{\#HistMethods}), valuation methods (\textit{\#ValMethods}), and forecasting methods (\textit{\#FcMethods}).
All regressions include report date, firm-year, analyst, and broker-year fixed effects, with standard errors clustered at the analyst and firm levels.
$t$-statistics are reported in parentheses below coefficient estimates.
$^{***}$, $^{**}$, and $^{*}$ denote significance at the 1\%, 5\%, and 10\% levels, respectively.
}
\vspace{1em}
\begin{tabular}{lccc}
\toprule
& \#HistMethods & \#ValMethods & \#FcMethods \\
\midrule
FactSet & 0.324$^{***}$ & 0.515$^{***}$ & 0.861$^{***}$ \\
& (7.90) & (5.70) & (6.98) \\
FactSet$\times$Post & 0.169$^{***}$ & 0.533$^{***}$ & 1.010$^{***}$ \\
& (2.81) & (4.56) & (5.38) \\
Team-Career & -0.115 & 0.050 & 0.241 \\
& (-0.67) & (0.17) & (0.60) \\
Firm-Experience & 0.007 & -0.102 & -0.048 \\
& (0.18) & (-1.29) & (-0.47) \\
IT-Experience & 0.034 & 0.128 & 0.256$^{**}$ \\
& (0.88) & (1.54) & (2.16) \\
Elite-Education & 0.122$^{***}$ & 0.183$^{**}$ & 0.283$^{**}$ \\
& (3.24) & (2.16) & (2.14) \\
Brokerage-Size & -0.098 & -0.080 & -0.217 \\
& (-0.87) & (-0.39) & (-0.87) \\
Team-Size & 0.024 & 0.203$^{***}$ & 0.187$^{*}$ \\
& (0.76) & (2.74) & (1.84) \\
Forecast-Frequency & -0.013 & 0.031 & -0.127 \\
& (-0.37) & (0.45) & (-1.33) \\
Firms-Followed & -0.019 & 0.000 & -0.180 \\
& (-0.28) & (0.00) & (-1.06) \\
\midrule
Report Date FE & Yes & Yes & Yes \\
Firm-Year FE & Yes & Yes & Yes \\
Analyst FE & Yes & Yes & Yes \\
Broker-Year FE & Yes & Yes & Yes \\
\midrule
N & 41,391 & 41,391 & 41,391 \\
Adjusted R$^{2}$ & 0.286 & 0.283 & 0.396 \\
\bottomrule
\end{tabular}
\end{table}

\begin{sidewaystable}
\centering
\scriptsize
\caption{Dynamic Effects of AI Platform Adoption}
\label{tab:dynamic_effects}
\caption*{\footnotesize
This table reports the dynamic effects of \textsc{FactSet} AI platform adoption on analyst report content.
The dependent variables are nine measures of report richness: information sources (\textit{\#TextSources}, \textit{\#FigureSources}, \textit{\#TableSources}), 
topic coverage (\textit{\#FirmTopics}, \textit{\#IndustryTopics}, \textit{\#MacroTopics}), and analytic methods (\textit{\#HistMethods}, \textit{\#ValMethods}, \textit{\#FcMethods}).
$Y2023Q1$--$Y2023Q4$ represent the four quarters before AI introduction, and $Y2024Q1$--$Y2024Q4$ represent the four quarters after AI introduction.
All regressions include report date, firm-year, analyst, and broker-year fixed effects, with standard errors clustered at the analyst and firm levels.
$t$-statistics are reported in parentheses below coefficient estimates.
$^{***}$, $^{**}$, and $^{*}$ denote significance at the 1\%, 5\%, and 10\% levels, respectively.
}
\resizebox{\textwidth}{!}{%
\begin{tabular}{l*{9}{>{\raggedleft\arraybackslash}p{1.4cm}}}
\toprule
& \#TextSources & \#FigureSources & \#TableSources & \#FirmTopics & \#IndustryTopics & \#MacroTopics & \#HistMethods & \#ValMethods & \#FcMethods \\
\midrule
FactSet$\times$Y2023Q1 & 0.465 & 0.116 & 0.499$^{*}$ & 0.586 & 0.100 & 0.155 & 0.055 & 0.097 & 0.255 \\
& (0.95) & (0.70) & (1.77) & (0.85) & (0.33) & (0.50) & (0.84) & (0.76) & (1.25) \\
FactSet$\times$Y2023Q2 & 0.252 & -0.024 & 0.029 & 0.309 & 0.062 & -0.049 & 0.002 & 0.113 & 0.035 \\
& (0.68) & (-0.16) & (0.12) & (0.62) & (0.25) & (-0.18) & (0.04) & (0.93) & (0.21) \\
FactSet$\times$Y2023Q3 & -0.143 & -0.085 & 0.073 & -0.123 & -0.284 & -0.298 & -0.013 & -0.071 & -0.266 \\
& (-0.30) & (-0.43) & (0.25) & (-0.19) & (-0.93) & (-0.99) & (-0.18) & (-0.53) & (-1.35) \\
FactSet$\times$Y2023Q4 & -0.647 & -0.212 & -0.456 & -0.507 & -0.412 & -0.662$^{**}$ & -0.145$^{**}$ & -0.073 & -0.203 \\
& (-1.49) & (-1.25) & (-1.61) & (-0.89) & (-1.63) & (-2.44) & (-2.41) & (-0.60) & (-1.23) \\
FactSet$\times$Y2024Q1 & 2.125$^{***}$ & 0.840$^{***}$ & 1.348$^{***}$ & 2.811$^{***}$ & 1.211$^{***}$ & 1.249$^{***}$ & 0.107 & 0.565$^{***}$ & 0.900$^{***}$ \\
& (3.80) & (3.49) & (3.29) & (3.81) & (3.79) & (3.47) & (1.38) & (3.77) & (3.99) \\
FactSet$\times$Y2024Q2 & 2.032$^{***}$ & 0.876$^{***}$ & 1.616$^{***}$ & 2.633$^{***}$ & 1.204$^{***}$ & 1.244$^{***}$ & 0.189$^{**}$ & 0.471$^{***}$ & 0.946$^{***}$ \\
& (3.04) & (3.00) & (3.30) & (3.06) & (3.08) & (2.98) & (2.03) & (2.75) & (3.44) \\
FactSet$\times$Y2024Q3 & 2.084$^{***}$ & 0.903$^{***}$ & 2.055$^{***}$ & 2.780$^{***}$ & 1.294$^{***}$ & 1.192$^{***}$ & 0.144 & 0.535$^{***}$ & 1.017$^{***}$ \\
& (2.98) & (3.52) & (4.21) & (3.21) & (3.35) & (2.99) & (1.64) & (3.14) & (3.85) \\
FactSet$\times$Y2024Q4 & 3.010$^{***}$ & 1.189$^{***}$ & 2.419$^{***}$ & 3.401$^{***}$ & 1.513$^{***}$ & 1.730$^{***}$ & 0.184 & 0.696$^{***}$ & 1.019$^{***}$ \\
& (3.16) & (3.20) & (3.61) & (2.82) & (2.94) & (3.03) & (1.48) & (2.92) & (2.78) \\
\midrule
Controls & Yes & Yes & Yes & Yes & Yes & Yes & Yes & Yes & Yes \\
Report Date FE & Yes & Yes & Yes & Yes & Yes & Yes & Yes & Yes & Yes \\
Firm-Year FE & Yes & Yes & Yes & Yes & Yes & Yes & Yes & Yes & Yes \\
Analyst FE & Yes & Yes & Yes & Yes & Yes & Yes & Yes & Yes & Yes \\
Broker-Year FE & Yes & Yes & Yes & Yes & Yes & Yes & Yes & Yes & Yes \\
\midrule
N & 41,391 & 41,391 & 41,391 & 41,391 & 41,391 & 41,391 & 41,391 & 41,391 & 41,391 \\
Adjusted R$^{2}$ & 0.373 & 0.398 & 0.416 & 0.382 & 0.386 & 0.423 & 0.287 & 0.283 & 0.396 \\
\bottomrule
\end{tabular}
}
\end{sidewaystable}

\begin{table}[htbp]
\centering
\setlength{\tabcolsep}{4pt}
\caption{AI Platform Adoption and Report Quality}
\label{tab:ai_report_quality}
\caption*{\footnotesize
This table reports the effect of \textsc{FactSet} AI platform adoption on report quality.
The dependent variables are forecast timeliness ($Prompt$), forecast accuracy ($Accuracy$), and abnormal trading volume ($AbnVol$).
All regressions include report date, firm-year, analyst, and broker-year fixed effects, with standard errors clustered at the analyst and firm levels.
$t$-statistics are reported in parentheses below coefficient estimates.
$^{***}$, $^{**}$, and $^{*}$ denote significance at the 1\%, 5\%, and 10\% levels, respectively.
}
\vspace{1em}
\begin{tabular}{lccc}
\toprule
& Prompt & Accuracy & AbnVol \\
\midrule
FactSet & -0.040$^{***}$ & 0.007 & 0.001 \\
& (-3.09) & (0.18) & (0.02) \\
FactSet$\times$Post & 0.033$^{**}$ & -0.157$^{***}$ & -0.127$^{***}$ \\
& (2.00) & (-2.96) & (-3.41) \\
Team-Career & 0.082 & -0.320 & 0.128 \\
& (1.47) & (-0.93) & (0.97) \\
Firm-Experience & -0.020 & 0.032 & 0.049$^{*}$ \\
& (-1.27) & (0.54) & (1.66) \\
IT-Experience & -0.005 & -0.024 & 0.021 \\
& (-0.45) & (-0.64) & (0.77) \\
Elite-Education & -0.013 & -0.006 & -0.050$^{*}$ \\
& (-1.33) & (-0.14) & (-1.83) \\
Brokerage-Size & -0.088$^{*}$ & -0.204 & -0.114 \\
& (-1.78) & (-1.14) & (-1.30) \\
Team-Size & 0.017$^{*}$ & 0.038 & 0.003 \\
& (1.81) & (0.91) & (0.14) \\
Forecast-Frequency & 0.004 & -0.040 & -0.012 \\
& (0.34) & (-0.68) & (-0.43) \\
Firms-Followed & 0.003 & 0.226$^{*}$ & -0.037 \\
& (0.10) & (1.93) & (-0.62) \\
\midrule
Report Date FE & Yes & Yes & Yes \\
Firm-Year FE & Yes & Yes & Yes \\
Analyst FE & Yes & Yes & Yes \\
Broker-Year FE & Yes & Yes & Yes \\
\midrule
N & 41,619 & 33,079 & 39,740 \\
Adjusted R$^{2}$ & 0.513 & 0.279 & 0.478 \\
\bottomrule
\end{tabular}
\end{table}

\begin{table}[htbp]
\centering
\setlength{\tabcolsep}{4pt}
\caption{Mechanism Tests}
\label{tab:mechanism_tests_p1}
\caption*{\footnotesize
This table reports mechanism tests for the decline in forecast accuracy following AI adoption.
Panel A examines the relationship between AI adoption, signal misalignment ($SignalSD$), and forecast accuracy. $SignalSD$ is the standard deviation of signal sentiment scores (positive=1, negative=-1, neutral=0) extracted from each report; higher values indicate greater signal inconsistency.
Panel B presents subsample analyses by attention constraints: \textit{Firms-Followed} (number of firms covered, split at median), \textit{Team-Size} (team size, split at median), and \textit{Busy-Day} (indicator for revising forecasts for two or more firms on the same day). Coeff. Diff. One-tailed $p$-values from interaction models are reported in brackets. All specifications include report date, firm-year, analyst, and broker-year fixed effects, with standard errors clustered at the analyst and firm levels. $t$-statistics in parentheses. $^{***}$, $^{**}$, and $^{*}$ denote significance at the 1\%, 5\%, and 10\% levels, respectively.
}
\vspace{1em}
\begin{tabular}{lcccc}
\toprule
\multicolumn{5}{c}{\textit{Panel A: Signal inconsistency}} \\
\midrule
& \multicolumn{2}{c}{SignalSD} & \multicolumn{2}{c}{Accuracy} \\
\midrule
FactSet & 0.033$^{***}$ & (5.89) & & \\
FactSet$\times$Post & 0.016$^{**}$ & (2.33) & & \\
SignalSD & & & -0.256$^{***}$ & (-3.48) \\
\midrule
Controls \& FEs & \multicolumn{2}{c}{Yes} & \multicolumn{2}{c}{Yes} \\
\midrule
N & \multicolumn{2}{c}{41,387} & \multicolumn{2}{c}{32,874} \\
Adjusted R$^{2}$ & \multicolumn{2}{c}{0.240} & \multicolumn{2}{c}{0.277} \\
\midrule
\multicolumn{5}{c}{\textit{Panel B: Cross-sectional tests}} \\
\midrule
& \multicolumn{2}{c}{Accuracy} & \multicolumn{2}{c}{SignalSD} \\
\cmidrule(lr){2-3} \cmidrule(lr){4-5}
& High & Low & High & Low \\
\midrule
\multicolumn{5}{l}{\textit{Firms-Followed}} \\
FactSet$\times$Post & -0.174$^{**}$ & 0.044 & 0.025$^{**}$ & 0.007 \\
& (-2.40) & (0.49) & (2.45) & (0.65) \\
Coeff. Diff. $p$-value & \multicolumn{2}{c}{[0.029$^{**}$]} & \multicolumn{2}{c}{[0.022$^{**}$]} \\
N & 17,565 & 14,906 & 22,179 & 18,596 \\
Adjusted R$^{2}$ & 0.295 & 0.296 & 0.254 & 0.240 \\
\midrule
\multicolumn{5}{l}{\textit{Team-Size}} \\
FactSet$\times$Post & -0.020 & -0.191$^{***}$ & 0.005 & 0.020$^{**}$ \\
& (-0.19) & (-2.74) & (0.44) & (2.36) \\
Coeff. Diff. $p$-value & \multicolumn{2}{c}{[0.069$^{*}$]} & \multicolumn{2}{c}{[0.044$^{**}$]} \\
N & 14,122 & 17,539 & 17,624 & 22,282 \\
Adjusted R$^{2}$ & 0.313 & 0.307 & 0.239 & 0.248 \\
\midrule
\multicolumn{5}{l}{\textit{Busy-Day}} \\
FactSet$\times$Post & -0.205$^{***}$ & -0.066 & 0.023$^{***}$ & 0.009 \\
& (-2.97) & (-0.77) & (2.83) & (0.88) \\
Coeff. Diff. $p$-value & \multicolumn{2}{c}{[0.092$^{*}$]} & \multicolumn{2}{c}{[0.051$^{*}$]} \\
N & 19,643 & 12,003 & 23,561 & 16,407 \\
Adjusted R$^{2}$ & 0.305 & 0.388 & 0.275 & 0.203 \\
\bottomrule
\end{tabular}
\end{table}

\begin{table}[htbp]
\centering
\setlength{\tabcolsep}{4pt}
\caption*{Table 9: Mechanism Tests (Continued)}
\label{tab:mechanism_tests_p2}
\caption*{\footnotesize
This table reports mechanism tests for the decline in forecast accuracy following AI adoption.
Panel C compares human analyst forecasts with machine learning (ML) forecasts using a random forest model trained on all information available to analysts. We repeat the random sample split 100 times for inference; reported coefficients and $t$-statistics are averages across all iterations. RMSE reports out-sample root mean squared error. ML Accuracy is measured following the same definition as for human analysts.
Panel D replicates the cross-sectional tests from Panel B using ML forecasts. All specifications include report date, firm-year, analyst, and broker-year fixed effects, with standard errors clustered at the analyst and firm levels, and Monte-Carlo $t$-statistics reported in parentheses. $^{***}$, $^{**}$, and $^{*}$ denote significance at the 1\%, 5\%, and 10\% levels, respectively.
}
\vspace{1em}
\begin{tabular}{lccc}
\toprule
\multicolumn{4}{c}{\textit{Panel C: ML forecasts}} \\
\midrule
& (1) Analyst Forecasts & (2) ML Forecasts & (1) -- (2) \\
\midrule
RMSE & 2.143 & 1.851 & 0.291$^{***}$ \\
& & & (4.98) \\
\midrule
& (1) Analyst Accuracy & (2) ML Accuracy & (1) -- (2) \\
\midrule
FactSet$\times$Post Coef. & -0.100$^{***}$ & 0.018 & -0.119$^{***}$ \\
& (-9.59) & (0.89) & (-5.23) \\
\midrule
\multicolumn{4}{c}{\textit{Panel D: Cross-sectional tests on ML forecasts}} \\
\midrule
& (1) ML Accuracy & (2) ML Accuracy & (1) -- (2) \\
\cmidrule(lr){2-3}
& High & Low & \\
\midrule
\multicolumn{4}{l}{\textit{Firms-Followed}} \\
FactSet$\times$Post Coef. & 0.025 & 0.001 & 0.024 \\
& (1.06) & (0.04) & (0.76) \\
\midrule
\multicolumn{4}{l}{\textit{Team-Size}} \\
FactSet$\times$Post Coef. & 0.073$^{**}$ & -0.002 & 0.075$^{*}$ \\
& (2.39) & (-0.08) & (1.92) \\
\midrule
\multicolumn{4}{l}{\textit{Busy-Day}} \\
FactSet$\times$Post Coef. & 0.031 & -0.005 & 0.036 \\
& (1.40) & (-0.15) & (1.04) \\
\bottomrule
\end{tabular}
\end{table}

\begin{table}[htbp]
\centering
\setlength{\tabcolsep}{4pt}
\caption{Alternative Explanations}
\label{tab:alt_explanations}
\caption*{\footnotesize
This table examines alternative explanations for the decline in forecast accuracy.
Panel A tests whether AI-assisted reports contain signals with lower earnings predictive power.
$SignalPos$ and $SignalNeg$ are the proportions of positive and negative signals extracted from each report.
Panel B tests whether AI-assisted reports differ in textual quality and redundancy.
$ContentTTR$ and $NounTTR$ are type-token ratios measuring lexical diversity (lower values indicate greater redundancy). $Fog$ and $FRE$ are readability measures.
Panel C tests whether the decline in accuracy is driven by analysts prioritizing speed over review.
$Prompt$ is an indicator for reports issued within three days after an earnings announcement.
All regressions include report date, firm-year (or firm FE in Panel A), analyst, and broker-year fixed effects, with standard errors clustered at the analyst and firm levels.
$t$-statistics are reported in parentheses below coefficient estimates.
$^{***}$, $^{**}$, and $^{*}$ denote significance at the 1\%, 5\%, and 10\% levels, respectively.
}
\vspace{1em}
\begin{tabular}{lcccc}
\toprule
\multicolumn{5}{c}{\textit{Panel A: Signal quality}} \\
\midrule
& \multicolumn{2}{c}{EpsChange} & \multicolumn{2}{c}{EpsChangeDum} \\
\midrule
SignalPos & 0.014$^{***}$ & (3.24) & 0.378$^{***}$ & (6.27) \\
SignalNeg & -0.025$^{***}$ & (-3.91) & -0.615$^{***}$ & (-6.64) \\
SignalPos$\times$FactSet$\times$Post & 0.001 & (0.07) & 0.081 & (0.54) \\
SignalNeg$\times$FactSet$\times$Post & -0.007 & (-0.37) & 0.027 & (0.13) \\
\midrule
Controls \& FEs & \multicolumn{2}{c}{Yes} & \multicolumn{2}{c}{Yes} \\
\midrule
N & \multicolumn{2}{c}{38,028} & \multicolumn{2}{c}{38,750} \\
Adjusted R$^{2}$ & \multicolumn{2}{c}{0.502} & \multicolumn{2}{c}{0.503} \\
\midrule
\multicolumn{5}{c}{\textit{Panel B: Textual output quality}} \\
\midrule
& ContentTTR & NounTTR & Fog & FRE \\
\midrule
FactSet & -0.049$^{***}$ & -0.049$^{***}$ & -0.299$^{**}$ & 0.453 \\
& (-6.41) & (-6.43) & (-2.58) & (0.75) \\
FactSet$\times$Post & -0.006 & -0.003 & 0.157 & 0.687 \\
& (-0.75) & (-0.38) & (1.20) & (1.20) \\
\midrule
Controls \& FEs & Yes & Yes & Yes & Yes \\
\midrule
N & 40,069 & 40,069 & 41,594 & 41,594 \\
Adjusted R$^{2}$ & 0.555 & 0.554 & 0.656 & 0.664 \\
\midrule
\multicolumn{5}{c}{\textit{Panel C: Speed-over-review}} \\
\midrule
& \multicolumn{2}{c}{Accuracy} & \multicolumn{2}{c}{AbnVol} \\
\midrule
Prompt & 0.162$^{***}$ & (3.57) & 0.448$^{***}$ & (10.42) \\
Prompt$\times$FactSet & -0.081 & (-1.19) & -0.035 & (-0.62) \\
Prompt$\times$FactSet$\times$Post & 0.028 & (0.27) & 0.042 & (0.76) \\
Prompt$\times$Post & 0.063 & (0.71) & -0.042 & (-0.93) \\
FactSet & 0.036 & (0.79) & 0.029 & (0.79) \\
FactSet$\times$Post & -0.171$^{***}$ & (-2.89) & -0.154$^{***}$ & (-3.87) \\
\midrule
Controls \& FEs & \multicolumn{2}{c}{Yes} & \multicolumn{2}{c}{Yes} \\
\midrule
N & \multicolumn{2}{c}{33,079} & \multicolumn{2}{c}{39,740} \\
Adjusted R$^{2}$ & \multicolumn{2}{c}{0.281} & \multicolumn{2}{c}{0.493} \\
\bottomrule
\end{tabular}
\end{table}

\begin{table}[htbp]
\centering
\setlength{\tabcolsep}{4pt}
\caption{Placebo Tests}
\label{tab:placebo_tests}
\caption*{\footnotesize 
This table conducts placebo tests using other platforms. We define \textit{OtherPlatforms} as an indicator of whether a report is developed using Bloomberg, Refinitiv, or Capital IQ. Panels A through D report the effect on information sources, topical coverage, analytic methods, and report quality, respectively.
All regressions include report date, firm-year, analyst, and broker-year fixed effects, with standard errors clustered at the analyst and firm levels.
$t$-statistics are reported in parentheses below coefficient estimates.
$^{***}$, $^{**}$, and $^{*}$ denote significance at the 1\%, 5\%, and 10\% levels, respectively.
}
\vspace{1em}
\begin{tabular}{lccc}
\toprule
\multicolumn{4}{c}{\textit{Panel A: Information sources}} \\
\midrule
& \#TextSources & \#FigureSources & \#TableSources \\
\midrule
OtherPlatforms$\times$Post & -0.351 & 0.079 & -0.008 \\
& (-0.89) & (0.44) & (-0.02) \\
\midrule
Controls \& FEs & Yes & Yes & Yes \\
\midrule
N & 41,391 & 41,391 & 41,391 \\
Adjusted R$^{2}$ & 0.382 & 0.406 & 0.430 \\
\midrule
\multicolumn{4}{c}{\textit{Panel B: Topic coverage}} \\
\midrule
& \#FirmTopics & \#IndustryTopics & \#MacroTopics \\
\midrule
OtherPlatforms$\times$Post & -0.333 & -0.009 & -0.079 \\
& (-0.64) & (-0.04) & (-0.30) \\
\midrule
Controls \& FEs & Yes & Yes & Yes \\
\midrule
N & 41,391 & 41,391 & 41,391 \\
Adjusted R$^{2}$ & 0.394 & 0.395 & 0.434 \\
\midrule
\multicolumn{4}{c}{\textit{Panel C: Analytic methods}} \\
\midrule
& \#HistMethods & \#ValMethods & \#FcMethods \\
\midrule
OtherPlatforms$\times$Post & 0.001 & -0.201$^{*}$ & 0.001 \\
& (0.02) & (-1.88) & (0.00) \\
\midrule
Controls \& FEs & Yes & Yes & Yes \\
\midrule
N & 41,391 & 41,391 & 41,391 \\
Adjusted R$^{2}$ & 0.291 & 0.304 & 0.409 \\
\midrule
\multicolumn{4}{c}{\textit{Panel D: Report quality}} \\
\midrule
& Prompt & Accuracy & AbnVol \\
\midrule
OtherPlatforms$\times$Post & 0.002 & 0.033 & 0.015 \\
& (0.12) & (0.71) & (0.52) \\
\midrule
Controls \& FEs & Yes & Yes & Yes \\
\midrule
N & 41,790 & 33,214 & 39,902 \\
Adjusted R$^{2}$ & 0.515 & 0.279 & 0.480 \\
\bottomrule
\end{tabular}
\end{table}

\clearpage

\appendix
\section{Appendix}
\subsection{Variable definition}\label{sec:variable_definition}
\setlength{\tablewidth}{520pt}
\begin{longtable}{p{0.15\tablewidth}
p{0.85\tablewidth}}
\toprule
Variables & Definition \\
\midrule
\multicolumn{2}{l}{\underline{\textit{Content richness}}} \\
\textit{\#TextSources} & Number of sources cited in textual discussions of the analyst report. \\ 
\textit{\#FigureSources} & Number of sources cited in figures of the analyst report. \\
\textit{\#TableSources} & Number of sources cited in tables of the analyst report. \\
\textit{\#FirmTopics} & Number of firm-specific topics discussed in the analyst report. \\        
\textit{\#IndustryTopics} & Number of industry-specific topics discussed in the analyst report. \\
\textit{\#MacroTopics} & Number of macro-economic topics discussed in the analyst report. \\
\textit{\#HistMethods} & Number of historical trend analysis methods used in the report. \\
\textit{\#ValMethods} & Number of valuation methods used in the report. \\
\textit{\#FcMethods} & Number of forecasting methods used in the report. \\
\multicolumn{2}{l}{\underline{\textit{Other report attributes}}} \\
\textit{Post} & An indicator of whether the report is issued after the introduction of \textsc{FactSet}'s AI-powered platform (December 14, 2023). \\
\textit{FactSet} & An indicator of whether \textsc{FactSet} is cited as a data source in the report. \textit{Bloomberg}, \textit{Refinitiv}, and \textit{CapitalIQ} are constructed similarly. \\
\textit{Prompt} & An indicator of whether the report is issued within three days following an earnings announcement. \\
\textit{Accuracy} & Forecast accuracy, measured as the median forecast error of all analysts covering the company in the prior 90 days less the analyst's own forecast error, scaled by the standard deviation of all forecasts in the prior 90 days. \\
\textit{AbnVol} & Daily abnormal trading volume (in percentage) over the $[-1,1]$ trading-day window around the report's release, measured as the market-adjusted number of shares traded scaled by outstanding shares over the $[-1,1]$ day window less the equivalent amount over the $[-40,-11]$ trading-day window. \\
\textit{SignalSD} & Signal inconsistency, measured as the standard deviation of signal scores ($+1$ for positive, $-1$ for negative, $0$ for neutral) within each report. \\
\\
\multicolumn{2}{l}{\underline{\textit{Analyst team attributes}}} \\
\textit{Team-Career} & Analyst general experience, measured as the number of years since the analyst team first appeared in the I/B/E/S database. \\
\textit{Firm-Experience} & Analyst firm-specific experience, measured as the number of years between the forecast data and the date when the analyst (team) first covered the firm. \\
\textit{IT-Experience} & Analyst IT work experience, which takes the value of one if a team member has worked in the IT industry, such as programming, software development, artificial intelligence, algorithm engineering, and machine learning, and otherwise zero. \\
\textit{Elite-Education} & Analyst educational experience, which takes the value of one if the lead analyst has obtained a degree at a QS World Top 100 University, and otherwise zero. \\
\textit{Brokerage-Size} & Brokerage firm size, measured as the number of analyst teams employed by the brokerage firm in year \textit{t}. \\
\textit{Team-Size} & Analyst team size, which equals the number of signatory analysts on the team. \\
\textit{Forecast-Frequency} & Forecast frequency, measured as the number of forecasts issued by the analyst (team) for the firm in year \textit{t} – 1. \\
\textit{Firms-Followed} & The number of firms followed by the analyst (team) in year \textit{t} – 1. \\
\bottomrule
\end{longtable}
\clearpage

\section{Online Appendix}

\subsection{Major financial data platforms and their AI-powered products}\label{app:ai_powered_platforms}
\setlength{\tablewidth}{520pt}
\begin{longtable}[htbp]{
p{0.10\tablewidth}
p{0.20\tablewidth}
p{0.15\tablewidth}
p{0.50\tablewidth}
}
\toprule
Vendor & Product & Launch Time & Key Features \\
\midrule
FactSet & FactSet Mercury & December 2023 & * GenAI knowledge agent for queries on financial fundamentals with audit trails \\
 &  &  & * Automated pitchbook creation \\
 &  &  & * Transaction benchmarking \& contextual stock recommendations \\
  & Intelligence Platform & November 2024 & * Powered by Mercury’s agentic orchestration for auditable answers and actionable insights \\
 &  &  & * Includes Enterprise AI Building Blocks for in-house AI deployment \\
\midrule
Bloomberg & BloombergGPT & March 2023 & * Financial LLM for sentiment analysis \& news classification \\
 & Earnings Call Summaries & January 2024 & * Key insight extraction from earnings calls with contextual links to terminal data \& supply chain analysis \\
 & Document Insights & April 2025 & * Natural language queries across 200M+ company documents \\
 &  &  & * Earnings call comparisons and summarization \\
 & ASKB (Agentic AI) & February 2026 & * Conversational AI interface for the Bloomberg Terminal \\
\midrule
Refinitiv & ModuleQ Integration & March 2024 & * \emph{ModuleQ dissolved in 2025 due to funding issues; product no longer active} \\
\midrule
Capital IQ & ChatIQ & November 2024 & * Natural language query interface for financial documents \\
 & ChatIQ Pro & October 2025 & * Multi-document ChatIQ enabling simultaneous analysis across multiple documents \\
\bottomrule
\end{longtable}
\clearpage

\subsection{Illustration of FactSet Mercury interface}\label{app:Mercury_demo}
\begin{figure}[H]
\centering
\begin{minipage}{0.65\textwidth}
  \centering
  \includegraphics[width=\linewidth]{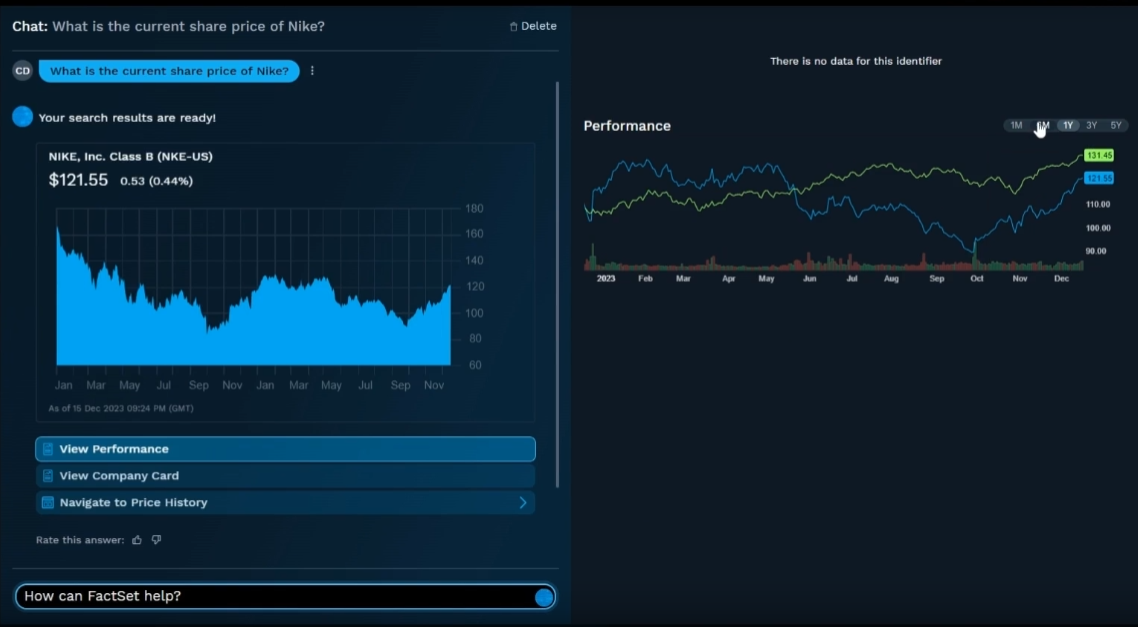}
  \subcaption{Panel A: Natural language query and data visualization}
\end{minipage}
\begin{minipage}{0.65\textwidth}
  \centering
  \includegraphics[width=\linewidth]{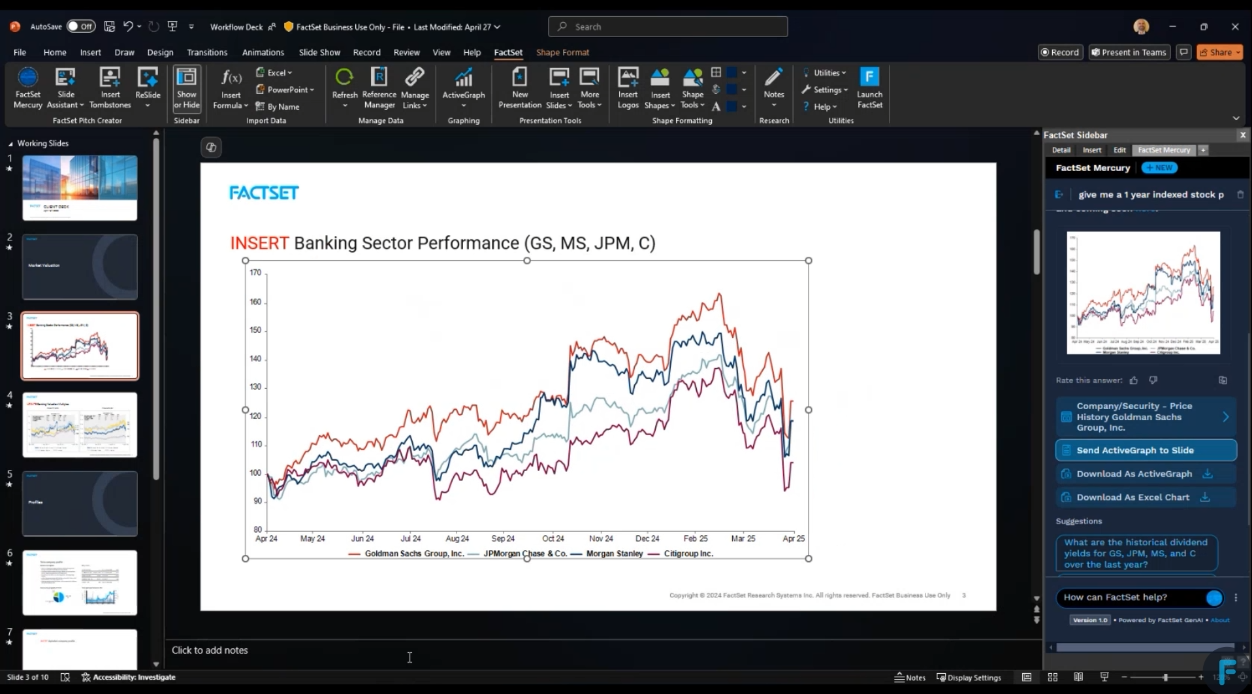}
  \subcaption{Panel B: Streamlined pitchbook creation}
\end{minipage}
\begin{minipage}{0.65\textwidth}
  \centering
  \includegraphics[width=\linewidth]{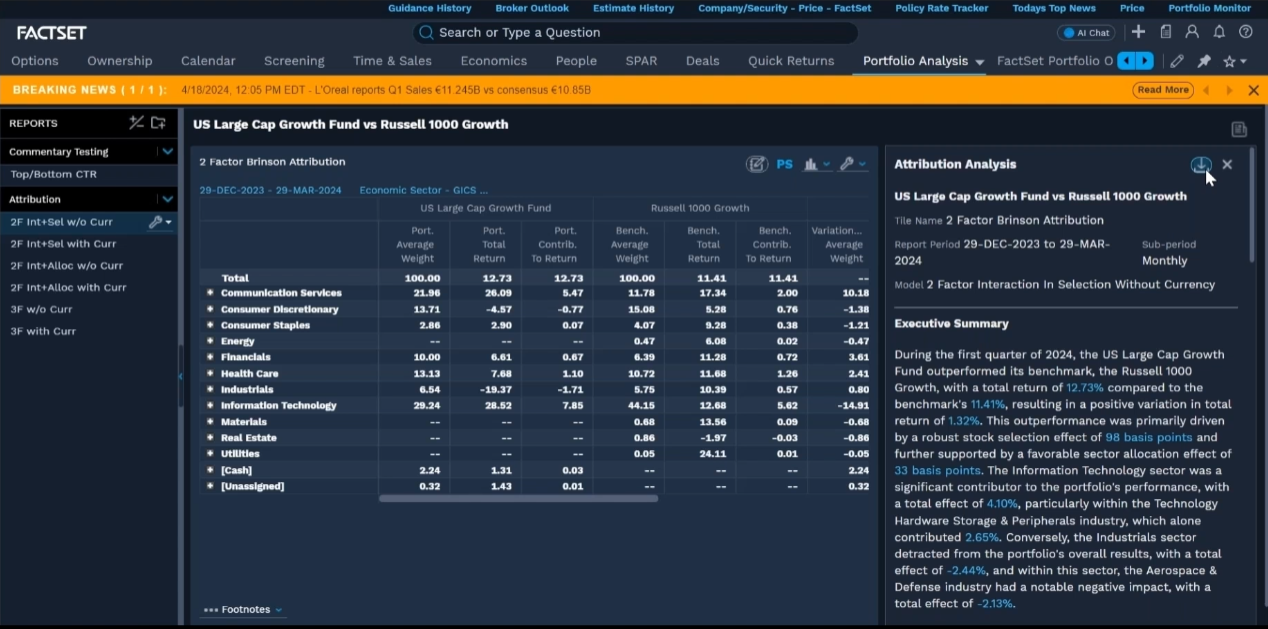}
  \subcaption{Panel C: Intelligent portfolio commentary}
\end{minipage}
\end{figure}
\clearpage

\subsection{Sample construction}\label{app:sample_construction}
The difficulties in identifying analyst names in the masked databases of I/B/E/S are well acknowledged. Ensuring accurate, one-on-one matches between the original reports and I/B/E/S forecasts is responsible for the loss of a substantial portion (in fact, 60\%) of the full sample of reports that we have gathered. This loss of sample is mainly attributable to two fundamental reasons: (i) detailed estimates published by certain brokers do not appear in the academic version of I/B/E/S, as they require licenses directly with each client to use their estimates;\footnote{For details, see \href{ibes manual}{https://wrds-www.wharton.upenn.edu/pages/support/manuals-and-overviews/i-b-e-s/ibes-estimates/wrds-overview-ibes/}.} and (ii) many reports do not provide an update of earnings forecasts (known as reiterating reports) and therefore cannot be matched to any observation in the I/B/E/S estimate file, which only adds an entry when a revision is issued. First, we provide a detailed account of our sample construction procedure, including how we assess the accuracy of our matches. Then, we replicate our main results on a random sample of the dropped reports and show that our main inference remains consistent across both the main and the excluded samples.

\begin{figure}[htbp]
\renewcommand{\thefigure}{B.3.1}
\caption{Sample Matching Procedure}
\label{fig:sample_matching}
\includegraphics[width=\linewidth]{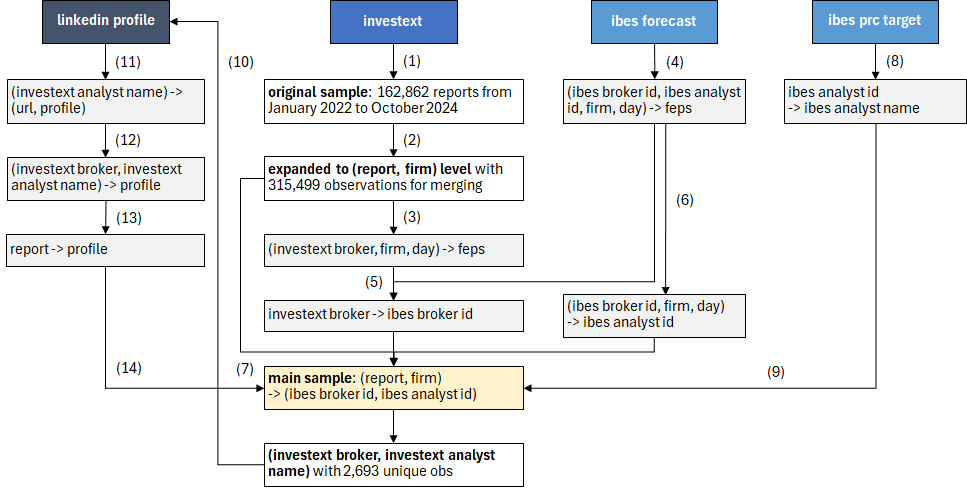}
\end{figure}

Our sample matching, illustrated in Figure~\ref{fig:sample_matching}, proceeds in the following steps.

\begin{enumerate}
\item[\textbf{Step 1}] We obtain 162,862 unique equity research reports published by all U.S. analysts covering all U.S. firms from Investext. To expedite download, we leave out reports that are fewer than five pages long, after we manually check a random sample and verify that such reports mostly only provide an estimate update with minimal discussion (boilerplate disclaimers and stock rating explanations usually take up four pages), which makes it difficult to develop meaningful measurements of content richness such as the number of information sources and topics.

\item[\textbf{Step 2}] Some reports cover more than one firm. We expand the sample to the report-firm level for merging, which yields 315,499 observations. Table~\ref{tab:report_firm_obs} reports the number of observations.

\begin{table}[htbp]
\renewcommand{\thetable}{B.3.1}
\centering\caption{Number of Report-Firm Observations}
\label{tab:report_firm_obs}
\begin{tabular}{lcc}
\toprule
Number of firms covered & Number of reports & Observations \\
\midrule
1 & 115,327 & 115,327 \\
2 & 10,289 & 20,578 \\
3 & 6,427 & 19,281 \\
4 & 4,054 & 16,216 \\
$\ge$5 & 11,101 & 144,097 \\
No matched firms (dropped) & 15,664 &  \\
\midrule
Total & 162,862 & 315,499 \\
\bottomrule
\end{tabular}
\end{table}

I/B/E/S no longer provides the brokerage translation file. Therefore, we match Investext broker name with I/B/E/S ID from the I/B/E/S forecast file based on the timing and content (i.e., earnings forecasts) of firm coverage as revealed by both datasets. The general idea is that if a broker from the Investext dataset and a broker from the I/B/E/S dataset has a substantially higher likelihood of issuing the same combination of forecasts (including four quaterly forecasts and one annual forecast) for the same company on the same day, relative to all possible broker pairs, then it is almost certain that these two IDs refer to the same brokerage firm. We implement this match in steps 3 to 5.

\item[\textbf{Step 3}] For each report in the Investext dataset, we obtain its broker name, firm, and publication date. Also, we obtain from the report raw text all segments of text (up to 500 characters) following the token ``EPS.'' If the analyst provides their earnings forecasts in the report, the forecast numbers are likely contained in one of those text segments. We do not attempt to capture the forecasts in \emph{all} reports, because a partial sample is sufficient to distinguish the correct match of brokers from other possible matches 

\item[\textbf{Step 4}] We obtain from the I/B/E/S forecast file all quarterly and annual forecasts (five in total) for each analyst-firm-day observations over the sample period.

\item[\textbf{Step 5}] For each observation from step 3 and each observation from step 4, if at least three forecast numbers from the latter are found in the text segments from the former, provided that both observations pertain to the same firm-day, append this (Investext broker, I/B/E/S broker ID) pair to a new table. Then, calculate the frequency of all broker pairs. If two IDs indeed refer to the same entity, one should notice that the frequency of this pair is substantially higher than the other pairs. Figure~\ref{fig:broker_match_example} illustrates the number of matches between JPMorgan (an Investext broker) and some brokers from I/B/E/S. The broker with ID 873 has 9,354 matches with JPMorgan (43\% of all matches) whereas others only have a $<5\%$ match rate. Therefore, JPMorgan is matched to I/B/E/S broker 873. In practice, for each Investext broker, we keep the I/B/E/S broker ID with the highest match rate, while requiring that the match rate is above 25\% and the number of matches is above 10: the 25\% cutoff ensures that the top candidate is clearly distinguished from alternatives, while the minimum of 10 matches excludes incidental co-occurrences.

\begin{figure}[htbp]
\renewcommand{\thefigure}{B.3.2}
\caption{Example of Broker Matching}
\label{fig:broker_match_example}
\includegraphics[width=\linewidth]{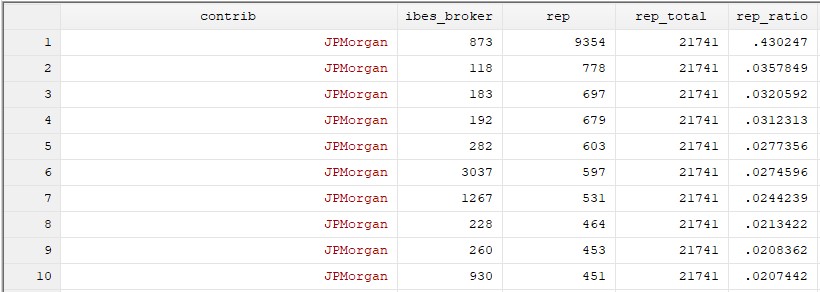}
\end{figure}

In the end, we are able to match 46 Investext brokers to 43 I/B/E/S broker IDs. There are three duplicates on the left side, because Roth Capital Partners, Inc. and Roth MKM are assigned to the same I/B/E/S ID (481) and SVB Securities, Leerink Partners, and SVB Leerink are assigned to the same I/B/E/S ID (2129) as well, for reasons that become apparent by looking at their names. These nuanced matching issues, which are uncovered purely by the algorithm described above, provide some level of reassurance regarding the accuracy of our matching procedure. In later steps, we cross-validate our matching results by using a third dataset.
Because of our short sample period and the exclusion of some reports in step 1, we will eventually need to match each report to an analyst in the I/B/E/S dataset to calculate variables such as career length and coverage frequency using the longer, fuller sample provided by I/B/E/S. I/B/E/S assigns a mask ID for each analyst (or lead analyst of a team) without revealing their name. However, with the (Investext broker, I/B/E/S broker) match obtained from step 5, we are able to match each report to an I/B/E/S analyst ID. This procedure is detailed in steps 6 to 7.

\item[\textbf{Step 6}] Brokerage firms usually appoint the same analyst (team) to cover the same company until they are eventually replaced. From the I/B/E/S file, we can determine, for each broker-firm pair, the day on which a given I/B/E/S analyst issues a forecast.

\item[\textbf{Step 7}] Taking our report-firm level sample from step 2, we first match Investext broker to I/B/E/S broker ID, and then based on broker ID, firm, and the day that the report is released, we can match each observation to an I/B/E/S analyst ID using the information obtained in step 6. Specifically, following \citet{huang2014evidence} and \citet{de2015analyst}, we assume that an Invetext report and an I/B/E/S forecast refer to the same report if they are registered in their respective database fewer than two days apart. The two-day gap is to account for the fact that sometimes either the report or the forecast can be registered one day earlier or later than the \emph{actual} publication date that is printed on the report's cover page. We confirm this pattern through a random manual check. Finally, to ensure comparability across the pre- and post-periods, we restrict the sample to analysts that appear in both periods and drop analyst-firm pairs where coverage is terminated prior to the introduction of the AI platform. The effect of each step of sample construction on the sample size is presented in Table~\ref{tab:sample_size_effect}. Eventually, our main sample consists of 46,853 observations. The loss of sample, as noted before, is attributable to proprietary estimates not included in the I/B/E/S academic version and the absence of a forecast revision in many reports. Prior studies treat reports issued without a revision as ``reiterating'' and impute zero as the forecast revision (i.e., no revision) \citep{huang2014evidence, de2015analyst}. These studies focus on forecast revision and mostly treat it as a control variable, so a lack of revision can reasonably be interpreted as zero.

Our study differs in that we also examine forecast accuracy as a primary outcome, which requires a contemporaneous updated forecast. Moreover, the choice to focus on reports with forecast revisions aligns with prior literature that examines analyst forecasts using the I/B/E/S dataset. Still, we acknowledge the potential selection bias that such a choice might introduce, particularly to the tests on report content richness. We properly address this later by replicating our main tests on a random sample of the excluded reports.
\begin{table}[htbp]
\renewcommand{\thetable}{B.3.2}
\centering
\caption{Sample Construction and Observations}
\label{tab:sample_size_effect}
\begin{tabular}{lc}
\toprule
Sample construction (steps 6-7) & Effect on sample size \\
\midrule
Sample from step 2 & 315,499 \\
Match Investext brokers to I/B/E/S broker IDs & (92,299) \\
Match reports to I/B/E/S forecasts and analyst IDs & (155,625) \\
Keep analysts that appear in both pre- and post-periods & (14,945) \\
Drop analyst-firm pairs where coverage is terminated in pre-period & (5,335) \\
Drop observations with main variables missing & (422) \\
\midrule
Main sample & 46,853 \\
\bottomrule
\end{tabular}
\end{table}
\newline In what follows, we use a third dataset to evaluate the accuracy of our matching procedure. The I/B/E/S price target file provides (i) I/B/E/S analyst ID, which is identical to the I/B/E/S analyst ID in the forecast file; and (ii) I/B/E/S analyst name, referring to the (lead) analyst and given as their last name plus the initial of their first name (e.g., ``BROOK  A''). We cannot use this information to match reports to I/B/E/S analyst IDs because the full name is not available and, in the case of multiple authors, we do not know which member of the analyst team the ID should be assigned to. However, it does allow us to verify matching accuracy ex post. We implement this verification in steps 8 to 9.

\item[\textbf{Step 8}] We obtain all pairs of I/B/E/S analyst IDs and names from the price target file.

\item[\textbf{Step 9}] Merge the I/B/E/S analyst name to the main sample, and verify if it matches one of the authors of the report (from Investext). The results reported in Table~\ref{tab:match_verification} suggest that our match is accurate 98\% of the time.
\begin{table}[htbp]
\renewcommand{\thetable}{B.3.3}
\centering
\caption{Matching Results}
\label{tab:match_verification}
\begin{tabular}{lc}
\toprule
Matching results & Observations \\
\midrule
I/B/E/S analyst name is in Investext authors & 45,323 \\
I/B/E/S analyst name is not in Investext authors & 741 \\
\emph{Accuracy} & \emph{98.39\%} \\
Either information missing & 789 \\
\midrule
Total (main sample) & 46,853 \\
\bottomrule
\end{tabular}
\end{table}
\newline After obtaining and verifying our main sample, we identify each analyst (including lead and associate analysts) and merge them to their LinkedIn profiles that we have scraped from the internet. Steps 10--14 detail our methods.

\item[\textbf{Step 10}] We expand the sample by the author(s) of each report and then collapse it to the (Investext broker, Investext analyst) level, which produces 1,982 unique observations.

\item[\textbf{Step 11}] Using analyst names, we scrape all related profiles (including name, educational background, and work experience) from LinkedIn as well as the URL that uniquely identifies each profile. In total, we obtain 3,084 profiles associated with 1,835 observations from step 10.

\item[\textbf{Step 12}] We evaluate each potential match based on four criteria: (i) Investext analyst first name exactly matches the name in the profile; (ii) Investext analyst last name exactly matches the name in the profile; (iii) the word ``analyst'' is mentioned in the profile; and (iv) the Investext broker name or its common abbreviation is mentioned in the profile. For each (Investext broker, Investext analyst) pair, we keep the profile with the highest number of criteria met, while requiring that at least two criteria are met. The matching results are reported in Table~\ref{tab:linkedin_match_results}. This process yields 1,901 profiles matched to (Investext broker, Investext analyst) pairs. 1,561 (82\%) of the profiles meet at least three criteria. 1,094 (58\%) meet all four criteria.
\begin{table}[htbp]
\renewcommand{\thetable}{B.3.4}
\centering
\caption{LinkedIn Profile Matching Results}
\label{tab:linkedin_match_results}
\begin{tabular}{lc}
\toprule
Matching results & Observations \\
\midrule
Sample from step 11 (1,835 analysts) & 3,084 \\
Drop unlikely matches (step 12) & (1,183) \\
Matched profiles (1,608 analysts) & 1,901 \\
\midrule
4 criteria met & 1,094 \\
3 criteria met & 467 \\
2 criteria met & 340 \\
\bottomrule
\end{tabular}
\end{table}

\item[\textbf{Step 13}] We aggregate profiles of individual analysts at the report level to obtain variables such as whether the analyst team includes an IT analyst.

\item[\textbf{Step 14}] Finally, the report-level measures obtained in step 13 are merged to the main sample.

\end{enumerate}

As noted above, much of our sample loss stems from the exclusion of reiterating reports. This raises a legitimate concern: if analysts use the FactSet AI platform differently when producing reiterating versus revising reports, our results may not generalize to the broader population of reports. To mitigate this concern, we replicate our main analysis of content richness on a random 10\% subsample of the reiterating reports. While we cannot run the full sample due to the significant cost of LLM-based content parsing, a 10\% random subsample provides sufficient statistical power to detect meaningful effects. If the positive effect of AI platform use on report richness persists in this subsample, it suggests that our main results are not an artifact of restricting the sample to revising reports.

To construct a matched sample of reiterating reports with I/B/E/S analyst IDs, we rely on the assumption that brokerage firms assign the same analyst team to cover a given firm until that team is replaced. Thus, for each broker-firm pair, we match each reiterating report to the analyst team that was actively issuing revising reports for that firm around the same period. We apply the same sample selection criteria as in our main tests, requiring analysts to appear in both the pre- and post-periods and dropping analyst-firm pairs whose coverage terminated before the Mercury launch. This yields a final sample of 10,984 observations.

We use the same controls and fixed effect structure, and cluster standard error by analyst and firm. Table~\ref{tab:excluded_reports} reports the results. The coefficients on $FactSet \times Post$ are positive and significant across all content richness dimensions except $\#HistMethods$. These findings indicate that the positive effect of FactSet's AI platform on report richness extends to reiterating reports as well.

\begin{table}[htbp]
\begin{adjustwidth}{-2cm}{-2cm}
\renewcommand{\thetable}{B.3.5}
\centering
\caption{Content Richness of Reiterating Reports}
\label{tab:excluded_reports}
\setlength{\tabcolsep}{1.5pt}
\scriptsize
\begin{tabular}{lccccccccc}
\toprule
& \#TextSources & \#FigureSources & \#TableSources & \#FirmTopics & \#IndustryTopics & \#MacroTopics & \#HistMethods & \#ValMethods & \#FcMethods \\
\midrule
FactSet & 1.497*** & 0.640*** & 1.345*** & 2.067*** & 0.671*** & 0.867*** & 0.309*** & 0.639*** & 0.825*** \\
& (4.95) & (4.46) & (7.27) & (5.72) & (3.10) & (4.31) & (4.41) & (6.65) & (6.79) \\
FactSet$\times$Post & 1.398*** & 0.622*** & 0.733*** & 1.158*** & 0.709*** & 0.731*** & 0.118 & 0.351** & 0.344* \\
& (3.42) & (3.58) & (2.63) & (2.77) & (2.78) & (2.74) & (1.14) & (2.52) & (1.80) \\
\midrule
Controls & Yes & Yes & Yes & Yes & Yes & Yes & Yes & Yes & Yes \\
Report Date FE & Yes & Yes & Yes & Yes & Yes & Yes & Yes & Yes & Yes \\
Firm-Year FE & Yes & Yes & Yes & Yes & Yes & Yes & Yes & Yes & Yes \\
Analyst FE & Yes & Yes & Yes & Yes & Yes & Yes & Yes & Yes & Yes \\
Broker-Year FE & Yes & Yes & Yes & Yes & Yes & Yes & Yes & Yes & Yes \\
\midrule
N & 10,984 & 10,984 & 10,984 & 10,984 & 10,984 & 10,984 & 10,984 & 10,984 & 10,984 \\
Adjusted R$^2$ & 0.458 & 0.431 & 0.467 & 0.502 & 0.427 & 0.498 & 0.345 & 0.440 & 0.464 \\
\bottomrule
\end{tabular}
\end{adjustwidth}
\end{table}
\clearpage

\subsection{Number of reports issued and analyst teams employed by brokerage firms included in the main sample}\label{app:brokerage}
\setlength{\tablewidth}{520pt}
\begin{longtable}[htbp]{*{1}{p{0.5\tablewidth}}
*{2}{p{0.25\tablewidth}}}
\toprule
Brokerage Firm & \#Reports Issued & \#Analyst Teams \\
\midrule
JPMorgan & 14,719 & 91 \\
Wells Fargo Securities, LLC & 8,664 & 63 \\
BMO Capital Markets & 3,212 & 33 \\
BTIG & 2,233 & 24 \\
Oppenheimer \& Co., Inc. & 2,194 & 32 \\
D.A. Davidson \& Company & 1,883 & 17 \\
Seaport Global Securities LLC & 1,508 & 19 \\
The Benchmark Company LLC & 887 & 16 \\
JMP Securities & 680 & 16 \\
Barrington Research Associates & 604 & 5 \\
Leerink Partners & 483 & 7 \\
Scotiabank GBM & 470 & 14 \\
CIBC Capital Markets & 461 & 17 \\
Needham \& Company Inc. & 219 & 14 \\
Berenberg & 209 & 13 \\
Susquehanna Financial Group LLLP & 201 & 3 \\
Truist Securities & 169 & 3 \\
Compass Point Research \& Trading LLC & 85 & 3 \\
Roth MKM & 82 & 3 \\
Janney Montgomery Scott LLC & 81 & 7 \\
Piper Sandler Companies & 52 & 13 \\
\bottomrule
\end{longtable}
\clearpage

\subsection{GPT prompt for parsing report content}\label{app:prompt_example}
We utilize the following prompt to instruct GPT-4o-mini to identify the information sources, topics, and analytic methods used in the analyst report.

\begin{lstlisting}
system_prompt = (
    "You are an expert in financial analysis and equity or bond research, specializing in analyzing public company reports."
    "Your task is to extract structured insights from financial analyst reports using a chain-of-thought approach."
)

user_prompt = (
    f"Read the provided segment of a financial analyst report and perform the following analysis step-by-step, using a chain-of-thought approach:\n\n"
    f"Provide only the final structured response. Do not include explanations, intermediate reasoning, or step-by-step analysis in your chain-of-thought--just return the structured output."
    f"## Document Segment\n"
    f"{content}\n\n"

    f"### 1. Information Sources\n"
    f"Identify all sources of information referenced in the report that analysts used for their analysis. Follow these steps for each source, categorizing them into tables, figures, and textual references:\n\n"

    f"#### 1.1 Table Sources\n"
    f"Analyze all tables used in the report:\n"
    f"1. Identify the **table type** (e.g., financial summary, revenue breakdown).\n"
    f"2. Extract the **table name**.\n"
    f"3. Determine the **data source** for the table.\n"
    f"4. Provide a **brief description** of the table (<=20 words).\n"
    f"5. **Prediction**: Based on the table, does it imply EPS (Earnings Per Share) is 'going up', 'going down', or 'unknown'? Explain why.\n"
    f"6. **Confidence**: Estimate the confidence in this prediction ('high', 'medium', 'low').\n"
    f"7. **Reasoning**: Explain why this prediction was made (e.g., 'revenue is increasing').\n"
    f"8. If any information is missing or unclear, just answer 'unknown'.\n"
    f"Format your response as a list of dictionaries, each entry in the list covers on table source:\n"
    f"[{{'table_type': '<content>', 'table_name': '<content>', 'source': '<content>', 'description': '<content>',"
    f"'prediction': '<content>', 'confidence': '<content>', 'reasoning': '<content>'}}]\n\n"

    f"#### 1.2 Figure Sources\n"
    f"Analyze all figures used in the report (you can refer to the figure name and its description):\n"
    f"1. Identify the **figure type** (e.g., histogram, heatmap, network graph).\n"
    f"2. Extract the **figure name**.\n"
    f"3. Determine the **data source** for the figure.\n"
    f"4. Provide a **brief description** of the figure (<=20 words).\n"
    f"5. **Prediction**: Based on the figure, does it imply EPS is 'going up', 'going down', or 'unknown'? Explain why.\n"
    f"6. **Confidence**: Estimate the confidence in this prediction ('high', 'medium', 'low').\n"
    f"7. **Reasoning**: Explain why this prediction was made (e.g., 'revenue is increasing').\n"
    f"8. If any information is missing or unclear, just answer 'unknown'.\n"
    f"Format your response as a list of dictionaries:\n"
    f"[{{'figure_type': '<content>', 'figure_name': '<content>', 'source': '<content>', 'description': '<content>',"
    f"'prediction': '<content>', 'confidence': '<content>', 'reasoning': '<content>'}}]\n\n"

    f"#### 1.3 Text Sources\n"
    f"Analyze all textual sources referenced in the report:\n"
    f"1. Identify the **source name**.\n"
    f"2. Provide a **brief description** of the source (<=20 words).\n"
    f"3. **Prediction**: Based on the source, does it imply EPS is 'going up', 'going down', or 'unknown'? Explain why.\n"
    f"4. **Confidence**: Estimate the confidence in this prediction ('high', 'medium', 'low').\n"
    f"5. **Reasoning**: Explain why this prediction was made (e.g., 'revenue is increasing').\n"
    f"6. If any information is missing or unclear, just answer 'unknown'.\n"
    f"Format your response as a list of dictionaries:\n"
    f"[{{'source': '<content>', 'description': '<content>', 'prediction': '<content>', 'confidence': '<content>',"
    f"'reasoning': '<content>'}}]\n\n"

    f"### 2. Information Topics\n"
    f"Identify key topics discussed at different levels of analysis. Follow these steps for each topic:\n\n"
    f"#### 2.1 Firm Level\n"
    f"List company-specific topics (e.g., market competition, financial performance):\n"
    f"1. Identify the **topic**.\n"
    f"2. Provide a **brief description** (<=20 words).\n"
    f"3. **Prediction**: Based on the topic, does it imply EPS is 'going up', 'going down', or 'unknown'? Explain why.\n"
    f"4. **Confidence**: Estimate the confidence in this prediction ('high', 'medium', 'low').\n"
    f"5. **Reasoning**: Explain why this prediction was made (e.g., 'revenue is increasing').\n"
    f"Format your response as a list of dictionaries:\n"
    f"[{{'topic': '<content>', 'description': '<content>', 'prediction': '<content>', 'confidence': '<content>',"
    f"'reasoning': '<content>'}}]\n\n"

    f"#### 2.2 Industry Level\n"
    f"List industry-wide topics affecting the company (e.g., supply chain, regulations):\n"
    f"Format your response as above.\n\n"

    f"#### 2.3 Macro Level\n"
    f"list macroeconomic topics affecting the company (e.g., monetary policy, inflation):\n"
    f"Format your response as above.\n\n"

    f"### 3. Methodology\n"
    f"Identify the analytical methods used in the report. Follow these steps for each method:\n\n"

    f"#### 3.1 Valuation Methods\n"
    f"Identify all valuation models and their data sources:\n"
    f"1. Identify the **data** used for the valuation.\n"
    f"2. Identify the **method** (e.g., DCF, multiples).\n"
    f"3. Provide a **brief description** of the method.\n"
    f"4. **Prediction**: Based on the method, does it imply EPS is 'going up', 'going down', or 'unknown'? Explain why.\n"
    f"5. **Confidence**: Estimate the confidence in this prediction ('high', 'medium', 'low').\n"
    f"6. **Reasoning**: Explain why this prediction was made (e.g., 'revenue is increasing').\n"
    f"Format your response as a list of dictionaries:\n"
    f"[{{'data': '<content>', 'method': '<content>', 'description': '<content>', 'prediction': '<content>',"
    f"'confidence': '<content>', 'reasoning': '<content>'}}]\n\n"

    f"#### 3.2 Historical Analysis Methods\n"
    f"Identify methods used for analyzing historical performance:\n"
    f"Format your response as above.\n\n"

    f"#### 3.3 Forecasting Methods\n"
    f"Identify forecasting models and their data sources:\n"
    f"Format your response as above.\n\n"
)
\end{lstlisting}
\clearpage

\subsection{Validating LLM outputs}\label{app:validate_llm}
LLM-based measures could suffer from hallucinations. We validate our LLM outputs by conducting manual check and benchmarking with non-LLM measures. We also further demonstrate measure stability by changing the length of segments (2,000 characters in the main design) that the model is allowed to parse and process each time.

\textit{Human verification.} First, we hire two undergraduate students fluent in English and familiar with equity research reports to independently evaluate the accuracy of our LLM outputs, based on 200 randomly selected segments from the full sample (2 were dropped because the LLM failed to respond to the request about those segments). The students are provided with a list of these segments and the original LLM output, and are given the following instructions:
\begin{itemize}
\item An LLM pipeline is used to identify and imply the distinct information sources, topics, and analytic methods used in segments of equity research reports. For each identified item, the model returns the type (text source, etc.), name, description, earnings prediction, and reasoning based on that item.
\item Please read the original segments and the corresponding LLM responses to determine if the response accurately reflects \emph{all} the information contained in each segment. An accurate extraction means that the model correctly identifies the type of the information, prints the name correctly, provides an accurate description, and, if applicable, draws clear and correct implications from that piece of information about firm earnings. You are to go through each item carefully and identify all inconsistencies in the response.
\item For each segment, provide your assessment of the response accuracy (Y or N) and a brief statement of the inconsistency, if any. Also, provide the number of sources, topics, and methods that you believe are incorrect in the responses, and the number of sources, topics, and methods that you believe are missing from the responses.
\end{itemize}

We collect the students' evaluations and summarize the results in Table~\ref{tab:llm_verification}. Student 1 appears to be more conservative in her evaluation than Student 2: of the 198 segments, Student 1 determines 157 to be 100\% correct (a 79.29\% accuracy rate), while Student 2 determines 182 to be 100\% correct (a 91.92\% accuracy rate). For a conservative summary, we determine a segment to be 100\% correctly identified only if both agree. Consequently, the segment-level accuracy rate of our LLM algorithm is 73.23\%. We partition the sample into reports published in 2023 and reports published in 2024, and find that they do not significantly differ in accuracy rates ($p = 0.277$). Similarly, when we partition based on whether FactSet is used, the difference is also insignificant ($p = 0.789$). These results indicate that the model's accuracy does not appear to drift across time or present bias toward reports developed using FactSet. This rate, however, inevitably understates the accuracy of our model because we deem the entire response ``inaccurate'' if the model makes one mistake, but in reality we find that the model is satisfactorily effective if examined on an item-by-item basis.

From the 198 segments, the model has identified 784 distinct sources, 670 topics, and 126 different analytic methods. We ask students to provide the number of items, for each category respectively, that they believe are incorrect or missing from the response. For a conservative summary, we take the maximum number of incorrect or missing items for each segment and category, so that we can assess the lower bound of our model's accuracy. We find that our model performs best in identifying sources (2.81\% incorrect or missing), and performs more poorly in correctly identifying topics (10.75\% incorrect or missing) and methods (8.73\% incorrect or missing). We examine the reasons provided by students for their decisions and find that the disagreement mostly arises in cases that are inherently ambiguous. For example, Student 1 claims in one response that a topic with name ``COVID impact,'' classified by the LLM as a firm topic because the segment mainly discusses how the firm is affected by the pandemic, should instead be classified as a macro topic, which also makes sense. This is less of a concern to us because the misclassification does not affect the total count of topics. In another case, Student 2 claims that the response to a report segment about Microsoft fails to include the company's ``ESG metrics'' in topics. In summary, despite the potential understatement of content richness, our model performs at a relatively high accuracy overall, alleviating concerns about AI hallucination.

\begin{table}[htbp]
\begin{adjustwidth}{-2cm}{-2cm}
\renewcommand{\thetable}{B.6.1}
\centering
\caption{LLM Output Verification}
\label{tab:llm_verification}
\setlength{\tabcolsep}{1.5pt}

\begin{tabular}{lcccccccccc}
\toprule
& \multicolumn{3}{c}{Student 1} & \multicolumn{3}{c}{Student 2} & \multicolumn{3}{c}{Conservative} \\
\cmidrule(lr){2-4} \cmidrule(lr){5-7} \cmidrule(lr){8-10}
\# Segments & \multicolumn{3}{c}{198} & \multicolumn{3}{c}{198} & \multicolumn{3}{c}{198} \\
\# Segments 100\% Correct & \multicolumn{3}{c}{157} & \multicolumn{3}{c}{182} & \multicolumn{3}{c}{145} \\
\% Correct & \multicolumn{3}{c}{79.29\%} & \multicolumn{3}{c}{91.92\%} & \multicolumn{3}{c}{73.23\%} \\
\midrule
\% Correct, 2023 & & & & & & & \multicolumn{3}{c}{70.60\%} \\
\% Correct, 2024 & & & & & & & \multicolumn{3}{c}{77.78\%} \\
Diff. $p$-value & & & & & & & \multicolumn{3}{c}{[0.277]} \\
\midrule
\% Correct, FactSet & & & & & & & \multicolumn{3}{c}{72.06\%} \\
\% Correct, Non-FactSet & & & & & & & \multicolumn{3}{c}{73.85\%} \\
Diff. $p$-value & & & & & & & \multicolumn{3}{c}{[0.789]} \\
\midrule
& Sources & Topics & Methods & Sources & Topics & Methods & Sources & Topics & Methods \\
\cmidrule(lr){2-4} \cmidrule(lr){5-7} \cmidrule(lr){8-10}
\# Total & 784 & 670 & 126 & 784 & 670 & 126 & 784 & 670 & 126 \\
\# Incorrect & 1 & 45 & 6 & 9 & 3 & 0 & 10 & 48 & 6 \\
\# Missing & 3 & 7 & 2 & 10 & 17 & 3 & 12 & 24 & 5 \\
\% Incorrect & 0.13\% & 6.72\% & 4.76\% & 1.15\% & 0.45\% & 0.00\% & 1.28\% & 7.16\% & 4.76\% \\
\% Missing & 0.38\% & 1.04\% & 1.59\% & 1.28\% & 2.54\% & 2.38\% & 1.53\% & 3.58\% & 3.97\% \\
\% Missing or Incorrect & 0.51\% & 7.76\% & 6.35\% & 2.42\% & 2.99\% & 2.38\% & 2.81\% & 10.75\% & 8.73\% \\
\bottomrule
\end{tabular}
\end{adjustwidth}
\end{table}

\textit{Alternative richness measure.} In a second approach, we develop an alternative measure of topic richness following \citet{huang2018analyst} and compare it with our measures. We apply latent Dirichlet allocation (LDA) to our sample of analyst reports to calculate the number of topics covered in each report. LDA takes a corpus of different documents as a whole and assumes that each document is generated in two steps. First, a topic is drawn from a probabilistic distribution of topics unique to this document; then, a word is drawn from a probabilistic distribution of words unique to the topic generated in the previous step. The document can therefore be represented by a joint probability of those words occurring in the document. The LDA model aims to find those document-topic and topic-word distributions that maximize the probability of observing the entire corpus. Then, we pass each sentence to the LDA model, which uses the estimated topic-word distributions and the document-topic prior to infer the posterior probability that the sentence belongs to each of the topics, and assigns the sentence to the most probable topic. The steps are described below:

\begin{itemize}
\item[\textbf{Step 1}] We use the entire sample of analyst reports from 2018 to 2024 as our main corpus, and treat each report as a separate document. Because many topics are industry-specific, we partition the sample by the covered firm's four-digit Global Industry Classification Standard (GICS) code, following \citet{huang2018analyst}. We train the LDA model and estimate the number of topics in each analyst report for each industry separately.

\item[\textbf{Step 2}] We preprocess the corpus and tokenize the text into a machine-readable format. Specifically,
(i) Convert all text to lowercase, and remove punctuation, numbers, and non-alphabetic characters;
(ii) Remove common English stop words (e.g., ``a'', ``the'') as well as additional contextual words that are unlikely to help identify economically meaningful topics (e.g., pronouns, adverbs);
(iii) Remove company names, tickers, broker names, and a set of high-frequency terms that are not topic-specific (e.g., ``company'', ``market'', ``year'');
(iv) Remove boilerplate sentences containing words such as ``disclaimer'', ``rating'', ``certification'', ``conflict of interest'', etc., which are not part of the analyst's substantive discussion; and,
(v) Lemmatize each remaining word to its base form using the WordNet lemmatizer; stemming is not performed because it overly conflates distinct financial terms.

\item[\textbf{Step 3}] After preprocessing, each report is represented as a list of lemmatized tokens. We map each unique token to an integer ID, and filter out tokens that appear in more than 80\% of the documents to reduce noise. Then, we convert each report into a bag-of-words vector, i.e., a list of (token\_id, frequency) pairs. This results in a document-term matrix for all reports covering the industry.

\item[\textbf{Step 4}] To choose the optimal number of latent topics, we follow the perplexity-based approach common in computational linguistics. For each industry, we randomly split the set of reports into a training set and a testing set. For a candidate number of topics $k$ ranging from 2 to 120 (e.g., 2, 4, 6, \dots, 120), we train an LDA model on the training set with standard hyperparameters. We then compute the perplexity of the trained model on the test set. Perplexity is defined as the exponential of the negative average log-likelihood of the test documents; lower perplexity indicates better generalization. Following \citet{huang2018analyst}, we calculate and plot the perplexity values against the corresponding number of topics $k$. The curve typically shows a sharp decline in perplexity as $k$ increases, followed by a flattening region where additional topics yield only marginal improvement. Consistent with \citet{huang2018analyst}, we find that the perplexity improvement diminishes significantly once the number of topics exceeds 60. Therefore, we set the number of topics $k = 60$ for every industry. This choice balances model fit with interpretability and avoids overfitting.

\item[\textbf{Step 5}] Using the full set of documents for each industry (no train-test split), we train a final LDA model with 60 topics, using the same hyperparameters as in Step 4. The resulting model provides estimates of the topic-word distributions and the document-topic priors.

\item[\textbf{Step 6}] For every sentence in every analyst report, we convert the sentence into a bag-of-words vector using the same dictionary. The trained LDA model infers a posterior probability distribution over the 60 topics for that sentence. We assign the sentence to the topic with the highest probability while requiring that the probability is above 0.2. Aggregating over all sentences in a report, we then obtain the number of topics covered in the report.
\end{itemize}

We report the results using the LDA approach in Table~\ref{tab:lda_topics}. Panel A presents the correlation between the LDA topics and the number of firm, industry, and macro topics obtained using our measures. The correlation is above 0.5 across all measures. In Panel B, we replace the dependent variable with the number of LDA topics and replicate the main regression. The coefficient on $FactSet \times Post$ is 2.333 and significantly positive ($t = 2.27$), suggesting a 16\% GenAI-driven increase in topical richness relative to the sample mean of 14.5 topics. This is of a similar magnitude to what we have obtained using our approach.

\begin{table}[htbp]
\begin{adjustwidth}{-1cm}{-1cm}
\renewcommand{\thetable}{B.6.2}
\centering
\caption{LDA Topic Modeling}
\label{tab:lda_topics}
\setlength{\tabcolsep}{4pt}

\textit{Panel A: Correlation}

\begin{tabular}{lc}
\toprule
& \#LDATopics \\
\midrule
\#FirmTopics & 0.577*** \\
\#IndustryTopics & 0.556*** \\
\#MacroTopics & 0.513*** \\
\bottomrule
\end{tabular}

\medskip

\textit{Panel B: Regression}

\begin{tabular}{lc}
\toprule
& \#LDATopics \\
\midrule
FactSet & 1.418*** \\
& (2.74) \\
FactSet$\times$Post & 2.333** \\
& (2.27) \\
\midrule
Controls & Yes \\
Report Date FE & Yes \\
Firm-Year FE & Yes \\
Analyst FE & Yes \\
Broker-Year FE & Yes \\
\midrule
N & 41,569 \\
Adjusted R$^2$ & 0.388 \\
\bottomrule
\end{tabular}
\end{adjustwidth}
\end{table}

\textit{Measure sensitivity.} LLM outputs could be mechanically driven by presentation formats, such as the layout of tables and figures, or by how the PDF is compiled by the publishers and then parsed by us. Our tight research design, which includes broker-year and analyst fixed effects (and, in robustness tests, broker-month and analyst-month fixed effects), absorbs any confounding effects resulting from broker- or analyst-initiated changes in presentation formats. To further demonstrate that our richness measures are insensitive to how PDF files are parsed, we randomly select 200 reports and partition them into segments of no more than 1,500 and 2,500 characters in length, respectively. We then follow the same procedure to develop richness measures based on each parsing and compare the results with our main variables, which are developed from 2,000-character parsing. We report the comparison in Table~\ref{tab:chunk_size}.

Panel A reports the mean of each variable under different chunk sizes. The table indicates that the average count of unique sources, topics, and methods tends to decrease as the chunk size increases. This is expected: with longer segments, the language model tends to ignore relatively trivial items, whereas with shorter segments these items resurface in the outputs. Despite this issue, the average counts do not differ dramatically. This difference likely reflects not the model's inability to provide consistent identification but rather the granularity of the identification. More importantly, we examine the correlation between the measures under different parsing chunk sizes. As long as they are highly correlated, our inferences should remain the same regardless of the parsing procedure. Panel B reports the results. The correlation for information sources between different chunk sizes ranges between 0.6 and 0.8, and for information topics between 0.8 and 0.9, indicating high alignment in measurement despite different parsing methods. The correlation for methods is safely above 0.7 with the exception of historical methods, where the correlation coefficient is around 0.4, indicating that the identification of historical trend analyses is more sensitive to presentation formats. Overall, our results provide support for the stability of our richness measures. Regarding the issue with historical methods, we provide a caveat in the footnote.

\begin{table}[htbp]
\begin{adjustwidth}{-1cm}{-1cm}
\renewcommand{\thetable}{B.6.3}
\centering
\caption{Content Richness with Alternative Chunk Sizes}
\label{tab:chunk_size}
\setlength{\tabcolsep}{3pt}

\textit{Panel A: Mean}

\begin{tabular}{lccc}
\toprule
Chunk size & 1,500 & 2,000 & 2,500 \\
\midrule
\#TextSources & 12.210 & 9.685 & 8.485 \\
\#FigureSources & 4.210 & 2.985 & 2.265 \\
\#TableSources & 8.285 & 7.135 & 6.125 \\
\#FirmTopics & 14.660 & 12.250 & 10.735 \\
\#IndustryTopics & 6.365 & 5.035 & 4.190 \\
\#MacroTopics & 6.540 & 5.335 & 4.525 \\
\#HistMethods & 1.225 & 1.180 & 1.010 \\
\#ValMethods & 4.120 & 3.635 & 3.260 \\
\#FcMethods & 3.815 & 3.115 & 2.725 \\
\bottomrule
\end{tabular}

\medskip

\textit{Panel B: Correlation}

\begin{tabular}{lccc}
\toprule
Correlation & 1,500 \& 2,000 & 2,000 \& 2,500 & 1,500 \& 2,500 \\
\midrule
\#TextSources & 0.888 & 0.884 & 0.838 \\
\#FigureSources & 0.692 & 0.732 & 0.671 \\
\#TableSources & 0.788 & 0.797 & 0.726 \\
\#FirmTopics & 0.920 & 0.917 & 0.897 \\
\#IndustryTopics & 0.878 & 0.835 & 0.836 \\
\#MacroTopics & 0.881 & 0.888 & 0.852 \\
\#HistMethods & 0.452 & 0.475 & 0.545 \\
\#ValMethods & 0.713 & 0.756 & 0.704 \\
\#FcMethods & 0.736 & 0.692 & 0.746 \\
\bottomrule
\end{tabular}
\end{adjustwidth}
\end{table}
\clearpage

\subsection{Validating FactSet usage and AI platform adoption}\label{app:validate_factset}
\begin{figure}[htbp]
\renewcommand{\thefigure}{B.7.1}
\caption{Example of Citations}
\label{fig:citation_examples}
\includegraphics[width=\linewidth]{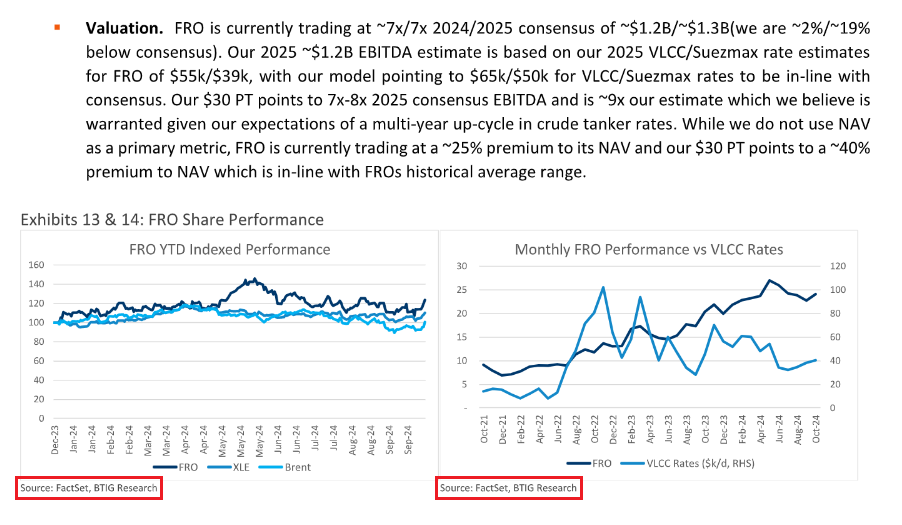}
\end{figure}

We make sure that we capture actual data citations, which would indicate actual usage of the platform, rather than mere mentions. To identify citations, we first use the PyPDF2 Python library to extract the raw text from analyst report files, which extracts PDF content line by line. We then convert each line of text to lowercase and retain only those lines that contain the word ``source,'' as analysts routinely provide data sources under tables, graphs, and statements in the format of ``Source: [source name(s)].'' Within these retained lines, we identify each line that also contains the word ``factset'' as an incidence of citation. Figure~\ref{fig:citation_examples} shows examples of such citations. Citations to other data vendors are identified in the same manner.

The remaining question is: Do post-period FactSet citations indicate actual AI use? As we do not directly observe analysts' usage of the GenAI interface: analysts could be using FactSet without using its AI interface. We address this issue in two ways.

\textit{Institutional background.} To substantiate our claim that FactSet use is related to GenAI use, we provide additional industry insights drawn from private interviews with a product manager and an AI engineer at FactSet, and a junior analyst who uses FactSet AI for research. We also draw on public disclosures and earnings calls from FactSet Research Systems Inc.

First, consistent with our interpretation, generative AI is not a peripheral feature but a core strategic differentiator for FactSet. The company's fiscal 2025 annual report explicitly identifies AI as one of three core strategic pillars, stating that ``AI is embedded across our offerings to enhance data discovery, automate routine workflows and improve the speed and accuracy of client insights'' (FactSet 10-K 2025, p. 5). In the company's Q4 2024 earnings call, CEO Phil Snow emphasized that FactSet's ``federated approach'' to AI, particularly through its conversational API (i.e., FactSet Mercury), has become a key competitive advantage in displacing incumbent providers.\footnote{See \href{factset_transcript}{https://seekingalpha.com/symbol/FDS/earnings/transcripts}.} Industry analysts have echoed this view, with JPMorgan estimating that FactSet's AI tools could boost EBITDA margins by 3-5\% over the next three years.\footnote{See \href{factset_report_1}{https://www.ainvest.com/news/factset-q4-revenue-beat-signal-ai-driven-financial-data-demand-2509/} and \href{factset_report_2}{https://www.ainvest.com/news/factset-dividend-aristocrat-attractive-valuation-strong-growth-catalysts-2509/}.} Our interviews with FactSet managers confirm that the Mercury platform is specifically designed for bankers to streamline workflows and generate reusable, AI-created charts, an area where FactSet has made substantial, targeted investments. Unlike generic AI tools that merely summarize publicly available information, FactSet Mercury is integrated with proprietary financial and market data, enabling it to deliver differentiated value that analysts cannot obtain from widely accessible models like ChatGPT or Gemini.

Second, these AI capabilities have generated significant user adoption and revenue. By the end of fiscal 2024, FactSet's AI product preview program had expanded to over 50 clients across banking, buy-side, and wealth management. Management has reported ``meaningful usage of each [AI product] by clients, which is starting to drive incremental ASV [(annual subscription value)] and improve retention,'' according to the earnings call transcript. These metrics confirm that AI is more than a marketing talking point but a driver of commercial outcomes. Our interview with the junior analyst who uses FactSet AI daily further corroborates this: the analyst reported that Mercury's ability to convert Bloomberg codes into FactSet codes and directly generate data queries and charts has made it a highly practical and frequently used feature. The analyst also noted that FactSet AI supports direct data integration into a knowledge base, enabling the analyst to ask specific questions and receive evidence-backed answers. In model-building workflows, where FactSet's raw disclosure data is considered ``best practice'' within the firm, AI has further accelerated the research process. These qualitative insights align with and reinforce our empirical findings.

\textit{Detection of AI writing.} To further substantiate the connection between FactSet usage and AI usage, we test whether FactSet-assisted reports exhibit a higher likelihood of AI writing. The idea is that if analysts use FactSet's AI interface, particularly its pitchbook creation tool, to directly generate part of their report content, we might observe that reports developed using FactSet exhibit more AI-like writing style after its integration, compared to reports developed without FactSet. Empirically, we measure reports' AI-like writing style and regress it on $FactSet \times Post$ in our standard model.

We first test whether reports developed using FactSet AI have lower perplexity and burstiness. Perplexity and burstiness are commonly used textual features to detect machine-generated text, employed in many settings including financial text analysis \citep{bertomeu2025impact, blankespoor2026generative_reporting}. Perplexity ($PPL$) is computed as the average negative log-likelihood (NLL) of all tokens in the entire report. Intuitively, lower perplexity indicates that the report's word choice and discourse patterns are more predictable. Machine-generated texts tend to follow these patterns and therefore exhibit lower perplexity than human-written texts. Burstiness ($BST$) is computed as the standard deviation of sentence-level perplexity. Higher burstiness indicates larger within-report variation in predictability (i.e., different sentences differ more in how "model-expected" they are). Machine-generated texts often have a smoother and more uniform flow, hence lower burstiness, compared to human-written texts.

In a second approach, we use GPTZero, a widely available AI detection tool,\footnote{For details, see \href{https://gptzero.me/}{https://gptzero.me/}.} to process a small random sample (approximately 2,000) of analyst reports in our sample, to determine whether generative AI is plausibly involved in the production of the reports. If so, $AI$ takes the value of one; otherwise, $AI$ takes the value of zero.

We report the results in Table~\ref{tab:ai_writing_style}. The coefficient on $FactSet \times Post$ is negative ($t = -1.77$) when the dependent variable is $BST$, and positive ($t = 1.89$) when the dependent variable is $AI$. This means that reports developed using FactSet exhibit a stronger AI writing style after the platform's AI integration.

We interpret these results as corroborative, rather than definitive, evidence that post-integration FactSet association is related to greater GenAI involvement. The test is necessarily one-sided: Mercury can affect data retrieval, visualization, and workflow without leaving an AI-writing footprint, so the absence of a detectable writing-style shift would not imply non-use. More importantly, our main treatment measure remains FactSet exposure after AI integration, not directly observed Mercury prompts or clicks.

\begin{table}[htbp]
\begin{adjustwidth}{-1cm}{-1cm}
\renewcommand{\thetable}{B.7.1}
\centering
\caption{AI Writing Style in FactSet-Assisted Reports}
\label{tab:ai_writing_style}
\begin{tabular}{lccc}
\toprule
& PPL & BST & AI \\
\midrule
FactSet & -3.033*** & 0.013 & -0.117 \\
& (-2.82) & (1.50) & (-1.07) \\
FactSet$\times$Post & 1.432 & -0.023* & 0.212* \\
& (1.46) & (-1.77) & (1.89) \\
Team-Career & 1.042 & 0.007 & -0.108 \\
& (0.49) & (0.19) & (-0.58) \\
Firm-Experience & 0.076 & -0.000 & -0.018 \\
& (0.13) & (-0.01) & (-0.31) \\
IT-Experience & -2.156** & -0.001 & 0.114** \\
& (-2.34) & (-0.08) & (2.36) \\
Elite-Education & 2.127*** & -0.003 & -0.056 \\
& (2.92) & (-0.32) & (-0.82) \\
Brokerage-Size & 4.111** & 0.020 & -0.224 \\
& (2.28) & (0.81) & (-1.57) \\
Team-Size & -0.474 & 0.004 & -0.002 \\
& (-0.79) & (0.58) & (-0.06) \\
Forecast-Frequency & 1.093* & -0.004 & 0.044 \\
& (1.96) & (-0.52) & (1.04) \\
Firms-Followed & -0.340 & 0.024* & -0.005 \\
& (-0.35) & (1.68) & (-0.04) \\
\midrule
Controls & Yes & Yes & Yes \\
Report Date FE & Yes & Yes & Yes \\
Firm-Year FE & Yes & Yes & Yes \\
Analyst FE & Yes & Yes & Yes \\
Broker-Year FE & Yes & Yes & Yes \\
\midrule
N & 41,556 & 40,833 & 1,009 \\
Adjusted R$^2$ & 0.618 & 0.355 & 0.726 \\
\bottomrule
\end{tabular}
\end{adjustwidth}
\end{table}
\clearpage

\subsection{Robustness tests}\label{app:robustness}

\begin{table}[htbp]
\renewcommand{\thetable}{B.8.1}
\centering
\caption{Alternative Measures of Forecast Accuracy}
\label{tab:accuracy_measures}
\setlength{\tabcolsep}{4pt}
\caption*{\footnotesize
This table reports results using alternative measures of forecast accuracy.
The dependent variable in the first column is forecast error deflated by the firm's beginning-of-year stock price, multiplied by 100 \citep{dechow2012analysts}.
The second column uses proportional forecast error, calculated as the analyst's forecast error scaled by the mean forecast error of all analysts following the same firm in the past 90 days, multiplied by $-1$ so that higher values indicate greater accuracy \citep{clement1999analyst}.
The third column measures accuracy as the maximum forecast error of all analysts following the same firm in the same year less the analyst's own forecast error, scaled by the range of all analysts' forecast errors \citep{clement2005financial}.
All regressions include report date, firm-year, analyst, and broker-year fixed effects, with standard errors clustered at the analyst and firm levels.
$t$-statistics are reported in parentheses below coefficient estimates.
$^{***}$, $^{**}$, and $^{*}$ denote significance at the 1\%, 5\%, and 10\% levels, respectively.
}
\begin{tabular}{lccc}
\toprule
& \multicolumn{1}{c}{Accuracy} & \multicolumn{1}{c}{Accuracy} & \multicolumn{1}{c}{Accuracy} \\
& \multicolumn{1}{c}{Price Deflated} & \multicolumn{1}{c}{Mean Deflated} & \multicolumn{1}{c}{Range Deflated} \\
\midrule
FactSet & 0.012 & -0.007 & 0.003 \\
& (0.28) & (-0.42) & (0.46) \\
FactSet$\times$Post & -0.149*** & -0.057** & -0.016** \\
& (-2.83) & (-2.32) & (-1.99) \\
\midrule
Controls & Yes & Yes & Yes \\
Report Date FE & Yes & Yes & Yes \\
Firm-Year FE & Yes & Yes & Yes \\
Analyst FE & Yes & Yes & Yes \\
Broker-Year FE & Yes & Yes & Yes \\
\midrule
N & 32,969 & 33,870 & 40,747 \\
Adjusted R$^2$ & 0.280 & 0.219 & 0.539 \\
\bottomrule
\end{tabular}
\end{table}

\begin{table}[htbp]
\centering
\renewcommand{\thetable}{B.8.2}
\caption{Report Length}
\label{tab:length_control}
\setlength{\tabcolsep}{4pt}
\caption*{\footnotesize
This table examines the sensitivity of our main results to controlling for report length.
Panel A includes the natural log of word count (\textit{Length}) as an additional control variable.
Panel B replaces the length control with length percentile fixed effects, comparing reports within the same length percentile.
The dependent variables are nine measures of report richness: information sources (\textit{\#TextSources}, \textit{\#FigureSources}, \textit{\#TableSources}), 
topic coverage (\textit{\#FirmTopics}, \textit{\#IndustryTopics}, \textit{\#MacroTopics}), and analytic methods (\textit{\#HistMethods}, \textit{\#ValMethods}, \textit{\#FcMethods}).
All regressions include report date, firm-year, analyst, and broker-year fixed effects, with standard errors clustered at the analyst and firm levels.
$t$-statistics are reported in parentheses below coefficient estimates.
$^{***}$, $^{**}$, and $^{*}$ denote significance at the 1\%, 5\%, and 10\% levels, respectively.
}
\scriptsize

\textit{Panel A: Control for report length}
\medskip
\begin{tabular}{lccccccccc}
\toprule
& \#TextSources & \#FigureSources & \#TableSources & \#FirmTopics & \#IndustryTopics & \#MacroTopics & \#HistMethods & \#ValMethods & \#FcMethods \\
\midrule
FactSet & -0.258 & 0.095 & 0.603** & -0.331 & -0.305* & -0.023 & 0.134*** & 0.045 & 0.135 \\
& (-0.83) & (0.70) & (2.33) & (-0.87) & (-1.66) & (-0.12) & (3.07) & (0.50) & (1.14) \\
FactSet$\times$Post & 1.476*** & 0.689*** & 1.216*** & 1.823*** & 0.891*** & 0.978*** & 0.094* & 0.341*** & 0.718*** \\
& (4.65) & (4.31) & (4.10) & (4.32) & (4.98) & (4.83) & (1.87) & (3.72) & (5.15) \\
Length & 5.716*** & 2.027*** & 3.357*** & 7.586*** & 3.574*** & 3.749*** & 0.547*** & 1.365*** & 2.103*** \\
& (17.42) & (13.87) & (12.84) & (18.21) & (18.30) & (15.32) & (12.90) & (16.48) & (16.10) \\
\midrule
Controls & Yes & Yes & Yes & Yes & Yes & Yes & Yes & Yes & Yes \\
Report Date FE & Yes & Yes & Yes & Yes & Yes & Yes & Yes & Yes & Yes \\
Firm-Year FE & Yes & Yes & Yes & Yes & Yes & Yes & Yes & Yes & Yes \\
Analyst FE & Yes & Yes & Yes & Yes & Yes & Yes & Yes & Yes & Yes \\
Broker-Year FE & Yes & Yes & Yes & Yes & Yes & Yes & Yes & Yes & Yes \\
\midrule
N & 41,367 & 41,367 & 41,367 & 41,367 & 41,367 & 41,367 & 41,367 & 41,367 & 41,367 \\
Adjusted R$^2$ & 0.618 & 0.556 & 0.567 & 0.663 & 0.628 & 0.632 & 0.377 & 0.437 & 0.575 \\
\bottomrule
\end{tabular}

\bigskip

\textit{Panel B: Length percentile fixed effects}
\medskip
\begin{tabular}{lccccccccc}
\toprule
& \#TextSources & \#FigureSources & \#TableSources & \#FirmTopics & \#IndustryTopics & \#MacroTopics & \#HistMethods & \#ValMethods & \#FcMethods \\
\midrule
FactSet & 0.677*** & 0.407*** & 1.076*** & 0.876*** & 0.275** & 0.593*** & 0.208*** & 0.260*** & 0.443*** \\
& (2.94) & (3.44) & (4.87) & (3.23) & (2.16) & (3.85) & (5.33) & (3.17) & (4.13) \\
FactSet$\times$Post & 0.911*** & 0.482*** & 0.902*** & 1.082*** & 0.542*** & 0.598*** & 0.043 & 0.220** & 0.516*** \\
& (3.27) & (3.34) & (3.19) & (2.75) & (3.38) & (3.34) & (0.90) & (2.48) & (4.00) \\
\midrule
Controls & Yes & Yes & Yes & Yes & Yes & Yes & Yes & Yes & Yes \\
Report Date FE & Yes & Yes & Yes & Yes & Yes & Yes & Yes & Yes & Yes \\
Firm-Year FE & Yes & Yes & Yes & Yes & Yes & Yes & Yes & Yes & Yes \\
Analyst FE & Yes & Yes & Yes & Yes & Yes & Yes & Yes & Yes & Yes \\
Broker-Year FE & Yes & Yes & Yes & Yes & Yes & Yes & Yes & Yes & Yes \\
Length \% FE & Yes & Yes & Yes & Yes & Yes & Yes & Yes & Yes & Yes \\
\midrule
N & 41,367 & 41,367 & 41,367 & 41,367 & 41,367 & 41,367 & 41,367 & 41,367 & 41,367 \\
Adjusted R$^2$ & 0.704 & 0.612 & 0.616 & 0.759 & 0.713 & 0.709 & 0.407 & 0.482 & 0.628 \\
\bottomrule
\end{tabular}
\end{table}

\begin{table}[htbp]
\centering
\renewcommand{\thetable}{B.8.3}
\caption{Entropy Balancing Robustness}
\label{tab:entropy_balancing}
\setlength{\tabcolsep}{4pt}
\caption*{\footnotesize
This table replicates our main results using entropy balancing to reweight the sample.
Entropy balancing ensures that the covariate distributions of FactSet and non-FactSet reports are balanced in their first moments (means), effectively approximating the effect of one-on-one propensity score matching.
The dependent variables are nine measures of report richness (top panel) and three measures of report quality (bottom panel).
All regressions include controls, report date, firm-year, analyst, and broker-year fixed effects, with standard errors clustered at the analyst and firm levels.
$t$-statistics are reported in parentheses below coefficient estimates.
$^{***}$, $^{**}$, and $^{*}$ denote significance at the 1\%, 5\%, and 10\% levels, respectively.
}
\scriptsize

\begin{tabular}{lccccccccc}
\toprule
& \#TextSources & \#FigureSources & \#TableSources & \#FirmTopics & \#IndustryTopics & \#MacroTopics & \#HistMethods & \#ValMethods & \#FcMethods \\
\midrule
FactSet & 2.058*** & 0.897*** & 2.004*** & 2.614*** & 1.073*** & 1.444*** & 0.362*** & 0.607*** & 0.997*** \\
& (7.10) & (7.25) & (8.91) & (7.57) & (6.27) & (7.97) & (8.41) & (6.49) & (7.71) \\
FactSet$\times$Post & 1.888*** & 0.804*** & 1.360*** & 2.611*** & 1.299*** & 1.330*** & 0.153** & 0.465*** & 0.839*** \\
& (3.67) & (3.72) & (3.67) & (4.02) & (4.42) & (4.37) & (2.36) & (3.70) & (4.05) \\
\midrule
Controls & Yes & Yes & Yes & Yes & Yes & Yes & Yes & Yes & Yes \\
Report Date FE & Yes & Yes & Yes & Yes & Yes & Yes & Yes & Yes & Yes \\
Firm-Year FE & Yes & Yes & Yes & Yes & Yes & Yes & Yes & Yes & Yes \\
Analyst FE & Yes & Yes & Yes & Yes & Yes & Yes & Yes & Yes & Yes \\
Broker-Year FE & Yes & Yes & Yes & Yes & Yes & Yes & Yes & Yes & Yes \\
\midrule
N & 41,391 & 41,391 & 41,391 & 41,391 & 41,391 & 41,391 & 41,391 & 41,391 & 41,391 \\
Adjusted R$^2$ & 0.432 & 0.470 & 0.491 & 0.441 & 0.440 & 0.484 & 0.357 & 0.345 & 0.472 \\
\bottomrule
\end{tabular}

\begin{tabular}{lccc}
\toprule
& Prompt & Accuracy & AbnVol \\
\midrule
FactSet & -0.037*** & 0.008 & -0.005 \\
& (-2.66) & (0.20) & (-0.14) \\
FactSet$\times$Post & 0.025 & -0.201*** & -0.143*** \\
& (1.32) & (-3.13) & (-3.16) \\
\midrule
Controls & Yes & Yes & Yes \\
Report Date FE & Yes & Yes & Yes \\
Firm-Year FE & Yes & Yes & Yes \\
Analyst FE & Yes & Yes & Yes \\
Broker-Year FE & Yes & Yes & Yes \\
\midrule
N & 41,619 & 33,079 & 39,740 \\
Adjusted R$^2$ & 0.571 & 0.379 & 0.534 \\
\bottomrule
\end{tabular}
\end{table}

\begin{table}[htbp]
\centering
\renewcommand{\thetable}{B.8.4}
\caption{Alternative Fixed Effects Specifications}
\label{tab:finer_fixed_effects}
\setlength{\tabcolsep}{4pt}
\caption*{\footnotesize
This table examines the robustness of our main results to alternative fixed effects structures.
Panel A includes month-level fixed effects for firms, analysts, and brokers, in addition to report date fixed effects.
Panel B replaces individual analyst and firm fixed effects with analyst-firm fixed effects, and replaces year-level fixed effects with firm-year and broker-year fixed effects.
The dependent variables are nine measures of report richness (top panels) and three measures of report quality (bottom panels).
All specifications include controls and the fixed effects as indicated, with standard errors clustered at the analyst and firm levels.
$t$-statistics are reported in parentheses below coefficient estimates.
$^{***}$, $^{**}$, and $^{*}$ denote significance at the 1\%, 5\%, and 10\% levels, respectively.
}
\scriptsize

\textit{Panel A: Month fixed effects}
\medskip
\begin{tabular}{lccccccccc}
\toprule
& \#TextSources & \#FigureSources & \#TableSources & \#FirmTopics & \#IndustryTopics & \#MacroTopics & \#HistMethods & \#ValMethods & \#FcMethods \\
\midrule
FactSet$\times$Post & 1.720** & 1.036*** & 1.496** & 2.562*** & 1.197*** & 1.394*** & 0.045 & 0.540** & 0.943*** \\
& (2.20) & (3.35) & (2.56) & (2.66) & (2.96) & (3.16) & (0.44) & (2.28) & (2.98) \\
\midrule
Controls & Yes & Yes & Yes & Yes & Yes & Yes & Yes & Yes & Yes \\
Report Date FE & Yes & Yes & Yes & Yes & Yes & Yes & Yes & Yes & Yes \\
Firm-Month FE & Yes & Yes & Yes & Yes & Yes & Yes & Yes & Yes & Yes \\
Analyst-Month FE & Yes & Yes & Yes & Yes & Yes & Yes & Yes & Yes & Yes \\
Broker-Month FE & Yes & Yes & Yes & Yes & Yes & Yes & Yes & Yes & Yes \\
\midrule
N & 28,763 & 28,763 & 28,763 & 28,763 & 28,763 & 28,763 & 28,763 & 28,763 & 28,763 \\
Adjusted R$^2$ & 0.332 & 0.396 & 0.390 & 0.341 & 0.351 & 0.383 & 0.233 & 0.217 & 0.369 \\
\bottomrule
\end{tabular}

\begin{tabular}{lccc}
\toprule
& Prompt & Accuracy & AbnVol \\
\midrule
FactSet$\times$Post & -0.011 & -0.184*** & -0.025 \\
& (-0.62) & (-2.90) & (-0.71) \\
\midrule
Controls & Yes & Yes & Yes \\
Report Date FE & Yes & Yes & Yes \\
Firm-Month FE & Yes & Yes & Yes \\
Analyst-Month FE & Yes & Yes & Yes \\
Broker-Month FE & Yes & Yes & Yes \\
\midrule
N & 28,934 & 21,623 & 27,657 \\
Adjusted R$^2$ & 0.704 & 0.694 & 0.824 \\
\bottomrule
\end{tabular}

\bigskip

\textit{Panel B: Analyst-firm fixed effects}
\medskip
\begin{tabular}{lccccccccc}
\toprule
& \#TextSources & \#FigureSources & \#TableSources & \#FirmTopics & \#IndustryTopics & \#MacroTopics & \#HistMethods & \#ValMethods & \#FcMethods \\
\midrule
FactSet$\times$Post & 2.242*** & 0.995*** & 1.749*** & 2.827*** & 1.382*** & 1.534*** & 0.164** & 0.507*** & 1.038*** \\
& (4.19) & (4.46) & (4.44) & (4.13) & (4.51) & (4.73) & (2.48) & (3.89) & (5.02) \\
\midrule
Controls & Yes & Yes & Yes & Yes & Yes & Yes & Yes & Yes & Yes \\
Report Date FE & Yes & Yes & Yes & Yes & Yes & Yes & Yes & Yes & Yes \\
Firm-Year FE & Yes & Yes & Yes & Yes & Yes & Yes & Yes & Yes & Yes \\
Analyst-Firm FE & Yes & Yes & Yes & Yes & Yes & Yes & Yes & Yes & Yes \\
Broker-Year FE & Yes & Yes & Yes & Yes & Yes & Yes & Yes & Yes & Yes \\
\midrule
N & 41,060 & 41,060 & 41,060 & 41,060 & 41,060 & 41,060 & 41,060 & 41,060 & 41,060 \\
Adjusted R$^2$ & 0.337 & 0.364 & 0.384 & 0.345 & 0.351 & 0.389 & 0.243 & 0.240 & 0.362 \\
\bottomrule
\end{tabular}

\begin{tabular}{lccc}
\toprule
& Prompt & Accuracy & AbnVol \\
\midrule
FactSet$\times$Post & 0.030* & -0.097* & -0.126*** \\
& (1.73) & (-1.67) & (-2.99) \\
\midrule
Controls & Yes & Yes & Yes \\
Report Date FE & Yes & Yes & Yes \\
Firm-Year FE & Yes & Yes & Yes \\
Analyst-Firm FE & Yes & Yes & Yes \\
Broker-Year FE & Yes & Yes & Yes \\
\midrule
N & 41,287 & 32,673 & 39,402 \\
Adjusted R$^2$ & 0.509 & 0.244 & 0.449 \\
\bottomrule
\end{tabular}
\end{table}

\begin{table}[htbp]
\centering
\renewcommand{\thetable}{B.8.5}
\caption{Broker- and Firm-Level Clustering}
\label{tab:broker_cluster}
\setlength{\tabcolsep}{4pt}
\caption*{\footnotesize
This table examines the robustness of our inference to alternative clustering of standard errors.
All coefficients are estimated using the same specification as our main tests, but standard errors are clustered at both the broker and firm levels (two-way clustering).
The dependent variables are nine measures of report richness (top panel) and three measures of report quality (bottom panel).
All regressions include controls, report date, firm-year, analyst, and broker-year fixed effects.
$t$-statistics are reported in parentheses below coefficient estimates.
$^{***}$, $^{**}$, and $^{*}$ denote significance at the 1\%, 5\%, and 10\% levels, respectively.
}
\scriptsize
\begin{tabular}{lccccccccc}
\toprule
& \#TextSources & \#FigureSources & \#TableSources & \#FirmTopics & \#IndustryTopics & \#MacroTopics & \#HistMethods & \#ValMethods & \#FcMethods \\
\midrule
FactSet & 1.721*** & 0.790*** & 1.760*** & 2.293*** & 0.921*** & 1.265*** & 0.324*** & 0.515*** & 0.861*** \\
& (3.54) & (3.91) & (4.24) & (3.85) & (3.46) & (4.09) & (5.16) & (4.10) & (4.32) \\
FactSet$\times$Post & 2.273*** & 0.977*** & 1.690*** & 2.874*** & 1.394*** & 1.506*** & 0.169** & 0.533*** & 1.010*** \\
& (3.36) & (3.51) & (3.19) & (3.56) & (4.29) & (3.85) & (2.46) & (3.89) & (4.62) \\
\midrule
Controls & Yes & Yes & Yes & Yes & Yes & Yes & Yes & Yes & Yes \\
Report Date FE & Yes & Yes & Yes & Yes & Yes & Yes & Yes & Yes & Yes \\
Firm-Year FE & Yes & Yes & Yes & Yes & Yes & Yes & Yes & Yes & Yes \\
Analyst FE & Yes & Yes & Yes & Yes & Yes & Yes & Yes & Yes & Yes \\
Broker-Year FE & Yes & Yes & Yes & Yes & Yes & Yes & Yes & Yes & Yes \\
\midrule
N & 41,391 & 41,391 & 41,391 & 41,391 & 41,391 & 41,391 & 41,391 & 41,391 & 41,391 \\
Adjusted R$^2$ & 0.372 & 0.398 & 0.415 & 0.382 & 0.386 & 0.422 & 0.286 & 0.283 & 0.396 \\
\bottomrule
\end{tabular}
\medskip
\begin{tabular}{lccc}
\toprule
& Prompt & Accuracy & AbnVol \\
\midrule
FactSet & -0.040*** & 0.007 & 0.001 \\
& (-3.09) & (0.27) & (0.03) \\
FactSet$\times$Post & 0.033* & -0.157** & -0.127*** \\
& (1.73) & (-2.80) & (-2.89) \\
\midrule
Controls & Yes & Yes & Yes \\
Report Date FE & Yes & Yes & Yes \\
Firm-Year FE & Yes & Yes & Yes \\
Analyst FE & Yes & Yes & Yes \\
Broker-Year FE & Yes & Yes & Yes \\
\midrule
N & 41,619 & 33,079 & 39,740 \\
Adjusted R$^2$ & 0.513 & 0.279 & 0.478 \\
\bottomrule
\end{tabular}
\end{table}

\begin{table}[htbp]
\centering
\renewcommand{\thetable}{B.8.6}
\caption{Alternative Treatment Definitions}
\label{tab:alternative_treatment}
\setlength{\tabcolsep}{4pt}
\caption*{\footnotesize
This table replicates our main tests using alternative treatment definitions to address potential measurement error in \textit{FactSet} citations.
Panel A defines treatment at the analyst level: \textit{FactSet} equals one for analysts who have cited FactSet at least once in the sample period, and zero otherwise.
Panel B defines treatment at the broker level: \textit{FactSet} equals one for brokerage firms whose analysts have cited FactSet at least once, and zero otherwise.
Switching analysts and brokers (i.e., those that cite FactSet in one year but not in another) are dropped from each panel, respectively. The dependent variables are nine measures of report richness (Panels A and B, top) and three measures of report quality (Panels A and B, bottom).
All regressions include report date and firm-year fixed effects, with standard errors clustered at the analyst and firm levels.
$t$-statistics are reported in parentheses below coefficient estimates.
$^{***}$, $^{**}$, and $^{*}$ denote significance at the 1\%, 5\%, and 10\% levels, respectively.
}
\scriptsize

\textit{Panel A: Treatment at analyst level}
\medskip
\begin{tabular}{lccccccccc}
\toprule
& \#TextSources & \#FigureSources & \#TableSources & \#FirmTopics & \#IndustryTopics & \#MacroTopics & \#HistMethods & \#ValMethods & \#FcMethods \\
\midrule
FactSet$\times$Post & 1.354*** & 0.406*** & 1.051*** & 1.692*** & 0.757*** & 0.891*** & 0.128*** & 0.340*** & 0.705*** \\
& (4.30) & (3.08) & (4.53) & (4.68) & (4.35) & (4.52) & (2.75) & (4.12) & (5.29) \\
\midrule
Controls & Yes & Yes & Yes & Yes & Yes & Yes & Yes & Yes & Yes \\
Report Date FE & Yes & Yes & Yes & Yes & Yes & Yes & Yes & Yes & Yes \\
Firm-Year FE & Yes & Yes & Yes & Yes & Yes & Yes & Yes & Yes & Yes \\
\midrule
N & 36,998 & 36,998 & 36,998 & 36,998 & 36,998 & 36,998 & 36,998 & 36,998 & 36,998 \\
Adjusted R$^2$ & 0.289 & 0.293 & 0.313 & 0.299 & 0.303 & 0.332 & 0.204 & 0.234 & 0.289 \\
\bottomrule
\end{tabular}

\begin{tabular}{lccc}
\toprule
& Prompt & Accuracy & AbnVol \\
\midrule
FactSet$\times$Post & 0.032* & -0.100** & -0.016 \\
& (1.94) & (-2.38) & (-0.71) \\
\midrule
Controls & Yes & Yes & Yes \\
Report Date FE & Yes & Yes & Yes \\
Firm-Year FE & Yes & Yes & Yes \\
\midrule
N & 37,210 & 29,639 & 35,586 \\
Adjusted R$^2$ & 0.475 & 0.272 & 0.477 \\
\bottomrule
\end{tabular}

\bigskip

\textit{Panel B: Treatment at broker level}
\medskip
\begin{tabular}{lccccccccc}
\toprule
& \#TextSources & \#FigureSources & \#TableSources & \#FirmTopics & \#IndustryTopics & \#MacroTopics & \#HistMethods & \#ValMethods & \#FcMethods \\
\midrule
FactSet$\times$Post & 1.345*** & 0.470*** & 1.205*** & 1.725*** & 0.714*** & 0.883*** & 0.117*** & 0.366*** & 0.649*** \\
& (4.45) & (3.75) & (5.05) & (4.83) & (4.21) & (4.52) & (2.61) & (4.67) & (4.95) \\
\midrule
Controls & Yes & Yes & Yes & Yes & Yes & Yes & Yes & Yes & Yes \\
Report Date FE & Yes & Yes & Yes & Yes & Yes & Yes & Yes & Yes & Yes \\
Firm-Year FE & Yes & Yes & Yes & Yes & Yes & Yes & Yes & Yes & Yes \\
\midrule
N & 39,009 & 39,009 & 39,009 & 39,009 & 39,009 & 39,009 & 39,009 & 39,009 & 39,009 \\
Adjusted R$^2$ & 0.278 & 0.289 & 0.306 & 0.288 & 0.292 & 0.322 & 0.195 & 0.228 & 0.280 \\
\bottomrule
\end{tabular}

\begin{tabular}{lccc}
\toprule
& Prompt & Accuracy & AbnVol \\
\midrule
FactSet$\times$Post & 0.034** & -0.074* & -0.024 \\
& (2.34) & (-1.73) & (-1.16) \\
\midrule
Controls & Yes & Yes & Yes \\
Report Date FE & Yes & Yes & Yes \\
Firm-Year FE & Yes & Yes & Yes \\
\midrule
N & 39,232 & 31,249 & 37,458 \\
Adjusted R$^2$ & 0.473 & 0.277 & 0.480 \\
\bottomrule
\end{tabular}
\end{table}

\begin{table}[htbp]
\centering
\renewcommand{\thetable}{B.8.7}
\caption{Dynamic Effects of AI Platform Adoption (Firm-Quarter FE)}
\label{tab:parallel_trends_add}
\setlength{\tabcolsep}{4pt}
\caption*{\footnotesize
This table reports dynamic effects of FactSet AI platform adoption on analyst report content, using firm-quarter fixed effects to absorb time-varying firm-level shocks.
The dependent variables are nine measures of report richness (top panel) and three measures of report quality (bottom panel).
$Y2023Q1$--$Y2023Q4$ represent the four quarters before AI introduction, and $Y2024Q1$--$Y2024Q4$ represent the four quarters after.
All regressions include controls, report date, firm-quarter, analyst, and broker-year fixed effects, with standard errors clustered at the analyst and firm levels.
$t$-statistics are reported in parentheses below coefficient estimates.
$^{***}$, $^{**}$, and $^{*}$ denote significance at the 1\%, 5\%, and 10\% levels, respectively.
}
\scriptsize

\begin{tabular}{lccccccccc}
\toprule
& \#TextSources & \#FigureSources & \#TableSources & \#FirmTopics & \#IndustryTopics & \#MacroTopics & \#HistMethods & \#ValMethods & \#FcMethods \\
\midrule
FactSet$\times$Y2023Q1 & 0.642 & 0.348* & 0.653** & 0.866 & 0.260 & 0.380 & 0.119* & 0.153 & 0.291 \\
& (1.39) & (1.91) & (2.14) & (1.37) & (0.87) & (1.20) & (1.68) & (1.15) & (1.43) \\
FactSet$\times$Y2023Q2 & 0.323 & 0.152 & 0.083 & 0.583 & 0.215 & 0.139 & 0.040 & 0.127 & 0.033 \\
& (0.80) & (0.92) & (0.32) & (1.19) & (0.83) & (0.48) & (0.63) & (0.94) & (0.18) \\
FactSet$\times$Y2023Q3 & 0.061 & 0.051 & 0.050 & -0.168 & -0.243 & -0.341 & 0.018 & -0.144 & -0.316 \\
& (0.11) & (0.24) & (0.16) & (-0.25) & (-0.74) & (-1.07) & (0.25) & (-0.94) & (-1.53) \\
FactSet$\times$Y2023Q4 & -0.488 & -0.050 & -0.196 & -0.358 & -0.339 & -0.576* & -0.085 & -0.023 & -0.133 \\
& (-0.98) & (-0.25) & (-0.62) & (-0.58) & (-1.20) & (-1.82) & (-1.22) & (-0.17) & (-0.70) \\
FactSet$\times$Y2024Q1 & 2.101*** & 0.868*** & 1.385*** & 2.844*** & 1.192*** & 1.191*** & 0.112 & 0.516*** & 0.820*** \\
& (3.74) & (3.55) & (3.39) & (3.89) & (3.65) & (3.28) & (1.46) & (3.38) & (3.62) \\
FactSet$\times$Y2024Q2 & 2.498*** & 1.181*** & 2.007*** & 3.199*** & 1.415*** & 1.536*** & 0.285*** & 0.487*** & 1.139*** \\
& (3.79) & (4.26) & (4.17) & (3.79) & (3.63) & (3.69) & (3.06) & (2.84) & (4.17) \\
FactSet$\times$Y2024Q3 & 2.098*** & 1.017*** & 2.057*** & 2.646*** & 1.285*** & 1.219*** & 0.155* & 0.451** & 1.025*** \\
& (3.00) & (3.90) & (4.22) & (3.10) & (3.36) & (3.07) & (1.80) & (2.48) & (3.89) \\
FactSet$\times$Y2024Q4 & 3.489*** & 1.191*** & 2.439*** & 4.230*** & 1.659*** & 1.813*** & 0.214* & 0.730*** & 1.176*** \\
& (3.79) & (3.25) & (3.60) & (3.76) & (3.35) & (3.26) & (1.71) & (3.11) & (3.08) \\
\midrule
Controls & Yes & Yes & Yes & Yes & Yes & Yes & Yes & Yes & Yes \\
Report Date FE & Yes & Yes & Yes & Yes & Yes & Yes & Yes & Yes & Yes \\
Firm-Quarter FE & Yes & Yes & Yes & Yes & Yes & Yes & Yes & Yes & Yes \\
Analyst FE & Yes & Yes & Yes & Yes & Yes & Yes & Yes & Yes & Yes \\
Broker-Year FE & Yes & Yes & Yes & Yes & Yes & Yes & Yes & Yes & Yes \\
\midrule
N & 37,098 & 37,098 & 37,098 & 37,098 & 37,098 & 37,098 & 37,098 & 37,098 & 37,098 \\
Adjusted R$^2$ & 0.353 & 0.389 & 0.401 & 0.364 & 0.372 & 0.407 & 0.280 & 0.269 & 0.385 \\
\bottomrule
\end{tabular}

\begin{tabular}{lccc}
\toprule
& Prompt & Accuracy & AbnVol \\
\midrule
FactSet$\times$Y2023Q1 & -0.001 & -0.030 & -0.007 \\
& (-0.04) & (-0.31) & (-0.14) \\
FactSet$\times$Y2023Q2 & 0.032 & -0.008 & -0.014 \\
& (1.43) & (-0.11) & (-0.34) \\
FactSet$\times$Y2023Q3 & 0.002 & 0.008 & -0.000 \\
& (0.08) & (0.11) & (-0.01) \\
FactSet$\times$Y2023Q4 & -0.001 & 0.046 & 0.032 \\
& (-0.02) & (0.62) & (0.73) \\
FactSet$\times$Y2024Q1 & 0.019 & -0.104 & -0.074* \\
& (0.87) & (-1.24) & (-1.67) \\
FactSet$\times$Y2024Q2 & 0.031 & -0.233*** & -0.067 \\
& (1.28) & (-3.09) & (-1.31) \\
FactSet$\times$Y2024Q3 & 0.028 & -0.233*** & -0.063 \\
& (1.08) & (-2.84) & (-1.31) \\
FactSet$\times$Y2024Q4 & 0.094*** & -0.203** & -0.023 \\
& (3.58) & (-2.19) & (-0.47) \\
\midrule
Controls & Yes & Yes & Yes \\
Report Date FE & Yes & Yes & Yes \\
Firm-Quarter FE & Yes & Yes & Yes \\
Analyst FE & Yes & Yes & Yes \\
Broker-Year FE & Yes & Yes & Yes \\
\midrule
N & 37,330 & 28,959 & 35,552 \\
Adjusted R$^2$ & 0.583 & 0.547 & 0.667 \\
\bottomrule
\end{tabular}
\end{table}

\end{document}